\pdfoutput=1
\documentclass[preprint,12pt]{elsarticle}

\usepackage[letterpaper,margin=2.2cm,footskip=0.9cm]{geometry}
\usepackage{amsmath,amssymb,bm}
\usepackage{graphicx}
\usepackage{placeins}
\usepackage{booktabs,tabularx,array,multirow}
\usepackage{subcaption}
\usepackage{xcolor}
\definecolor{linkblue}{RGB}{26,93,171}
\usepackage[colorlinks=true,linkcolor=linkblue,citecolor=linkblue,urlcolor=linkblue]{hyperref}
\AtBeginDocument{\hypersetup{colorlinks=true,linkcolor=linkblue,citecolor=linkblue,urlcolor=linkblue}}
\usepackage[capitalise,noabbrev]{cleveref}

\graphicspath{{figures/}{./}}

\newcolumntype{L}{>{\raggedright\arraybackslash}X}
\newcolumntype{P}[1]{>{\raggedright\arraybackslash}p{#1}}
\makeatletter
\def\ps@pprintTitle{%
  \let\@oddhead\@empty
  \let\@evenhead\@empty
  \def\@oddfoot{\scriptsize\itshape Preprint submitted to \ifx\@journal\@empty Elsevier\else\@journal\fi\hfill\today}%
  \let\@evenfoot\@oddfoot}
\makeatother

\newcommand{\Kn}{\mathrm{Kn}}
\newcommand{\dd}{\mathrm{d}}
\newcommand{\bx}{\bm{x}}
\newcommand{\bv}{\bm{v}}
\newcommand{\bc}{\bm{c}}
\newcommand{\bu}{\bm{u}}
\newcommand{\M}{\mathcal{M}}
\newcommand{\Rset}{\mathbb{R}}
\newcommand{\Res}{\mathcal{R}}
\newcommand{\Nth}{N_{\theta}}
\newcommand{\Bmix}{B}

\newcommand{\Pran}{\mathrm{Pr}}

\makeatletter
\newcommand{\sref}[1]{\@ifundefined{suppref@#1}{\textbf{??}\typeout{SUPPREF UNDEFINED: #1}}{\textcolor{linkblue}{\@nameuse{suppref@#1}}}}
\makeatother
\expandafter\def\csname suppref@app:optimisers\endcsname{S8.7}
\expandafter\def\csname suppref@app:settings_boltz\endcsname{S8.11}
\expandafter\def\csname suppref@app:settings_shock\endcsname{S8.10}
\expandafter\def\csname suppref@fig:S_f_all\endcsname{S5}
\expandafter\def\csname suppref@fig:S_molecules_M5\endcsname{S3}
\expandafter\def\csname suppref@fig:boltz_M5_diff\endcsname{S4}
\expandafter\def\csname suppref@fig:mono_errors\endcsname{S1}
\expandafter\def\csname suppref@fig:n2_errors\endcsname{S2}
\expandafter\def\csname suppref@supp:acceptance_details\endcsname{S8.6}
\expandafter\def\csname suppref@supp:boltz_M5\endcsname{S9.10}
\expandafter\def\csname suppref@supp:boltz_M8M10\endcsname{S10.3}
\expandafter\def\csname suppref@supp:boltz_details\endcsname{S9.8}
\expandafter\def\csname suppref@supp:boltz_twoscale\endcsname{S9.9}
\expandafter\def\csname suppref@supp:collision_split\endcsname{S6.3}
\expandafter\def\csname suppref@supp:collision_supp\endcsname{S6}
\expandafter\def\csname suppref@supp:disc_limits\endcsname{S11.2}
\expandafter\def\csname suppref@supp:disc_times\endcsname{S11.1}
\expandafter\def\csname suppref@supp:dvm\endcsname{S5}
\expandafter\def\csname suppref@supp:failures\endcsname{S3}
\expandafter\def\csname suppref@supp:form_details\endcsname{S8.3}
\expandafter\def\csname suppref@supp:inherited\endcsname{S8.1}
\expandafter\def\csname suppref@supp:loss_details\endcsname{S8.4}
\expandafter\def\csname suppref@supp:models_details\endcsname{S8.1}
\expandafter\def\csname suppref@supp:mono\endcsname{S1}
\expandafter\def\csname suppref@supp:mono_al\endcsname{S10.1}
\expandafter\def\csname suppref@supp:mono_m20_attempts\endcsname{S10.2}
\expandafter\def\csname suppref@supp:mono_runs\endcsname{S9.1}
\expandafter\def\csname suppref@supp:mono_txx\endcsname{S9.2}
\expandafter\def\csname suppref@supp:n2\endcsname{S2}
\expandafter\def\csname suppref@supp:n2_blind\endcsname{S9.4}
\expandafter\def\csname suppref@supp:n2_history\endcsname{S9.3}
\expandafter\def\csname suppref@supp:n2_param\endcsname{S9.7}
\expandafter\def\csname suppref@supp:n2_points\endcsname{S9.5}
\expandafter\def\csname suppref@supp:n2_weak\endcsname{S9.6}
\expandafter\def\csname suppref@supp:network_details\endcsname{S8.2}
\expandafter\def\csname suppref@supp:optimiser_details\endcsname{S8.8}
\expandafter\def\csname suppref@supp:points_details\endcsname{S8.5}
\expandafter\def\csname suppref@supp:reduced_checks\endcsname{S8.14}
\expandafter\def\csname suppref@supp:refs_details\endcsname{S8.12}
\expandafter\def\csname suppref@supp:threed\endcsname{S8.9}
\expandafter\def\csname suppref@supp:val_classical\endcsname{S12.3}
\expandafter\def\csname suppref@supp:val_hs\endcsname{S12.2}
\expandafter\def\csname suppref@supp:val_n2\endcsname{S12.6}
\expandafter\def\csname suppref@supp:val_vdf\endcsname{S12.5}
\expandafter\def\csname suppref@supp:validation\endcsname{S12}
\expandafter\def\csname suppref@supp:video\endcsname{S7}
\expandafter\def\csname suppref@tab:B_boltz\endcsname{S8}
\expandafter\def\csname suppref@tab:ablation\endcsname{S4}
\expandafter\def\csname suppref@tab:boltz_M5_alloc\endcsname{S13}
\expandafter\def\csname suppref@tab:bounded\endcsname{S2}
\expandafter\def\csname suppref@tab:envelope_control\endcsname{S1}
\expandafter\def\csname suppref@tab:information_full\endcsname{S12}
\expandafter\def\csname suppref@tab:mono_free\endcsname{S3}
\expandafter\def\csname suppref@tab:n2_free\endcsname{S5}
\expandafter\def\csname suppref@tab:set_mono\endcsname{S6}
\expandafter\def\csname suppref@tab:units\endcsname{S11}
}

\journal{Computer Methods in Applied Mechanics and Engineering}

\begin{document}

\begin{frontmatter}

\title{Physics-informed neural networks for strong normal shocks in kinetic relaxation models and the hard-sphere Boltzmann equation}

\author[umass]{Ehsan Roohi\corref{cor}}
\ead{roohie@umass.edu}
\cortext[cor]{Corresponding author}
\affiliation[umass]{organization={Department of Mechanical and Industrial Engineering, University of Massachusetts Amherst},
            city={Amherst}, state={MA}, country={USA}}

\begin{abstract}
Physics-informed neural networks (PINNs) for kinetic equations have mostly been applied near equilibrium and judged by density, velocity and temperature. Here a network trained on the kinetic equation alone, with no reference data in the loss, solves steady normal shock waves and is assessed in eight fields: density, velocity, temperature, heat flux, stress, a third-order moment and two fourth-order moments. The network corrects a mixture of the two end-state Maxwellians by an unbounded, signed factor with exact inflow; the loss, the residual in three norms with conservation terms, is minimised by Adam and a self-scaled quasi-Newton method with continuation in the Mach number. For the hard-sphere Boltzmann equation at Mach 8, in a comparison made once after reference-free tests, the largest relative $L^2$ difference from direct simulation Monte Carlo (DSMC) over the eight fields is 0.19 per cent; at Mach 10 it is 0.34 per cent. 
The distribution function itself agrees with the reference over six decades
in velocity space. For the Bhatnagar--Gross--Krook (BGK) model, a blind
prediction at Mach 50 gives density, velocity and temperature to about
0.1 per cent and exposes a loss of accuracy confined to the two fourth-order
moments. Its origin is identified as a passive reduced distribution that
carries little weight in the loss; with a larger weight on that distribution
and a longer optimisation, both moments are within 3.2 per cent of the
discrete-velocity solution up to Mach 70 and 7.1 per cent at Mach 100. The Shakhov model is solved to
Mach 20 and a Rykov-type model of nitrogen to Mach 15. An exact heat-flux
balance, evaluated from the network alone, gives a reference-free measure of
accuracy: it flagged the runs at higher Mach number that pass the three
acceptance tests but have an inaccurate heat flux and thickness, without
recourse to a reference, though not with a sharp threshold.
\end{abstract}


\end{frontmatter}

\section{Introduction}\label{sec:intro}

\begin{figure}[tbp]
\centering
\includegraphics[width=0.8\textwidth]{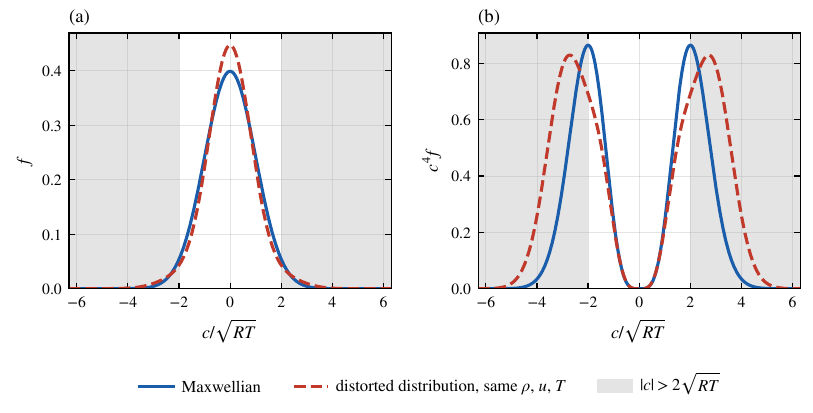}
\caption{Two distributions with the same density, velocity and temperature, after the corresponding figure of \cite{Roohi2026KineticShockPINN}: a Maxwellian (solid) and the Maxwellian multiplied by $1+0.04\,\mathrm{He}_4(c)$ (dashed), with $c$ the molecular velocity in units of $\sqrt{RT}$, $R$ the specific gas constant, and $\mathrm{He}_4$ the Hermite polynomial of degree four. (a) The distributions; (b) the integrand $c^4f$ of the fourth-order moments. The shaded band is $|c|>2\sqrt{RT}$: it holds 4.6 per cent of the molecules of the Maxwellian and 55 per cent of its fourth-order integral. The fourth-order integrals of the two distributions differ by 32 per cent.}
\label{fig:twodist}
\end{figure}

A gas far from local equilibrium is described by the distribution $f$ of its molecular velocities, which obeys the Boltzmann equation or a model of it \cite{Cercignani1988,Sone2007,Bird1994}. The quantities of engineering interest are moments of $f$, that is, integrals over the molecular velocity. Density, velocity and temperature are the lowest moments; the stress and the heat flux come next; and the balance equations of the stress and the heat flux contain a third-order moment and two fourth-order moments, which regularised moment theories add to Grad's thirteen variables \cite{Grad1949,StruchtrupTorrilhon2003,Struchtrup2005,GuEmerson2009R26}. The higher the order of a moment, the more it depends on the fast molecules: in a Maxwellian the fastest tenth of the molecules carries 28 per cent of the energy and one half of the fourth-order integral. Two distributions can therefore have the same density, velocity and temperature and differ by tens of per cent at fourth order (\cref{fig:twodist}). A numerical solution of a kinetic equation should be judged by the moments that the application needs, not by the bulk fields alone \cite{Roohi2026KineticShockPINN}.

The structure of a shock wave has tested the description of a gas for more than a century: from the steepening of finite-amplitude waves and the jump conditions \cite{Riemann1860Luftwellen,Rankine1870Waves}, through the continuum profiles of Rayleigh, Taylor and Becker, whose thickness of the order of a mean free path put the continuum description in doubt \cite{Rayleigh1910Aerial,Taylor1910Discontinuous,Becker1922Stosswelle,Thomas1944Becker,GilbargPaolucci1953Structure}, to the kinetic descriptions opened by Mott-Smith and Grad \cite{MottSmith1951Shock,Grad1952Profile} and the numerical solutions reviewed below.

The normal shock wave is therefore the standard test of a kinetic solver. Inside a strong shock the distribution is bimodal \cite{MottSmith1951Shock}: a cold beam, the supersonic upstream gas, coexists with the hot subsonic gas of the downstream side. At Mach 20 the thermal speed of the hot gas is eleven times that of the beam, and ahead of the shock 3 per cent of the molecules carry 99 per cent of the fourth-order moment. The values of $f$ at the velocities that contribute to the moments span six decades. Direct simulation Monte Carlo (DSMC), discrete-velocity methods and deterministic solvers of the Boltzmann equation resolve this structure and provide the references used here \cite{Bird1994,Mieussens2000,XuHuang2010,Tcheremissine2006}.

Mott-Smith represented the distribution inside a shock as a weighted sum of the upstream and downstream
Maxwellians~\cite{MottSmith1951Shock}, and Takata, Aoki and Cercignani analysed the velocity distribution in the limit
of an infinitely strong shock~\cite{TakataAokiCercignani2000Infinite}. The distribution was measured inside shocks in argon at Mach 7~\cite{HoltzMuntz1983Argon} and in helium at Mach
25~\cite{PhamVanDiep1989Hypersonic} and computed by molecular dynamics~\cite{ValentiniSchwartzentruber2009MD}; the DSMC method~\cite{Bird1970Strong,Bird1994} was validated against such
measurements~\cite{ErwinPhamVanDiepMuntz1991Validation} and carried to Mach 50 for Maxwell molecules and hard
spheres~\cite{FisckoChapman1989Hypersonic,LumpkinChapman1992Burnett}, results with which bimodal fluid models were later
compared~\cite{SolovchukSheu2011Bimodal}.
Deterministic solutions of the Boltzmann equation for hard spheres that agree with DSMC were given by
Ohwada~\cite{Ohwada1993Shock} up to Mach 3, by Tcheremissine and co-workers up to Mach 20, at a cost growing like
$M^2$~\cite{Tcheremissine2006,DoduladTcheremissine2013Shock}, and by Malkov et al.\ at Mach 4 and
8~\cite{Malkov2015Deterministic}. Fourier spectral methods for the collision
operator~\cite{PareschiPerthame1996Fourier,PareschiRusso2000Spectral,BobylevRjasanow1999Fast,GambaTharkabhushanam2009Spectral}
and the fast algorithms built on
them~\cite{MouhotPareschi2006Fast,FilbetMouhotPareschi2006NlogN,Wu2013FastSpectral,GambaHaackHauckHu2017Fast} made the
collision integral affordable on velocity grids; the fast spectral method is the one used here (\cref{sec:boltz_residual}). For
the relaxation models of Bhatnagar, Gross and Krook (BGK)~\cite{Bhatnagar1954}, of Shakhov~\cite{Shakhov1968} and the
ellipsoidal model~\cite{Holway1966ESBGK}, the transverse velocities of a planar flow can be integrated
out~\cite{Chu1965Reduced}, and discrete-velocity methods (DVM) solve the remaining
equation~\cite{YangHuang1995Rarefied,Mieussens2000}. Liepmann, Narasimha and Chahine computed the BGK shock by
iteration from the Navier--Stokes profile~\cite{LiepmannNarasimhaChahine1962}. Shakhov solved BGK and Shakhov shocks up
to Mach 10 for a viscosity proportional to $T$, the law used here~\cite{Shakhov1969Shock}, and Aristov computed Shakhov
shocks up to Mach 100, with a viscosity proportional to $T^{0.5}$ and $T^{0.68}$, on velocity grids that follow the
local state~\cite{Aristov1977VariableGrids}. Because the collision frequency of these models does not depend on the
molecular velocity, the coordinate $\int(p/\mu)\,\dd x$ maps a solution for one viscosity law onto any
other~\cite{LiepmannNarasimhaChahine1962,SegalFerziger1972Shock}, so that these results also apply to the law used here.
They are graphical, and none reports third- or fourth-order moments
above Mach 10. Unified and discrete unified gas-kinetic schemes extend such solvers across flow
regimes~\cite{XuHuang2010,HuangXuYu2012UGKS,GuoXuWang2013DUGKS}. Reviews are given
in~\cite{DimarcoPareschi2014,RjasanowWagner2005}, and kinetic models for shock waves are compared
in~\cite{FeiLiuLiuZhang2020ShockBenchmark}. Two Fokker--Planck models were compared with DSMC in supersonic Couette flow
and around a cylinder at Mach 2~\cite{JunPfeifferMieussensGorji2019FP}: the ellipsoidal model, which has an H-theorem,
was the more accurate in the shock region, and the cubic model, which has the correct Prandtl number, near the continuum
and in the wall heat flux; the authors concluded that both properties matter, a trade-off that the BGK and Shakhov
models show against DSMC of Maxwell molecules (\cref{sec:validation}). For a diatomic gas, whose rotational energy relaxes more slowly than its
translational energy, Rykov proposed a model with separate translational and rotational
temperatures~\cite{Rykov1975Model,LarinaRykov2010Diatomic}. Wu et al.\ wrote it in the two-distribution form used here and built on it a model with velocity-dependent relaxation
for non-vibrating polyatomic gases~\cite{Wu2015Polyatomic}; against DSMC at Mach 4 to 7 the Rykov model predicted the
density well but raised the temperature too early ahead of the shock, which they traced to its velocity-independent
relaxation time. With a collision number that depends on the temperatures, Rykov, Titarev and Shakhov reproduced the
measured density and rotational temperature of nitrogen shocks to Mach 12.9~\cite{RykovTitarevShakhov2008Diatomic}, and
Kosuge and Aoki
solved the shock of a polyatomic gas with large bulk viscosity with an ellipsoidal statistical
model~\cite{KosugeAoki2018BulkViscosity}. Density profiles and the rotational distribution of nitrogen shocks were
measured in~\cite{Alsmeyer1976Density,RobbenTalbot1966Nitrogen}; DSMC with classical trajectories for every collision,
without an adjustable rotational parameter, reproduced these measurements~\cite{Koura1997Rotational,Koura2002Rotational},
and molecular dynamics gave a rotational collision number rising from 3.2 at 100\,K to 6.9 at
2000\,K~\cite{ValentiniZhangSchwartzentruber2012Rotational}.

Grad's moment method~\cite{Grad1949} and the closure hierarchies that followed~\cite{Levermore1996} describe a rarefied
gas by a finite set of moments. The regularised 13-moment equations~\cite{StruchtrupTorrilhon2003} were applied to
the shock structure in~\cite{TorrilhonStruchtrup2004}, and the regularised 26-moment equations~\cite{GuEmerson2009R26}
add the third- and fourth-order moments that appear in the balance laws of the stress and the heat flux
(accounts in~\cite{Struchtrup2005,Torrilhon2016}). Because the heat-flux balance contains fourth-order moments, a
solution can be accurate in density, velocity and temperature and still be wrong in the quantities that close the
hierarchy; the networks are therefore assessed up to fourth order.

Networks that minimise the residual of a differential equation go back to Lagaris et al.~\cite{Lagaris1998ANN}; the
present form, physics-informed neural networks (PINNs), is due to Raissi et al.~\cite{Raissi2019} (reviews
in~\cite{Karniadakis2021PIML,Lu2021DeepXDE}). PINNs for the Euler equations with shocks were studied
in~\cite{MaoJagtapKarniadakis2020HighSpeed}, and a data-free finite-volume convolutional network solved the steady
Euler equations around a cylinder at Mach numbers from 2 to 15~\cite{Yano2026DataFree}. Reported failure modes are an
imbalance between loss terms with stiff gradient flow~\cite{WangTengPerdikaris2021Gradient,WangYuPerdikaris2022NTK}
and loss landscapes made ill-conditioned by the differential
operator~\cite{Krishnapriyan2021Failure,Rathore2024Landscape}. Remedies include adaptive loss
weights~\cite{WangTengPerdikaris2021Gradient,McClennyBragaNeto2023SelfAdaptive}, curriculum and causal
training~\cite{Krishnapriyan2021Failure,WangSankaranPerdikaris2024Causality}, adaptive collocation
points~\cite{Lu2021DeepXDE,Wu2023Sampling} and trial functions that satisfy the boundary conditions
exactly~\cite{Lagaris1998ANN,SukumarSrivastava2022Exact,Lu2021HardConstraints}. Adam~\cite{KingmaBa2015Adam} is usually
followed by the limited-memory Broyden--Fletcher--Goldfarb--Shanno method (L-BFGS)~\cite{LiuNocedal1989LBFGS}; much smaller errors are obtained with modified quasi-Newton
methods~\cite{UrbanStefanouPons2025}, in particular self-scaled BFGS and Broyden
updates~\cite{OrenLuenberger1974SelfScaling,Kiyani2025Optimizer}, with natural-gradient
methods~\cite{MullerZeinhofer2023Natural} and with the Levenberg--Marquardt
method~\cite{Levenberg1944,Marquardt1963,Taylor2022Optimizer,Bonfanti2024Nonlinear,Shahab2026Shallow}.

Lou et al.\ formulated a PINN for the BGK equation on discrete velocity sets~\cite{LouMengKarniadakis2021PINNBGK}. Their
forward problems use no data and include a low-speed Couette flow up to a Knudsen number $\Kn=5$ and a cavity at a Reynolds number of
100, assessed in the velocity; a rarefied cavity at $\Kn=0.1$ to 10 was solved as an inverse problem with interior
velocity data from a discrete unified gas-kinetic solution. Linearised BGK flows were solved by networks
in~\cite{DeFlorio2021ThermalCreep} and, with a discrete-velocity formulation and adaptive weights, for Couette,
heat-transfer, duct and cavity flows in~\cite{ZhangMaLouZhang2023PINNDVM}; reduced BGK microflows
in~\cite{ZhangWang2024Microflows}. Asymptotic-preserving networks keep the diffusive limit of multiscale kinetic
equations~\cite{JinMaWu2021APNNTransport,JinMaWu2023APNNKinetic} and have been extended to uncertainty quantification
and bi-fidelity settings~\cite{ChenDimarcoPareschi2025SAPNN,LiuZhuZhu2026BiAPNN}. Li et al.\ solved the BGK and the
quadratic Boltzmann collision models with a neural sparse representation in velocity, for unsteady wave and Sod
problems~\cite{Li2024NeuralSparse}, and separable PINNs solved the BGK model for smooth and Riemann problems in up to
three space dimensions at $\Kn$ from 0.01 to 1~\cite{Oh2025SPINNBGK}; both compare density, velocity and temperature.
Ko et al.\ argued that minimising the standard $L^2$ residual does not guarantee accurate moments and proposed a loss
weighted towards large velocities~\cite{Ko2026WeightedBGK}; error estimates for Boltzmann PINNs near equilibrium are
given in~\cite{Abdo2024ErrorEstimates}. For shocks, Kim and Jun trained a network on a five-moment projection of an
ellipsoidal Fokker--Planck model with a Grad-type closure and obtained an argon shock at Mach 10 without data, assessed
up to the stress and the heat flux~\cite{KimJun2026FPPINN}. The predecessor of the present work solved BGK and Shakhov
shocks with and without moment supervision and reported moments up to fourth order~\cite{Roohi2026KineticShockPINN}.

A separate line of work learns parts of the kinetic description from data: moment
closures~\cite{HanMaMaE2019Hydrodynamic}, among them a closure of the BGK equation with one velocity dimension that,
trained on kinetic solutions at lower Mach numbers, reproduces discrete-velocity shock profiles of density, velocity and
temperature at Mach 11 and 21~\cite{LiDongWang2023Closure}; the collision
operator~\cite{XiaoFrank2021Accelerate,XiaoFrank2023RelaxNet,Miller2022Collision,Lee2024OperatorLearning,Lee2026FourierSpecNet},
including Fourier neural operators trained on spectral solutions of the spatially homogeneous BGK and Boltzmann
equations with soft conservation constraints~\cite{HuQi2026FNO}, a data-trained, spatially homogeneous counterpart of
the equation-only, spatially inhomogeneous solutions of this paper; differentiable kinetic
solvers~\cite{Xiao2025Differentiable}; operator surrogates for rarefied flows~\cite{RoohiShojaSaniAzghadi2026PoF},
among them a physics-enforced operator network for the density profile of the polyatomic shock with large bulk
viscosity of~\cite{KosugeAoki2018BulkViscosity}, trained with a data loss and penalties for monotonicity and the
far-field gradient, which is a surrogate of known solutions rather than a solution of the kinetic equation;
and collision models in particle
methods~\cite{BallMacArtSirignano2026MLDMS,RoohiShojaSaniStefanov2026PoF}. The two closest works differ from the present one as follows. Kim and Jun~\cite{KimJun2026FPPINN} use no data and reach a
normal shock at Mach 10, but for a five-moment projection of a Fokker--Planck model with a closure, not for the
distribution function, and assess it up to the stress and the heat flux. Li et al.~\cite{Li2024NeuralSparse} represent the
distribution and include the quadratic Boltzmann collision operator, for unsteady wave and shock-tube problems assessed
in density, velocity and temperature. The contribution of the present work is specific: for the steady normal shock, a loss that contains only the kinetic equation, no kinetic solve or characteristic
correction after training, validation against discrete-velocity and DSMC solutions of the same model in eight fields
that include a third-order and two fourth-order moments, and the models and Mach numbers of \cref{tab:information}
(BGK to Mach 100, Shakhov to Mach 20, a nitrogen model to Mach 15, hard-sphere Boltzmann to Mach 10). We have not found
an earlier network solution with these properties; the data-trained closure of~\cite{LiDongWang2023Closure} (Mach 11 and
21, bulk fields) and the network for the continuum Euler equations of~\cite{Yano2026DataFree} address different problems.

In a previous paper \cite{Roohi2026KineticShockPINN} the author compared PINNs for BGK and Shakhov shocks trained with and without moment data and assessed them through the heat flux, the stress and higher moments. Its networks trained on the equation alone reached Mach 5 for the Shakhov model and Mach 7 for the BGK model; at Mach 8 and 12 a kinetic solve started from the network (a characteristic correction) was required. Those networks multiplied a Rankine--Hugoniot mixture by a bounded correction, weighted the residual by the fixed mixture, resolved the transverse velocities with $192\times64$ to $512\times144$ velocity nodes and used 187 prescribed points in the quasi-Newton stage. The present paper asks what has to change for a network trained on the equation alone to solve strong shocks without such a correction, in monatomic and diatomic gases.

The method rests on four choices (\cref{sec:inherited} states which parts come from \cite{Roohi2026KineticShockPINN}). The first is the form of the distribution: the network gives an unbounded, signed correction to a mixture of the two end-state Maxwellians, multiplied by the probability that a molecule has collided since it entered the domain, so that the inflow distributions are exact. The correction that a reference solution requires is measured over the whole phase plane and compared with the one the network learns. The second is the loss, which contains only the kinetic equation: its residual in three norms that weight the cold beam, the sparse hot molecules and the fast tail differently, the constancy of the fluxes, low-order moments of the residual and a condition that fixes the shock position. The third is the set of collocation points and velocity nodes, both concentrated where the shock and the cold beam need them, with separate test points never used in training. The fourth is the optimiser: Adam, a self-scaled quasi-Newton method, and continuation in the Mach number.

\begin{figure}[tp]
\centering
\includegraphics[width=\textwidth]{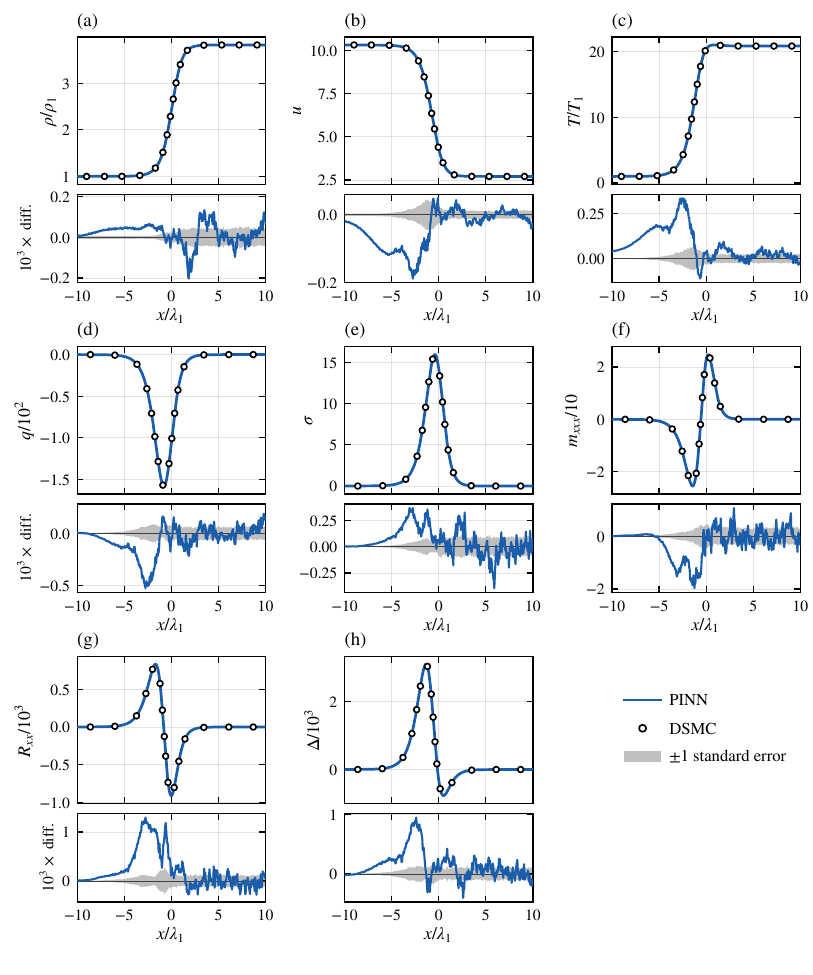}
\caption{Boltzmann equation for hard spheres at Mach 8, solved by the network (\cref{sec:boltz_M8M10}); the comparison was made once, after the reference-free tests. (a) Density $\rho$, (b) velocity $u$, (c) temperature $T$, (d) heat flux $q$, (e) stress $\sigma$, (f) third-order moment $m_{xxx}$, (g) and (h) fourth-order moments $R_{xx}$ and $\Delta$ (defined in \cref{sec:models}; units in which the upstream density and temperature give $\rho_1=RT_1=1$); line: PINN; symbols: DSMC. Under each panel: PINN minus DSMC, divided by the scale of the field (the jump for $\rho$, $u$ and $T$, the largest absolute value for the other fields) and multiplied by $10^3$; the grey band is plus and minus one standard error of the DSMC. The coordinate $x$ is in upstream mean free paths $\lambda_1$ from the density midpoint.}
\label{fig:boltz_M8}
\end{figure}

With these choices a network trained on the equation alone solves the BGK shock to Mach 100, the Shakhov shock to Mach 20 and a Rykov-type model of nitrogen to Mach 15, and the hard-sphere Boltzmann equation at Mach 2, 5, 8 and 10. All eight reported fields, including the two fourth-order moments $R_{xx}$ and $\Delta$, are compared with solutions of the same model. For the BGK model the two fourth-order moments are within 3.2 per cent of the discrete-velocity solution at Mach 70 and 7.1 per cent at Mach 100, after their error was traced to the weight of one reduced distribution in the loss and that weight and the length of the optimisation were increased (\cref{sec:mono_m50}); the Shakhov model is accurate to Mach 20, and above it the runs pass the acceptance tests while the heat flux is 4 to 18 per cent off, which the residual of the exact heat-flux balance indicates without a reference (\cref{sec:mono_shk_limit}). For the Boltzmann equation of hard spheres at Mach 8 each of the eight fields is within 0.19 per cent of DSMC in the relative $L^2$ norm over $|x|\le17\,\lambda_1$, with $x$ measured from the density midpoint and $\lambda_1$ the upstream mean free path, in a comparison made once after the reference-free tests, and on $|x|\le10\,\lambda_1$ the pointwise differences are at most $2.0\times10^{-3}$ of the scale of each field (\cref{fig:boltz_M8}); at Mach 10 the norms are within 0.34 per cent in a comparison that was not blind (\cref{sec:boltz_M8M10}).

A run is accepted only if it passes tests that need no reference: the constancy of the fluxes, the residual at test points and the change under a finer velocity rule. \Cref{tab:information} states, for every result, whether its reference was available during development. The failure modes specific to kinetic problems are documented with indicators that need no reference: a shock that collapses between two collocation points, a domain that cuts off the upstream-running molecules, and a velocity cube that cannot resolve the cold beam.

\Cref{sec:models,sec:method,sec:protocol} state the models and problems, the method and the reference solutions. \Cref{sec:mono,sec:n2,sec:boltz} report the shocks of the monatomic relaxation models, of nitrogen (with the evidence for the loss, the points and the optimiser) and of the Boltzmann equation, \cref{sec:validation} compares them with published solutions and measurements, and \cref{sec:discussion} discusses cost, failures and limits. The supplementary material contains the reference solvers, the evaluation of the collision integral, implementation details, the settings of every run and further results.

\section{Kinetic models and test problems}\label{sec:models}

\subsection{Kinetic equations}\label{sec:equations}

A dilute gas is described by the distribution $f(\bx,\bv)$ of its molecules over position $\bx$ and molecular velocity $\bv$. All flows considered here are steady, so that
\begin{equation}
  \bv\cdot\nabla_{\bx}f=\mathcal{C}[f],
  \label{eq:kinetic}
\end{equation}
where the collision term $\mathcal{C}$ is the Boltzmann integral or a relaxation model of it \cite{Cercignani1988,Sone2007}. All variables are dimensionless, the gas constant is absorbed in the temperature (units in \cref{sec:units}), and the notation of the relaxation models is that of \cite{Roohi2026KineticShockPINN}. Density $\rho$, velocity $\bu$ and temperature $T$ are the moments
\begin{equation}
  \rho=\int_{\Rset^3}f\,\dd\bv,\qquad
  \rho\bu=\int_{\Rset^3}\bv f\,\dd\bv,\qquad
  3\rho T=\int_{\Rset^3}|\bc|^2f\,\dd\bv,\qquad \bc=\bv-\bu ,
  \label{eq:bulk}
\end{equation}
with $\bc$ the peculiar velocity, and the pressure is $p=\rho T$.

\paragraph{Boltzmann equation for hard spheres}
The collision term is the integral
\begin{equation}
  \mathcal{C}[f]=Q(f,f)=\frac{1}{\sqrt2}\int_{\Rset^3}\int_{S^2}|\bv-\bv_*|\,
  \bigl[f(\bv')f(\bv_*')-f(\bv)f(\bv_*)\bigr]\,\frac{\dd\bm{\Omega}}{4\pi}\,\dd\bv_*,
  \label{eq:boltzmann}
\end{equation}
in which $\bv'=\tfrac12(\bv+\bv_*)+\tfrac12|\bv-\bv_*|\bm{\Omega}$ and $\bv_*'=\tfrac12(\bv+\bv_*)-\tfrac12|\bv-\bv_*|\bm{\Omega}$ are the velocities after a collision and $\bm{\Omega}$ runs over the unit sphere $S^2$. The factor $1/\sqrt2$ is that of hard spheres of diameter $d$ when lengths are measured in the upstream mean free path $\lambda_1=1/(\sqrt2\,\pi d^2n_1)$, with $n_1$ the upstream number density.

\paragraph{BGK and Shakhov models}
The relaxation models replace $Q$ by
\begin{equation}
  \mathcal{C}[f]=\frac{f^*-f}{\tau},
  \label{eq:relaxation}
\end{equation}
with a target $f^*$ built from a few moments of $f$ and a collision time $\tau$. The BGK model \cite{Bhatnagar1954} takes the local Maxwellian,
\begin{equation}
  f^*=\M[\rho,\bu,T],\qquad
  \M[\rho,\bu,T](\bv)=\frac{\rho}{(2\pi T)^{3/2}}\exp\Bigl(-\frac{|\bv-\bu|^2}{2T}\Bigr),
  \label{eq:bgk_target}
\end{equation}
and has Prandtl number one. The Shakhov model \cite{Shakhov1968} corrects the target with the heat flux $\bm{q}=\tfrac12\int\bc\,|\bc|^2f\,\dd\bv$,
\begin{equation}
  f^*=\M[\rho,\bu,T]\Bigl[1+(1-\Pran)\,\frac{\bm{q}\cdot\bc}{5\rho T^2}\Bigl(\frac{|\bc|^2}{T}-5\Bigr)\Bigr],
  \label{eq:shakhov_target}
\end{equation}
so that the stress relaxes at the rate $1/\tau$ and the heat flux at the rate $\Pran/\tau$, where $\Pran$ is the Prandtl number; $\Pran=2/3$ is the monatomic value and $\Pran=1$ returns the BGK model. The Shakhov target is negative for fast molecules with $\bm{q}\cdot\bc<0$, and so is the solution of the model there; this decides the form of the correction in \cref{sec:form}. The collision time follows from the viscosity $\mu=\mu_1T^{\omega}$,
\begin{equation}
  \tau=\frac{\mu}{p}=\frac{\mu_1T^{\omega-1}}{\rho}.
  \label{eq:tau_law}
\end{equation}

\paragraph{Rykov-type model of nitrogen}
For a diatomic gas with two rotational degrees of freedom (ratio of specific heats $\gamma=7/5$) we use the model of Rykov \cite{Rykov1975Model} in the two-distribution form of Wu and co-workers \cite{Wu2015Polyatomic}: $F(\bx,\bv)$ is the distribution of the molecules and $H(\bx,\bv)$ that of their rotational energy, and
\begin{equation}
  \bv\cdot\nabla_{\bx}F=\frac{1}{\tau}\Bigl[(F_t-F)+\frac{F_r-F_t}{Z}\Bigr],\qquad
  \bv\cdot\nabla_{\bx}H=\frac{1}{\tau}\Bigl[(H_t-H)+\frac{H_r-H_t}{Z}\Bigr].
  \label{eq:rykov}
\end{equation}
Density and velocity are the moments \eqref{eq:bulk} of $F$. The translational and rotational temperatures and heat fluxes are
\begin{equation}
\begin{aligned}
  3\rho T_{\mathrm{tr}}&=\int|\bc|^2F\,\dd\bv, & \rho T_{\mathrm{rot}}&=\int H\,\dd\bv,\\
  \bm{q}_{\mathrm{tr}}&=\frac12\int\bc\,|\bc|^2F\,\dd\bv, & \bm{q}_{\mathrm{rot}}&=\int\bc\,H\,\dd\bv,
\end{aligned}
\label{eq:rykov_moments}
\end{equation}
and $T=(3T_{\mathrm{tr}}+2T_{\mathrm{rot}})/5$ is the overall temperature. With $\M(\Theta)$ the Maxwellian of density $\rho$, velocity $\bu$ and temperature $\Theta$, the four targets are
\begin{equation}
\begin{aligned}
  F_t&=\M(T_{\mathrm{tr}})\Bigl[1+\frac{\bm{q}_{\mathrm{tr}}\cdot\bc}{15\rho T_{\mathrm{tr}}^2}\Bigl(\frac{|\bc|^2}{T_{\mathrm{tr}}}-5\Bigr)\Bigr],\\
  F_r&=\M(T)\Bigl[1+\omega_0\frac{\bm{q}_{\mathrm{tr}}\cdot\bc}{15\rho T^2}\Bigl(\frac{|\bc|^2}{T}-5\Bigr)\Bigr],\\
  H_t&=T_{\mathrm{rot}}F_t+(1-\delta_{\mathrm{R}})\,\frac{\bm{q}_{\mathrm{rot}}\cdot\bc}{\rho T_{\mathrm{tr}}}\,\M(T_{\mathrm{tr}}),\\
  H_r&=T\,F_r+\omega_1(1-\delta_{\mathrm{R}})\,\frac{\bm{q}_{\mathrm{rot}}\cdot\bc}{\rho T}\,\M(T).
\end{aligned}
\label{eq:rykov_targets}
\end{equation}
Elastic collisions drive $F$ towards $F_t$ in the time $\tau$, and translation and rotation exchange energy in the time $Z\tau$. The collision time is that of the translational mode, $\tau=T_{\mathrm{tr}}^{\omega-1}/\rho$. The constants, the same in the network code and in the reference solvers, are $Z=3.1616$, $\delta_{\mathrm{R}}=0.7519$, $\omega_0=0.4946$, $\omega_1=1.9574$ and $\omega=0.74$; $\delta_{\mathrm{R}}$ is the constant written $\delta$ by Wu and co-workers. With $Z=3.1616$ the rotational energy relaxes towards the translational temperature in $\tfrac53Z\tau=5.3\,\mu/p$, a rotational collision number $Z_{\mathrm{rot}}=20Z/(3\pi)=6.7$ in units of $\pi\mu/(4p)$, which corresponds to $Z_{\mathrm{DSMC}}\approx5$ in the convention of Wu and co-workers; $\omega_0$ and $\omega_1$ follow from $Z$ through the Mason--Monchick conductivities as in \cite{Wu2015Polyatomic}, and the model has the bulk viscosity $\mu_b=\tfrac{4Z}{15}\mu=0.84\,\mu$ and $\Pran=0.707$. Wu and co-workers used $Z=2.226$ for their own nitrogen shocks; the consequences of a constant $Z$ are discussed in \cref{sec:validation}. The heat conductivities that these constants imply are derived in Supplementary Section~\sref{supp:models_details}.

\subsection{Units}\label{sec:units}

In the shock calculations with relaxation models the upstream state gives $\rho_1=T_1=1$, velocities are measured in $\sqrt{RT_1}$ and lengths in $\ell_0=\mu(T_1)/(\rho_1\sqrt{RT_1})$, so that $\mu_1=1$ in \eqref{eq:tau_law}. The BGK and Shakhov calculations use $\omega=1$, hence $\tau=1/\rho$; one BGK sequence uses a constant $\tau=1$, the convention of the BGK calculations of the previous paper \cite{Roohi2026KineticShockPINN}. The nitrogen calculations have $\omega=0.74$. The hard-sphere calculations measure lengths in $\lambda_1$. With the hard-sphere viscosity $\mu=1.016034\,(5\pi/32)\,\rho\,\bar c\,\lambda$, $\bar c=\sqrt{8RT/\pi}$ \cite{ChapmanCowling1970}, one has $\ell_0=0.7959\,\lambda_1$ and $\omega=1/2$, so that a Shakhov calculation with $\mu_1=0.7959$ and $\omega=1/2$ has the length unit and the viscosity of the Boltzmann calculation. Supplementary Table~\sref{tab:units} summarises the units and collision terms of all calculations.

\subsection{Reduced distributions}\label{sec:reduced}

In a one-dimensional flow $f$ depends on the streamwise molecular velocity $v$ and on the transverse speed $r=(v_y^2+v_z^2)^{1/2}$. The transport operator $v\,\partial_x$ does not act on $r$, every target of the relaxation models is a Maxwellian in $r$ times a polynomial in $r^2$, and the moments that enter the targets weight the distribution with $1$ and $r^2$ only. Integration over the transverse plane therefore gives closed equations \cite{Chu1965Reduced} for
\begin{equation}
\begin{aligned}
  g&=\int_0^\infty f\,2\pi r\,\dd r, & k&=\int_0^\infty r^2f\,2\pi r\,\dd r,\\
  l&=\int_0^\infty r^4f\,2\pi r\,\dd r, & h&=\int_0^\infty H\,2\pi r\,\dd r,
\end{aligned}
\label{eq:reduced}
\end{equation}
($f$ is replaced by $F$ for nitrogen, and $h$ exists only there), namely
\begin{equation}
  v\,\frac{\partial\varphi}{\partial x}=\frac{\varphi^*-\varphi}{\tau},\qquad
  \varphi\in\{g,k,l\}\ \text{or}\ \{g,k,h,l\},
  \label{eq:reduced_kinetic}
\end{equation}
with the targets $\varphi^*$ of \ref{app:reduced}. The reduction is exact. The distributions $g$, $k$ and $h$ carry the density, the velocity, the temperatures, the heat fluxes and the stress, and their equations are closed; $l$ is needed only for the fourth-order moments and does not feed back. The unknowns are thus functions of $(x,v)$, and a one-dimensional velocity rule of about fifty nodes replaces the many thousand nodes of a three-dimensional one. No such reduction exists for \eqref{eq:boltzmann}, which couples all velocity components: there the unknown remains $f(x,v,r)$ on a three-dimensional velocity grid (\cref{sec:models_diff}).

\subsection{Normal shock wave}\label{sec:shockproblem}

The shock is computed in its own frame on $x\in[-L,L]$, with the supersonic gas entering from the left. The upstream state is $(\rho_1,u_1,T_1)=(1,M\sqrt{\gamma},1)$ for the Mach number $M$, and the downstream state follows from the Rankine--Hugoniot relations,
\begin{equation}
  \rho_2=\frac{(\gamma+1)M^2}{(\gamma-1)M^2+2},\qquad
  u_2=\frac{u_1}{\rho_2},\qquad
  T_2=\frac{2\gamma M^2-(\gamma-1)}{(\gamma+1)\,\rho_2},
  \label{eq:rh}
\end{equation}
with $\gamma=5/3$ for the monatomic gas and $7/5$ for nitrogen, whose two temperatures are equal in both end states. Molecules enter at $x=-L$ with $v>0$ and the upstream Maxwellian, and at $x=L$ with $v<0$ and the downstream one. The steady problem is invariant under translation, and the position of the shock is fixed by $\rho(0)=(\rho_1+\rho_2)/2$. The half-length is $L=15M$ for the monatomic gas (75 at Mach 5, 300 at Mach 20) and 30, 50, 60, 60, 120, 120 and 120 for nitrogen at Mach 2, 5, 7, 8, 10, 12 and 15.

A steady normal shock in an unbounded gas has no external length and therefore no free Knudsen number: in units of $\ell_0$ or $\lambda_1$ its structure depends only on $M$ and on the molecular model. The relevant Knudsen number is local, the upstream mean free path over the density thickness $\delta_\rho=(\rho_2-\rho_1)/\max_x(\dd\rho/\dd x)$; its values for the present shocks are listed in Supplementary Section~\sref{supp:models_details}.

\begin{figure}[tp]
\centering
\includegraphics[width=\textwidth]{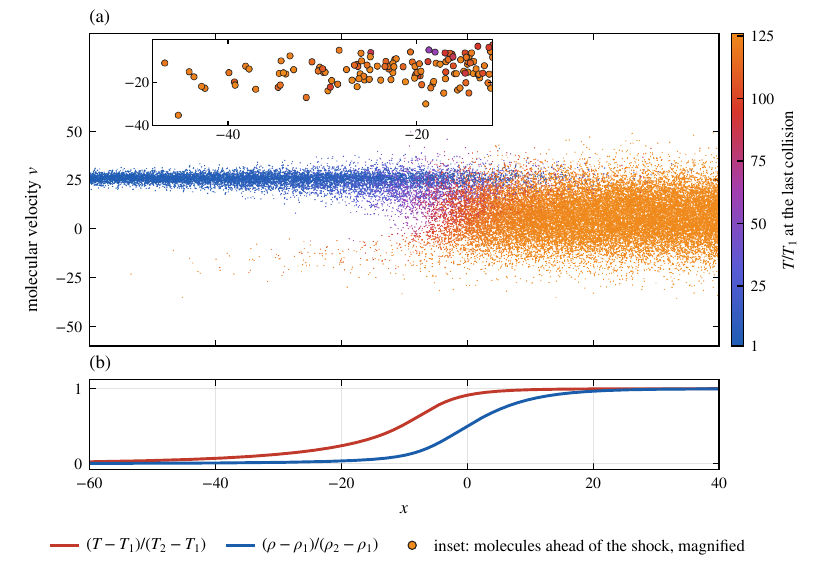}
\caption{The BGK shock at Mach 20 seen through its molecules. (a) Molecules in the plane of position $x$ and molecular velocity $v$ (the molecules at three instants of the stationary state), coloured by the temperature of the gas where each molecule last collided, from $T_1$ (blue, the incoming gas) to $T_2$ (orange, the downstream gas). The molecules are a stochastic realisation of the BGK equation with the converged fields of the discrete-velocity solution: each flies freely and, at the rate $1/\tau$, takes a new velocity from the local Maxwellian; their velocity distribution 20 units ahead of the shock agrees with that solution within 1.8 per cent (relative $L^2$, statistical). (b) Density and temperature of the discrete-velocity solution, scaled between their end values. The framed inset is a magnified view of the region $-48<x<-12$, $-40<v<0$ ahead of the shock, with the molecules that move upstream drawn larger; its axes carry the same coordinates as the panel. Supplementary Video 1 shows the molecules in motion, the distribution at a moving position, and the correction of the mixture of \cref{sec:form} (slow-motion version: \url{https://www.youtube.com/watch?v=9hSQn6-cPD8}).}
\label{fig:molecules}
\end{figure}

\Cref{fig:molecules} shows the kinetic structure of a strong shock. Upstream all molecules belong to a narrow beam at $v\approx u_1$; downstream they form a wide Maxwellian around $u_2$ whose thermal speed exceeds $u_2$, so that some hot molecules move upstream. Those that cross the shock run ahead into the incoming gas until they collide, the slow ones first; they form the tongue at negative velocities that reaches tens of units ahead of the density rise, and they heat the gas there, which is why the temperature rises long before the density. A neural representation of $f$ must therefore resolve the beam of width $\sqrt{T_1}$, the hot gas of width $\sqrt{T_2}$ (eleven times wider at Mach 20) and the sparse upstream-running molecules, which 20 units ahead of the shock are 3 per cent of the gas and carry 99 per cent of its fourth-order moment.

\subsection{Moments reported}\label{sec:moments}

Every solution is assessed through its moments up to fourth order. Besides $\rho$, $u$ and $T$, these are the heat flux and the normal stress,
\begin{equation}
  q=\frac12\int c_x|\bc|^2f\,\dd\bv,\qquad
  \sigma=\int\Bigl(c_x^2-\frac{|\bc|^2}{3}\Bigr)f\,\dd\bv=P_{xx}-\rho T,
  \label{eq:moments}
\end{equation}
and the third- and fourth-order moments that regularised moment theories add to Grad's thirteen variables \cite{Struchtrup2005,GuEmerson2009R26},
\begin{equation}
\begin{gathered}
  m_{xxx}=\int\Bigl(c_x^3-\frac35c_x|\bc|^2\Bigr)f\,\dd\bv,\qquad
  \Delta=\int|\bc|^4f\,\dd\bv-15\rho T^2,\\
  R_{xx}=\int|\bc|^2\Bigl(c_x^2-\frac{|\bc|^2}{3}\Bigr)f\,\dd\bv-7T\sigma .
\end{gathered}
\label{eq:higher_moments}
\end{equation}
All five vanish in a Maxwellian. For nitrogen $f$ is replaced by $F$ and $T$ by $T_{\mathrm{tr}}$ in \eqref{eq:moments} and \eqref{eq:higher_moments}, which then define $q_{\mathrm{tr}}$, $\sigma$, $m_{xxx}$, $R_{xx}$ and $\Delta$, and $T_{\mathrm{rot}}$ and $q_{\mathrm{rot}}$ of \eqref{eq:rykov_moments} are reported in addition. The formulas in terms of the reduced distributions, as implemented, are in \ref{app:reduced}.

\section{Method}\label{sec:method}

\begin{figure}
\centering
\includegraphics[width=\textwidth]{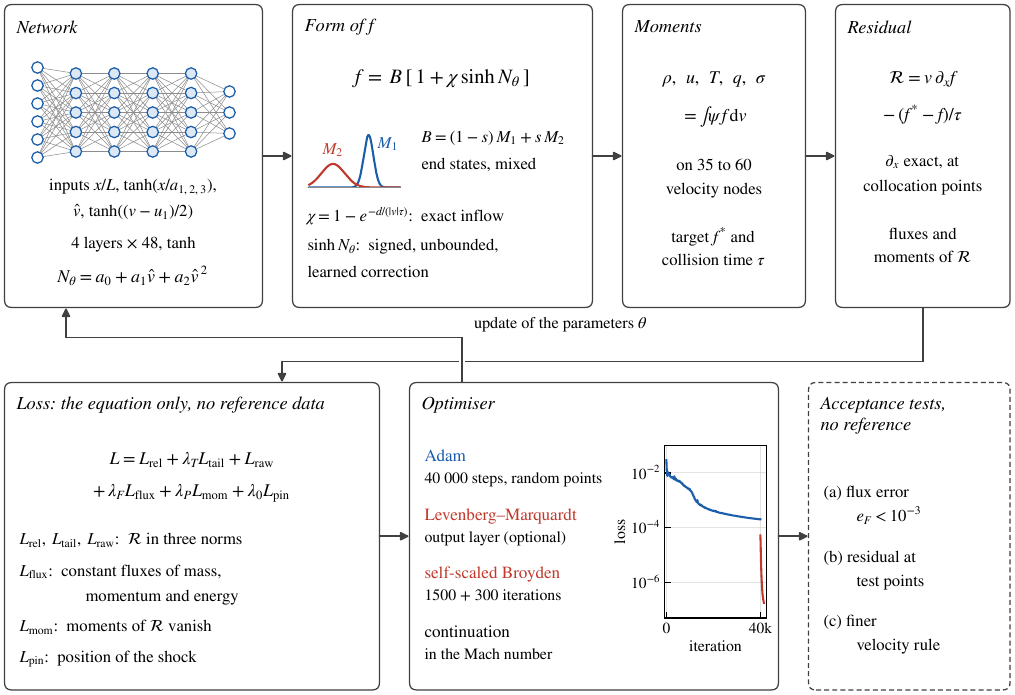}
\caption{Overview of the method: the network and its inputs, the form of $f$, the moments, the residual of the kinetic equation, the loss, the stages of the training and the acceptance tests without a reference, which are applied after the training and are not part of the loss (dashed frame); the inset shows the loss against the iteration for the BGK shock at Mach 5. The four choices of \cref{sec:method} are the form of $f$, the loss, the collocation points with the velocity nodes, and the optimiser. The velocity rule has 35 to 60 nodes, and 119 for the Shakhov shock at Mach 20 (\cref{sec:points}).}
\label{fig:schematic}
\end{figure}

\Cref{fig:schematic} summarises the method, and \cref{fig:algorithm} shows one optimisation step as a flowchart, for the relaxation models and for the Boltzmann equation. A small network gives an exponent $\Nth(x,v)$; the distribution is an analytic base times a signed function of this exponent; its moments, computed on a set of velocity nodes, define the target of the collision model; and the residual of the kinetic equation at a set of collocation points enters a loss that contains no reference data. Four choices decide whether minimising this loss gives the solution: the form of $f$ (\cref{sec:form}), the loss (\cref{sec:loss}), the collocation points and velocity nodes (\cref{sec:points}) and the optimiser (\cref{sec:optimiser}). They are described for the reduced relaxation models of \cref{sec:reduced}; \cref{sec:models_diff} states what changes with the collision model. The evidence for the form is given in \cref{sec:design_form}; the evidence for the loss, the points and the optimiser comes from the nitrogen calculations and is given in \cref{sec:design_terms,sec:design_points,sec:design_opt}.

\paragraph{Inherited and new elements}\label{sec:inherited}
The mixture with a trained width and the inflow factor (\cref{eq:mixture,eq:gate}), the three norms of the residual, the flux and pin terms, the moments of the residual, the default weights and the sequence of Adam and the self-scaled Broyden method are taken from the earlier paper \cite{Roohi2026KineticShockPINN}. New are the unbounded, signed correction $\sinh\Nth$; the one-dimensional velocity space with a stretched velocity rule; the local Maxwellian and the current solution in the envelope; a tail weight of fourth instead of eighth power; nested collocation sets with test points; the three weights changed from Mach 5 upward; continuation in the Mach number by the quasi-Newton method alone; and the treatment of the hard-sphere collision integral (details in Supplementary Section~\sref{supp:inherited}). Which new element is decisive has not been isolated. The earlier bounded form, tested with the present loss, points and optimiser, was accepted up to Mach 12 (\cref{sec:form}); this test does not identify what limited the earlier method.

\subsection{Network}\label{sec:network}

The network maps six inputs, built from the position $x$ and the molecular velocity $v$, to one exponent $\Nth^{(m)}(x,v)$ for each reduced distribution $\varphi_m$, $m\in\{g,k,l\}$ for the monatomic gas and $m\in\{g,k,h,l\}$ for nitrogen. The inputs are
\begin{equation}
  \frac{x}{L},\qquad \tanh\frac{x}{2.5},\qquad \tanh\frac{x}{8},\qquad \tanh\frac{x}{25},\qquad
  \hat v=\frac{v-v_c}{v_s},\qquad \tanh\frac{v-u_1}{2},
  \label{eq:inputs}
\end{equation}
with $v_c=(u_1+u_2)/2$ and $v_s=3\sqrt{T_2}$. The saturating functions of $x$ resolve the shock layer and its precursor on lengths that do not depend on $L$; $\hat v$ is of order one across the hot downstream distribution, and the last input varies only across the cold upstream beam. Because $v_c$, $v_s$ and $u_1$ follow the Rankine--Hugoniot states, a network trained at one Mach number is an approximate solution at a neighbouring one, which the continuation of \cref{sec:optimiser} uses. Four hidden layers of 48 neurons with hyperbolic-tangent activation feed a linear output layer, which returns three coefficients for each reduced distribution:
\begin{equation}
  \Nth^{(m)}(x,v)=a^{(m)}_0+a^{(m)}_1\hat v+a^{(m)}_2\hat v^2 ,
  \label{eq:exponent}
\end{equation}
where each $a^{(m)}_i$ is a function of the six inputs. A change of the velocity or the temperature of a Maxwellian is a quadratic change of its exponent, so that \eqref{eq:exponent} represents it without large weights. With the width $w$ of the base (\cref{sec:form}) the network has 7834 parameters for the monatomic gas and 7981 for nitrogen. The derivative $\partial\Nth/\partial x$ in the residual is propagated analytically with the forward pass, and all network codes compute in double precision; the initialisation and the parameter counts are given in Supplementary Section~\sref{supp:network_details}.

\subsection{Form of the distribution}\label{sec:form}

The network does not output the distribution. Each reduced distribution is written as
\begin{equation}
  \varphi_m(x,v)=\Bmix_m(x,v)\,\bigl[1+\chi(x,v)\sinh\Nth^{(m)}(x,v)\bigr],
  \label{eq:form}
\end{equation}
that is, $f=\Bmix\,[1+\chi\sinh\Nth]$ for the distribution itself. The base is a mixture of the equilibrium distributions $\varphi_{1,m}$ and $\varphi_{2,m}$ of the two Rankine--Hugoniot states, as in Mott-Smith \cite{MottSmith1951Shock} and in \cite{Roohi2026KineticShockPINN},
\begin{equation}
  \Bmix_m(x,v)=[1-s(x)]\,\varphi_{1,m}(v)+s(x)\,\varphi_{2,m}(v),\qquad
  s(x)=\frac{\Lambda(x/w)-\Lambda(-L/w)}{\Lambda(L/w)-\Lambda(-L/w)},
  \label{eq:mixture}
\end{equation}
with the logistic function $\Lambda(z)=(1+e^{-z})^{-1}$. The weight $s$ of the downstream state is zero at $x=-L$ and one at $x=L$, and its centre is at $x=0$, where the pin term of the loss (\cref{sec:loss}) places the shock. The width $w$ (initial value 3) is the only trained parameter of the base; it is optimised with the network on the same loss. The inflow factor
\begin{equation}
  \chi(x,v)=1-\exp\Bigl[-\frac{d_{\mathrm{in}}(x,v)}{|v|\,\tau_{\mathrm{in}}}\Bigr],\qquad
  (d_{\mathrm{in}},\tau_{\mathrm{in}})=
  \begin{cases}(x+L,\ \tau_1), & v\ge0,\\ (L-x,\ \tau_2), & v<0,\end{cases}
  \label{eq:gate}
\end{equation}
where $\tau_1$ and $\tau_2$ are the collision times of the upstream and downstream states, is the probability that a molecule of velocity $v$ has collided since it entered the domain, evaluated with the collision time of the state through which it entered.

The incoming distributions are exact: at $x=-L$, $s=0$ and, for $v\ge0$, $\chi=0$, so that $\varphi_m=\varphi_{1,m}$; at $x=L$, $\varphi_m=\varphi_{2,m}$ for $v<0$. The loss therefore needs no boundary term. The value $\Nth=0$ gives the mixture, which has the correct end states but is not a solution: ahead of a strong shock it contains slow hot molecules that are absent from the gas (\cref{fig:form}). The factor $1+\chi\sinh\Nth$ maps $\Nth$ one-to-one onto the real line. For large positive $\Nth$ it grows like $\chi e^{\Nth}/2$, as needed on the flanks of the upstream beam, which is heated and widened ahead of the shock. For negative $\Nth$ it passes through zero, so that molecules can be removed from the base over several decades and the distribution can change sign. The Shakhov and nitrogen models need the sign, because their solutions are negative in the fast tail; in an exploratory run that is not documented further, the positive form $\Bmix\exp(\chi\Nth)$ left the heat flux of the nitrogen shock at Mach 2 6 per cent off. In the code $|\Nth|\le40$.

The earlier networks \cite{Roohi2026KineticShockPINN} used a bounded correction, $f=\Bmix\,[1+\chi\,6\tanh(\Nth/6)]$, so that $f/\Bmix\le7$. \Cref{eq:form} removes this bound. Repeated with the bounded form as its one change, fixed in advance, the BGK continuation of \cref{sec:mono} ($\tau=1/\rho$) was accepted by tests (a) and (b) of \cref{sec:acceptance} at Mach 5, 7, 8, 10 and 12, with every field within 1.6 per cent of the DVM. Its $R_{xx}$ and $\Delta$ are 2.0 to 4.3 times less accurate than with the signed form at Mach 5, 10 and 12, and within the scatter between equal runs at Mach 7 and 8, while its $\rho$, $q$ and $\sigma$ are more accurate at Mach 10 and 12; the bounded network learned a base 1.5 to 1.7 times wider. This single-seed test stops at Mach 12; it does not show that the unbounded form is necessary, nor what limited the earlier method, which also differs in its loss, points and optimiser (Supplementary Section~\sref{supp:mono} and Table~\sref{tab:bounded}).

\subsection{The correction required by the BGK shock at Mach 20}\label{sec:design_form}

\begin{figure}[tp]
\centering
\includegraphics[width=0.8\textwidth]{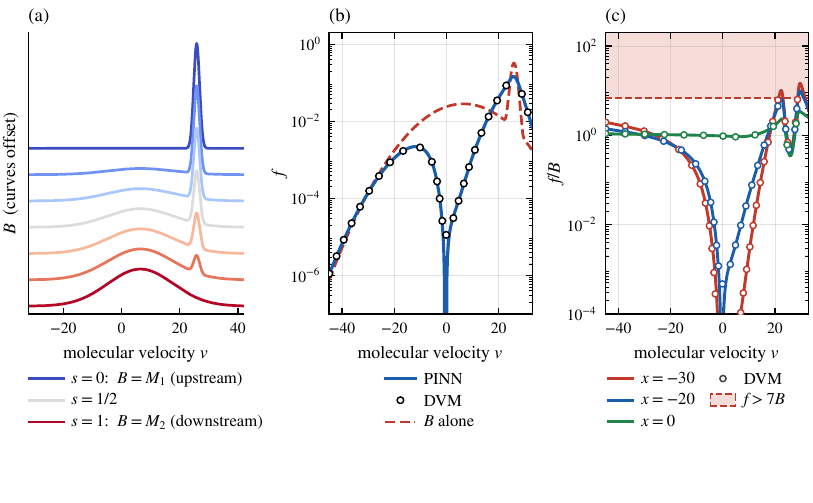}
\caption{The form of the distribution at Mach 20 (BGK). (a) The mixture $\Bmix$ from the upstream beam to the downstream Maxwellian (curves offset). (b) Reduced distribution 20 length units ahead of the density midpoint: PINN (line), DVM (symbols) and $\Bmix$ alone (dashed). (c) The factor $f/\Bmix$ at $x=-30$, $-20$ and $0$: PINN (lines) and DVM (symbols); the dashed line and the shading mark $f>7\Bmix$ over the base of this network, beyond the largest factor of the bounded form of the earlier paper (a bounded network trains a wider base, \cref{sec:form}).}
\label{fig:form}
\end{figure}

The evidence for the form is taken from the BGK shock at Mach 20 reported in \cref{sec:mono}; the DVM solution of the same model is used only after training, and lengths (in units of $\ell_0$) are measured from the density midpoint. The mixture (\cref{fig:form}a) has the correct end states but is far from the solution in between. It contains the whole downstream Maxwellian, including slow and forward-moving hot molecules; ahead of the shock these do not exist, because a molecule moving downstream there has come from the cold gas, and a slow one moving upstream collides before it travels far. At the position of \cref{fig:form}(b) the mixture alone would give a density of 1.592 where the DVM has 1.093 (46 per cent too high) and a normal stress $\int c^2g\,\dd v$ of 247.6 against 63.6 (almost four times too high). With the factor $1+\chi\sinh\Nth$ the network differs from the DVM at this position by 0.22 per cent in the relative $L^2$ norm of $g$ over the velocities that carry the fourth-order moment (0.26 per cent over those that carry the density); its density and normal stress are 0.06 per cent above the DVM values and its $\int c^4g\,\dd v$ is 0.006 per cent below.

\begin{figure}[tp]
\centering
\includegraphics[width=\textwidth]{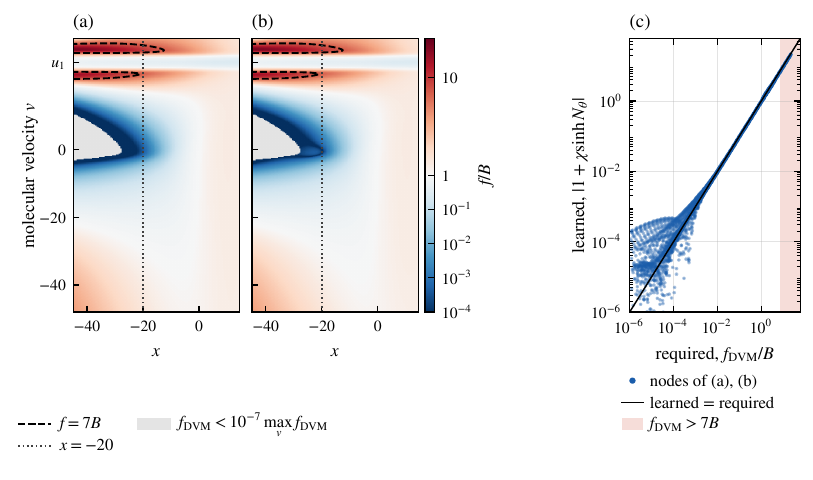}
\caption{The factor $f/\Bmix$ on the $(x,v)$ plane of the Mach 20 BGK shock: (a) required by the DVM, $f_{\mathrm{DVM}}/\Bmix$; (b) learned by the PINN; (c) learned against required at the 41\,080 nodes of the maps, with the line of equality. The colour scale is logarithmic and white at $f=\Bmix$; grey marks nodes where $f_{\mathrm{DVM}}$ is below $10^{-7}$ of its maximum over $v$. Dashed contour: $f=7\Bmix$ over the base of this network; dotted line: the position of \cref{fig:form}(b); the tick $u_1$ marks the upstream velocity; shading in (c): required factor above 7.}
\label{fig:factormap}
\end{figure}

\Cref{fig:factormap} compares the factor required by the DVM, $g_{\mathrm{DVM}}/\Bmix$, with the learned factor on the plane of position and velocity: 120 positions from 45 lengths ahead of the midpoint to 15 behind it and, at each, the velocities at which $g_{\mathrm{DVM}}$ exceeds $10^{-7}$ of its maximum, 41\,080 nodes in all. The required factor runs from $1.1\times10^{-6}$ to 21.1. The small values lie around zero velocity ahead of the shock; the large values lie on the two flanks of the cold beam, which is heated and widened before the density rises, whereas the mixture keeps it cold. At the density midpoint the factor stays between 0.35 and 3.5. Where the required factor is above about $10^{-3}$ the network reproduces it: the median of $|\text{learned}/\text{required}-1|$ over all nodes is 0.30 per cent, 94.6 per cent of the nodes agree within 5 per cent, and the nodes that carry 99 per cent of the fourth-order moment agree within 1.3 per cent.

Below $10^{-3}$, on 3.6 per cent of the nodes, which hold $5.9\times10^{-6}$ of the molecules, the learned factor scatters, and on 0.89 per cent of all nodes it is slightly negative, an error of the network that changes no moment at 20 lengths ahead by more than 0.02 per cent; a continued quasi-Newton stage reduces the scatter (Supplementary Sections~\sref{supp:mono} and \sref{supp:form_details}). On the flanks of the beam the factor required over this base exceeds 7, on velocities that hold up to 17.6 per cent of the density but at most 0.2 per cent of the fourth-order moment.

\subsection{Loss}\label{sec:loss}

Let $x_i$, $i=1,\dots,N_x$, be the collocation points and $v_j$, $w_j$ the velocity nodes and weights of \cref{sec:points}, and let $\langle\cdot\rangle$ denote the mean over the collocation points. The residual of the reduced equations \eqref{eq:reduced_kinetic} is
\begin{equation}
  \Res^{(m)}_{ij}=v_j\,\frac{\partial\varphi_m}{\partial x}(x_i,v_j)-\frac{\varphi^*_m(x_i,v_j)-\varphi_m(x_i,v_j)}{\tau(x_i)} .
  \label{eq:residual}
\end{equation}
The moments that define $\varphi^*$ and $\tau$ are computed from $\varphi$ itself on the same nodes, so that $\Res$ is a nonlinear functional of the network. The squared residual is summed over the velocities in three norms, each relative to a different measure of the local population. The first two use the envelope
\begin{equation}
  \mathcal{E}_{m}=\Bmix^{(3)}_m+M^{\mathrm{loc}}_m+|\varphi_m| ,
  \label{eq:envelope}
\end{equation}
the sum of the mixture with the fixed width $w=3$, the equilibrium distribution of the current local state and the magnitude of the current solution; we write $\mathcal{D}_{mij}=\mathcal{E}_{mij}+10^{-7}\max_j\mathcal{E}_{mij}$. The relative, tail-weighted and raw norms are
\begin{align}
  \mathcal{L}_{\mathrm{rel}}&=\Bigl\langle\sum_m a_m\,
     \frac{\sum_j w_j\,\mathcal{E}_{mij}\,\bigl(\Res^{(m)}_{ij}/\mathcal{D}_{mij}\bigr)^2}{\sum_j w_j\,\mathcal{E}_{mij}}\Bigr\rangle,
     \label{eq:lrel}\\
  \mathcal{L}_{\mathrm{tail}}&=\Bigl\langle\sum_m a_m\,
     \frac{\sum_j w_j\,t_{ij}\,\mathcal{E}_{mij}\,\bigl(\Res^{(m)}_{ij}/\mathcal{D}_{mij}\bigr)^2}{\sum_j w_j\,t_{ij}\,\mathcal{E}_{mij}}\Bigr\rangle,
     \label{eq:ltail}\\
  \mathcal{L}_{\mathrm{raw}}&=\Bigl\langle\sum_m a_m\sum_j w_j\,\bigl(\Res^{(m)}_{ij}/S_m\bigr)^2\Bigr\rangle,
     \label{eq:lraw}
\end{align}
with the tail weight $t_{ij}=1+[(v_j-u(x_i))^2/(5T_2)]^2$ and the scales $(S_g,S_k,S_h,S_l)=\rho_2\,(1,\allowbreak\ 2T_2,\allowbreak\ T_2,\allowbreak\ 8T_2^2)$. The raw norm is dominated by the velocities where the distribution is large and controls the bulk fields. The relative norm divides the squared residual by the envelope, so that an error counts roughly in proportion to the molecules it affects, including the sparse hot molecules ahead of the shock. The tail norm adds a weight growing as the fourth power of the peculiar velocity, for the molecules that carry the heat flux and the fourth-order moments; a result of Ko et al.~\cite{Ko2026WeightedBGK} for weighted residuals of the BGK equation motivates this weight but does not prove the convergence of the present loss (Supplementary Section~\sref{supp:loss_details}). The coefficients $a_m$ are one by default.

\begin{figure}[tp]
\centering
\includegraphics[width=0.8\textwidth]{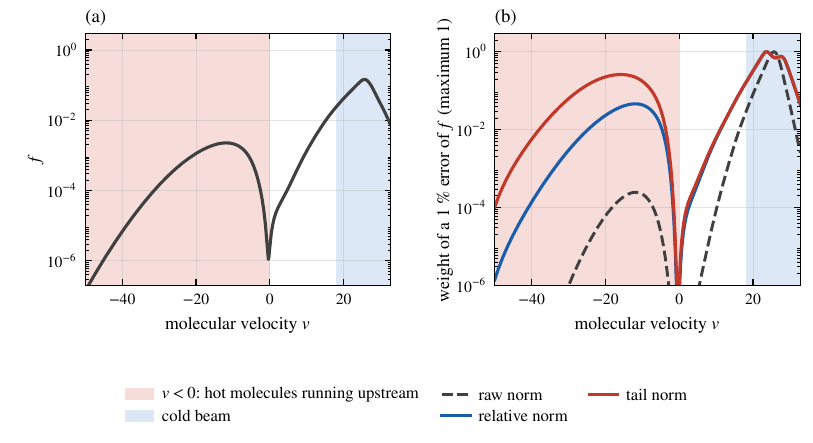}
\caption{Weight of the three norms of the residual for the BGK shock at Mach 20, 20 length units ahead of the density midpoint. (a) Reduced distribution. (b) Contribution of the same relative error of $f$, 1 per cent at every velocity, to the raw, the relative and the tail norm, each curve divided by its maximum; envelope and tail weight as computed by the training code for the trained network. Red shading: $v<0$, the hot molecules running upstream (3 per cent of the density, 99 per cent of the fourth-order moment); blue shading: the cold beam.}
\label{fig:norms}
\end{figure}

\Cref{fig:norms} shows why one norm is not enough. At the position of \cref{fig:form}(b) the hot molecules that run upstream are 3.0 per cent of the gas and carry 98.9 per cent of $\int c^4g\,\dd v$, but they receive only 0.06 per cent of the raw norm and 8 per cent of the relative norm, against 36 per cent of the tail norm. This is an argument from the weights; what the terms do in training is tested in \cref{sec:design_terms}.

Further terms impose conservation at each collocation point and fix the origin. With $J_n(x)$, $n=1,2,3$, the fluxes of mass, momentum and energy of the network and $J^{\mathrm{RH}}_n$ their upstream values,
\begin{equation}
  \mathcal{L}_{\mathrm{flux}}=\Bigl\langle\sum_{n=1}^{3}\Bigl(\frac{J_n(x_i)}{J^{\mathrm{RH}}_n}-1\Bigr)^2\Bigr\rangle .
  \label{eq:lflux}
\end{equation}
The low-order velocity moments $P_k$ of the residual, made dimensionless with downstream scales, are the local balances of mass, momentum, energy, stress and heat flux (five for the monatomic gas; seven for nitrogen, with the rotational heat flux and the rotational energy), and
\begin{equation}
  \mathcal{L}_{\mathrm{mom}}=\Bigl\langle\sum_k P_k(x_i)^2\Bigr\rangle .
  \label{eq:lmom}
\end{equation}
Because the kinetic equation does not determine where the shock lies, the density at the origin is pinned,
\begin{equation}
  \mathcal{L}_{\mathrm{pin}}=\Bigl[\frac{\rho(0)-\bar\rho}{\bar\rho}\Bigr]^2,\qquad \bar\rho=\frac{\rho_1+\rho_2}{2}.
  \label{eq:lpin}
\end{equation}
The fluxes and the $P_k$ are written out in \ref{app:reduced}. The loss is
\begin{equation}
  \mathcal{L}=\mathcal{L}_{\mathrm{rel}}+\lambda_T\,\mathcal{L}_{\mathrm{tail}}+\mathcal{L}_{\mathrm{raw}}
  +\lambda_F\,\mathcal{L}_{\mathrm{flux}}+\lambda_P\,\mathcal{L}_{\mathrm{mom}}+\lambda_0\,\mathcal{L}_{\mathrm{pin}},
  \label{eq:loss}
\end{equation}
with the default weights $\lambda_T=1$, $\lambda_F=30$, $\lambda_P=0.1$ and $\lambda_0=30$. It contains the kinetic equation, the analytic end states through \eqref{eq:form}, and conservation; no discrete-velocity or Monte Carlo data enter it.

The envelope and the tail weight carry no gradient. In the Adam stage they are recomputed at every step; in a full-batch stage they are evaluated once at its start and then frozen, so that the loss is a fixed function of the parameters during the line searches. Loss values of different stages are therefore not comparable.

From Mach 5 upward three changes are made in the quasi-Newton stages, each found by locating the loss of a stalled nitrogen run (\cref{sec:design_terms}): (i) $\lambda_T=0.1$, because with $\lambda_T=1$ the tail term is dominated by the few hot molecules far upstream; (ii) $a_l=(T_1/T_2)^2$, because the residual of $l$, a passive distribution that enters no collision target and no other equation, made up four fifths of $\mathcal{L}_{\mathrm{rel}}$ and $\mathcal{L}_{\mathrm{tail}}$ at Mach 7; (iii) $\lambda_P=100$ and $\lambda_F=3000$ in a final stage of 300 iterations, because the moments of the residual were about $10^{-3}$ of the loss and a one-signed error of the rotational-energy balance accumulated along the shock. These weights were chosen at Mach 5 to 8 with the reference available (\cref{tab:information}). Above Mach 20 the weight $(T_1/T_2)^2$ of (ii) turned out to be too small for the fourth-order moments of the BGK model and is replaced by \eqref{eq:al_rule} (\cref{sec:mono_m50}).

\subsection{Collocation points and velocity nodes}\label{sec:points}

\begin{figure}[tp]
\centering
\includegraphics[width=0.8\textwidth]{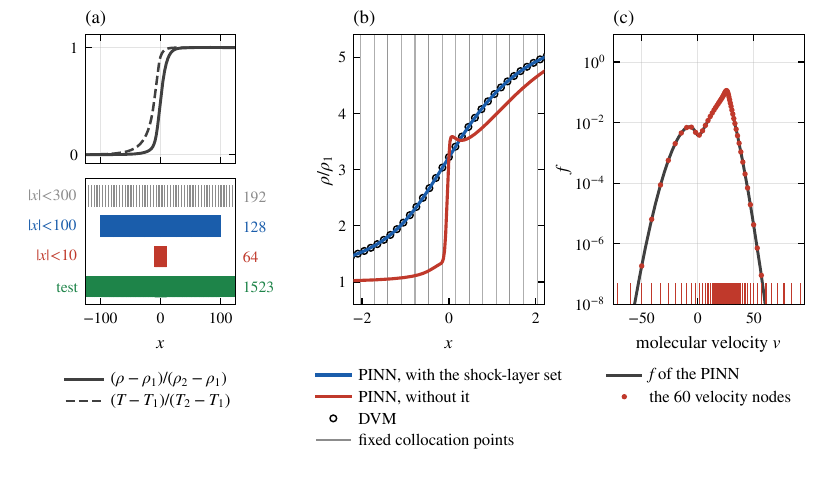}
\caption{Collocation points and velocity nodes. (a) Mach 20 (BGK): normalised density and temperature and, below, the fixed collocation points of the quasi-Newton stages in their three nested sets and the 1523 test points, which are never used in training; the view is $|x|<125$ of the domain $|x|<300$. (b) Nitrogen, Mach 7, trained from scratch without and with the shock-layer set (\cref{sec:design_points}); symbols: DVM; thin vertical lines: the fixed collocation points. (c) The 60 velocity nodes on the reduced distribution of the Mach 20 BGK shock at $x=-10$; the nodes are repeated as ticks on the axis.}
\label{fig:points}
\end{figure}

\paragraph{Collocation points}
The equation is enforced at collocation points in nested sets that follow the shock (\cref{fig:points}a). In the quasi-Newton stages the points are fixed: the midpoints of 128 equal cells of the whole domain, of 128 cells of the core $|x|<L/3$, of 32 cells of each end interval $5L/6<|x|<L$ and of 64 cells of the shock layer $|x|<10$, 384 points in all. At Mach 20 ($L=300$) their spacing is 0.31 in the layer, 1.6 in the core and at the ends, and 4.7 in between. In the Adam stage new points are drawn at every step, one at a random position in each cell of a partition of the same intervals (144 points per step). A further set of 1523 points in the same intervals is never used in training; these test points serve the test of \cref{sec:acceptance}. Supplementary Section~\sref{supp:points_details} gives the partitions.

\paragraph{Velocity nodes}
The distribution combines a narrow upstream beam of unit thermal speed with a hot part $\sqrt{T_2}$ wide, eleven units at Mach 20. The nodes are therefore stretched away from the beam,
\begin{equation}
  v_j=u_1+d\,\sinh\frac{\Delta v\,\xi_j}{d},\qquad w_j=\Delta v\,\cosh\frac{\Delta v\,\xi_j}{d},\qquad \xi_j=\xi_0+j+\varsigma ,
  \label{eq:vnodes}
\end{equation}
with $d=4$ and $\Delta v=0.5$, on the interval $[u_2-7.5\sqrt{T_2},\ \max(u_1+7,\ u_2+7.5\sqrt{T_2})]$, whose lower end fixes $\xi_0$; the spacing is $\Delta v$ at the beam and grows like $[1+(v-u_1)^2/d^2]^{1/2}$ away from it. This is the midpoint rule in the variable $\xi$, in which the integrands are smooth and decay rapidly, so that it is spectrally accurate for any offset $\varsigma$. The quasi-Newton stages use $\varsigma=1/2$; in the Adam stage $\varsigma$ is drawn uniformly from $[0,1)$ at every step, so that the equation is enforced at all velocities. The rule has 39 to 60 nodes for the monatomic gas between Mach 5 and 20 and 35 to 51 for nitrogen between Mach 5 and 15, against 337 nodes of a uniform rule with the same spacing at Mach 20 (\cref{fig:points}c); its accuracy is checked in Supplementary Section~\sref{supp:points_details}. For the Shakhov shock at Mach 20 the spacing at the beam had to be halved (\cref{sec:mono_m20}). At Mach 2 a uniform rule with $\Delta v=0.25$ is used, without the shock-layer set.

\subsection{Optimiser and continuation in the Mach number}\label{sec:optimiser}

At the first Mach number of a continuation (Mach 5) the training has three stages. The first is Adam \cite{KingmaBa2015Adam}, for 40\,000 steps on random collocation points and velocity offsets, with the learning rate decaying geometrically from $10^{-3}$ to $10^{-5}$. The second, optional, stage is a Levenberg--Marquardt solve \cite{Levenberg1944,Marquardt1963} for the output layer: with the hidden layers fixed, the loss on the fixed points is a sum of squares in the 441 or 588 output weights, minimised by at most eight damped Gauss--Newton iterations. Both stages use the default weights; their settings are given in Supplementary Section~\sref{supp:optimiser_details}. The third stage is a dense self-scaled Broyden method \cite{OrenLuenberger1974SelfScaling,UrbanStefanouPons2025,Kiyani2025Optimizer} on all parameters, with fixed collocation points, fixed velocity nodes and the frozen envelope: a quasi-Newton method with a full approximation of the inverse Hessian and a strong-Wolfe line search (Supplementary Section~\sref{app:optimisers}). It needs the exact gradient of the function it minimises, which is why the envelope is frozen. A stage ends after a number of iterations fixed in advance, or earlier after six consecutive failed searches, at a gradient norm below $10^{-13}$ or at its time limit; there is no tolerance on the loss. At Mach 5 this stage runs for 1500 iterations with the weights (i) and (ii) of \cref{sec:loss}, followed by 300 iterations with (iii).

Every higher Mach number is reached by continuation: the quasi-Newton method alone, started from the parameters of the Mach number below as they were before its final stage, for 800 iterations (600 to 800 for nitrogen), followed by the final 300. The inputs \eqref{eq:inputs} and the base \eqref{eq:mixture} are rescaled with the new Rankine--Hugoniot states, and $L$ and $a_l$ follow their rules. The sequences are $5\to7\to8\to10\to12\to15\to20$ for the BGK and Shakhov models and $5\to7\to8\to10\to12\to15$ for nitrogen. The monatomic sequences used this procedure from the start; the nitrogen sequence reached it at Mach 8, and its solutions at Mach 5 and 7 carry earlier stages whose weights were later changed (Supplementary Section~\sref{app:settings_shock}). Adam from scratch at Mach 8, and Adam across a large step in Mach number, fail (\cref{sec:design_opt}).

\subsection{Acceptance tests without a reference}\label{sec:acceptance}

Three tests use no reference solution. (a) The largest relative deviation of the fluxes of mass, momentum and energy from their upstream values, $e_F$, is below $10^{-3}$; it is evaluated on positions spaced by 0.05 (0.025 for nitrogen) across the whole domain, with a uniform velocity rule of step 0.25. (b) The ratio $r_{\mathrm{rel}}$ of $\mathcal{L}_{\mathrm{rel}}$ at the test points to its value at the training points, both with the envelope of the final solution, is at most ten. (c) The fields change by less than $10^{-3}$ of their downstream scales when the moments are evaluated on a finer and wider velocity rule. The points and velocity rules of (b) and (c) for each family of runs are given in Supplementary Section~\sref{supp:acceptance_details}.

Which tests decide whether a run is accepted was fixed in advance for each family of runs. For the monatomic sequences (\cref{sec:mono}), the nitrogen runs from scratch and the network with the Mach number as an input (\cref{sec:n2}), tests (a) and (b) were fixed before any training, and (c) is recorded. For the nitrogen sequence only test (a) was evaluated as specified: test (b) was written down after the runs to Mach 12, and at Mach 15 it was specified but, through an omission in the evaluation script, not evaluated. For the calculations with three velocity dimensions all three tests were fixed in advance. A run that misses one of its tests is not accepted, whatever its distance from a reference, and is marked so in every table in which it appears: the Shakhov run at Mach 15 misses (a), and the second Boltzmann run at Mach 3 misses (c). \Cref{tab:information} lists the status of every result.

The tests are necessary but not sufficient. One Shakhov run at Mach 20 passed (a) and (b) and was 46 per cent off in $\Delta$ (\cref{sec:mono}), and a good run can miss (b) because of a single collocation point next to the inflow boundary (\cref{sec:design_terms}); the ratio of (b) is close to one for good and poor runs alike. The typical failures of this problem (a shock collapsed between two collocation points, a quadrature too coarse for the beam, a domain too short for the upstream-running molecules) are detected by the tests before any reference is consulted. Shakhov runs at Mach 30 to 50 also pass the three tests and are 4 to 18 per cent off in the heat flux (\cref{sec:mono_shk_limit}). The residuals of the exact stress and heat-flux balances discriminate better (\cref{sec:mono_m20,sec:mono_shk_limit}); they were evaluated after the fact for the monatomic runs and were not among the tests fixed in advance.

\subsection{Other collision models}\label{sec:models_diff}

\begin{figure}[tp]
\centering
\includegraphics[width=\textwidth]{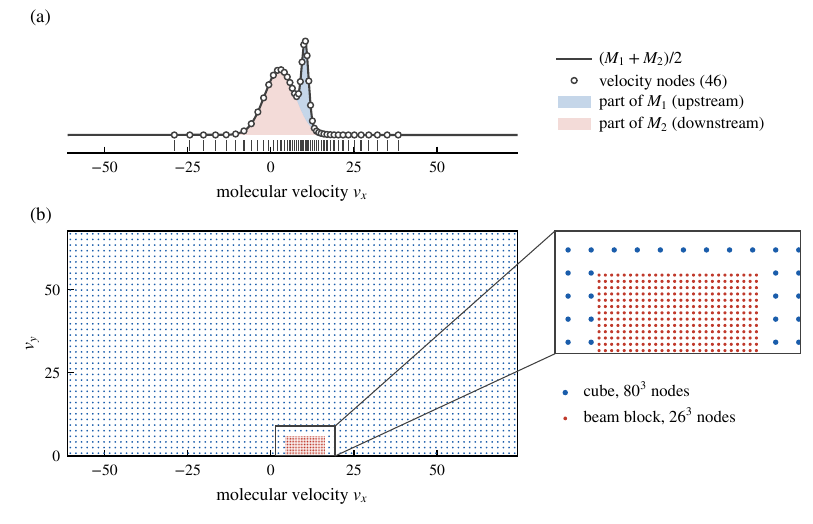}
\caption{Velocity nodes at Mach 8 on one velocity scale. (a) Relaxation models: the stretched one-dimensional rule (46 nodes for the monatomic gas), drawn on the mixture $(\M_1+\M_2)/2$, whose parts from $\M_1$ and $\M_2$ are shaded, with the nodes repeated as ticks. (b) Two-scale three-dimensional nodes in a plane of constant $v_z$: a uniform cube of $80^3$ nodes for the hot gas and a block of $26^3$ nodes around the upstream beam, enlarged on the right; 529\,576 nodes in all. These nodes are used for the Boltzmann equation at Mach 8 and 10 (\cref{sec:boltz_M8M10}).}
\label{fig:grids}
\end{figure}

The form \eqref{eq:form}, the six terms of \eqref{eq:loss} and the optimisers are common to the four collision models; what changes is what the collision term needs. The BGK target needs $\rho$, $u$ and $T$; the Shakhov target also needs $q$ and is negative in the fast tail. Both use three reduced distributions and five moments of the residual. The nitrogen model adds the distribution $h$, the temperature $T_{\mathrm{rot}}$, the heat flux $q_{\mathrm{rot}}$, the second relaxation time $Z\tau$, the rotational enthalpy in the energy flux and two rotational balances in $\mathcal{L}_{\mathrm{mom}}$. For these three models about fifty velocity nodes suffice (\cref{fig:grids}a).

The Boltzmann integral needs $f$ itself in three velocity dimensions (\ref{app:collision}). The unknown is $f(x,v,r)$ in the form \eqref{eq:form}; the network receives $r^2/v_s^2$ as a seventh input and returns a fourth coefficient, which multiplies $r^2/v_s^2$ in \eqref{eq:exponent}. The residual $v\,\partial_xf-Q(f,f)$ is collocated on a cube of velocity nodes, on which $Q$ is evaluated by the fast spectral method \cite{MouhotPareschi2006Fast} (Supplementary Section~\sref{supp:collision_supp}). The loss keeps its six terms, with the three-dimensional peculiar velocity $|\bc|^2=(v-u)^2+r^2$ in the tail weight and in the five $P_k$ (Supplementary Section~\sref{supp:threed}); the weights are $\lambda_T=0.1$ and the default $\lambda_F$, $\lambda_P$ and $\lambda_0$, and there is no passive distribution and no final stage. A uniform cube needs a step of about 0.73 at any Mach number, and the number of directions needed for the collisions between the beam and the hot gas grows with the square of the Mach number. Two devices therefore take the cold beam out of the velocity mesh. At Mach 3 and 5 the beam is split off as the upstream Maxwellian times a polynomial, and its collisions are evaluated in closed form ($48^3$ nodes and $12\times24$ directions at Mach 3, $64^3$ nodes and $16\times32$ directions at Mach 5). For stronger shocks the nodes are placed on two scales (\cref{fig:grids}b), a cube for the hot part and a block around the beam, and $f=(1-s)\M_1\mathcal{F}_b+s\,\M_2\mathcal{F}_h$, where $\M_1$ and $\M_2$ are the end-state Maxwellians and each factor $\mathcal{F}$ has the form $1+\chi\sinh\Nth$ with its own network (15\,177 parameters in all).

Each Boltzmann stage is preceded by a Shakhov stage on the same velocity nodes. It tests the discretisation against the DVM solution of the same model and, with the hard-sphere viscosity of \cref{sec:units}, provides the starting point of the Boltzmann stage in the same length unit. The Boltzmann solutions at Mach 2, 3 and 5, the Shakhov solutions on the two-scale nodes at Mach 8 and 10 and the Boltzmann solutions that start from them are reported in \cref{sec:boltz}; stages above Mach 10 are not reported (\cref{sec:boltz_progress}).

\section{Reference solutions and comparison}\label{sec:protocol}

Every network is compared with a solution of the same kinetic model computed by a conventional method: DVM solutions for the relaxation models and DSMC for hard spheres. This section summarises the references, defines the measure of the difference, and states when a reference was available while the method was developed.

\subsection{Discrete-velocity references for the shocks}\label{sec:refs_shock}

The discrete-velocity solvers, their algorithms and their convergence records are described in Supplementary Section~\sref{supp:dvm}; only their accuracy is summarised here. For the BGK and Shakhov shocks the reference is a steady solver of the reduced equations \eqref{eq:reduced_kinetic}. It iterates on the macroscopic fields and, for given fields, integrates each velocity exactly along $x$ in the optical depth, with fourth-order accuracy in the spatial step $\Delta x$. The solutions used here have $\Delta x=0.025$ and a velocity step of 0.25, on domains from $[-190,80]$ at Mach 5 to $[-710,260]$ at Mach 20 and $[-3530,1260]$ at Mach 100. Their fluxes are constant to $2.5\times10^{-10}$ or better, and their estimated numerical error is between $2\times10^{-10}$ and $4\times10^{-8}$ in the measure \eqref{eq:error}.

For nitrogen every comparison of \cref{sec:n2} uses a second fourth-order solver for the four reduced distributions, written independently for the tests at Mach 10 and 15 and then run at every Mach number of the nitrogen tables ($\Delta x=0.025$, or 0.05 at Mach 2, 5 and 7). Its estimated error is $10^{-9}$ to $3\times10^{-8}$ between Mach 8 and 15, and at Mach 2, 5 and 7 its two finest grids differ by at most $6\times10^{-6}$. The method was developed against a first, pseudo-time solver at Mach 2, 5, 7, 8 and 12, whose error is larger, up to $5\times10^{-4}$ in the heat fluxes, the stress and the higher moments; only the differences quoted from the development of the method (\cref{sec:design_terms,sec:design_points,sec:design_opt}) are against it. The error of the fourth-order references, including their interpolation to the evaluation points, is negligible against every difference reported for a network (Supplementary Section~\sref{supp:refs_details}).

\subsection{Monte Carlo reference for hard spheres}\label{sec:refs_dsmc}

For hard spheres the networks are compared with DSMC calculations \cite{Bird1994} made for this work (Supplementary Section~\sref{supp:collision_supp}): the no-time-counter scheme in cells of $\lambda_1/160$ with 16\,000 simulated molecules per $\lambda_1$, and 48 independent ensembles, which give the standard errors. The fluxes are constant to $7\times10^{-5}$, $8\times10^{-5}$, $5\times10^{-5}$ and $4\times10^{-5}$ at Mach 2, 3, 5 and 8. At Mach 10 the reference is a calculation with the cell and the time step halved and 24 ensembles; over the shock it differs from the calculation with the standard cell by less than $10^{-3}$ of the jumps. The profiles agree with the three-decimal table of Ohwada \cite{Ohwada1993RIMS} at Mach 1.1 to within $1.3\times10^{-3}$, and with the table of Dodulad and Tcheremissine \cite{DoduladTcheremissine2013Shock} at Mach 3 and 8 to between $2\times10^{-4}$ and $1.5\times10^{-3}$ of the downstream values; halving the cell and the time step at Mach 5 changes them by less than $10^{-3}$ of their scales. The standard errors are of order $10^{-4}$ of the scale of each field at Mach 5 and reach $2\times10^{-3}$ for $m_{xxx}$ at Mach 2, so that the statistical error of the reference is part of every difference reported in \cref{sec:boltz}.

\subsection{Measure of the difference}\label{sec:error}

For a field $a$ the difference between a network $a_\theta$ and a reference $a_{\mathrm{ref}}$ is the relative $L^2$ norm
\begin{equation}
  \varepsilon_a=\Biggl[\frac{\sum_{x_i\in I}\bigl(a_\theta(x_i)-a_{\mathrm{ref}}(x_i)\bigr)^2}{\sum_{x_i\in I}b_{\mathrm{ref}}(x_i)^2}\Biggr]^{1/2},
  \label{eq:error}
\end{equation}
where $b_{\mathrm{ref}}=a_{\mathrm{ref}}-a_{\mathrm{ref},1}$ for $\rho$, $u$, $T$, $T_{\mathrm{tr}}$ and $T_{\mathrm{rot}}$, with $a_{\mathrm{ref},1}$ the upstream value, and $b_{\mathrm{ref}}=a_{\mathrm{ref}}$ for all other fields. The sums run over evaluation points spaced uniformly by 0.05 (0.025 for nitrogen) inside a window $I$, and the reference is interpolated linearly to them. For the density, the velocity and the temperatures the difference is thus measured against the change of the field across the shock, which is stricter than the norm of the field used in the previous paper \cite{Roohi2026KineticShockPINN}; the other fields vanish in equilibrium and are measured against their own norm. The density thickness $\delta_\rho$ of \cref{sec:shockproblem} is formed with the largest centred difference quotient of $\rho$ on the evaluation points of each solution (for the fourth-order nitrogen references, with the largest slope of a quartic through five nodes).

The reference is translated so that its density midpoint $(\rho_1+\rho_2)/2$ lies at $x=0$, the origin that \eqref{eq:lpin} imposes on the network; in the calculations with three velocity dimensions the profile of the network is translated by the same rule. No shift and no amplitude is fitted. The window is $|x|\le\min(2L/3,40)$ for nitrogen, that is 20, 33.3 and 40 at Mach 2, 5 and above. For the BGK and Shakhov shocks it is the interval in which the reference density differs from both end states by more than $10^{-4}$ of the jump, extended by 10 on each side, so that it contains the precursor. For the calculations with three velocity dimensions it is $|x|\le40$, except for hard spheres at Mach 2, 5, 8 and 10, where it is $|x|\le20$, 18, 17 and $18\,\lambda_1$, the extent of the DSMC profiles. The nitrogen window excludes most of the far precursor, where the networks are least accurate (\cref{sec:n2_weak}); the nitrogen differences are therefore not directly comparable with the monatomic ones and are favourable to the network.

Where a difference is of the size of the error of the reference, as for the bulk fields of hard spheres at Mach 2 (\cref{sec:boltz_results}), it does not measure the error of the network.

\subsection{Use of the references during development}\label{sec:information}

No reference enters a loss, the choice of points, the acceptance of an optimiser step or a stopping rule in any calculation of this paper. Every reported result is the state at the end of a stage whose number of iterations or time limit was fixed in advance, not the state closest to a reference. The choice of the reported run was, however, influenced by a reference in two cases: the Shakhov continuation at Mach 20 was started because the comparison had shown its predecessor to be wrong (\cref{sec:mono_m20}), and the Shakhov run at Mach 15 was kept after the comparison (\cref{sec:mono_chain}). In one case a reference was consulted before a run ended: the Boltzmann stage at Mach 10 was compared at iteration 550 while it continued unchanged to the 800 iterations fixed in advance (\cref{sec:boltz_M8M10}). \Cref{tab:information} gives, for every result, whether its reference was available during development, what was fixed in advance and how the result fared in the tests of \cref{sec:acceptance}; Supplementary Table~\sref{tab:information_full} adds the optimiser of each result.

\begin{table}
\centering
\scriptsize
\setlength{\tabcolsep}{3pt}
\caption{Status of every result. Reference available: whether the reference of the same case could influence the training procedure, the weights or the choice of the reported run. Tests: the reference-free tests of \cref{sec:acceptance}; n.e., not evaluated.}
\label{tab:information}
\renewcommand{\arraystretch}{1.0}
\begin{tabularx}{\textwidth}{@{}>{\raggedright\arraybackslash}p{0.30\textwidth}LL>{\raggedright\arraybackslash}p{0.17\textwidth}@{}}
\toprule
Result & Reference available & Fixed in advance & Tests \\
\midrule
BGK, $\tau=1/\rho$ and $\tau=1$, Mach 5--20; Shakhov, Mach 5--12 & no (weights chosen on nitrogen) & procedure, tests, comparison & (a), (b) passed \\
BGK, $\tau=1/\rho$, Mach 30, 40 & examined when $L$, inputs and width were chosen & the four changes, tests, comparison & (a), (b) passed \\
BGK, $\tau=1/\rho$, Mach 50 & no; prediction hashed before the reference was read & the four changes, tests, comparison & (a), (b) passed \\
BGK, $\tau=1/\rho$, Mach 70, $a_l=(T_1/T_2)^2$ & reference not blind & the continuation, tests, comparison & (a), (b) passed \\
BGK, $\tau=1/\rho$, Mach 100, $a_l=(T_1/T_2)^2$ & no; prediction hashed before the reference was read & the continuation, tests, comparison & (a), (b) passed \\
BGK, $\tau=1/\rho$, Mach 60, both weights, 1100 and 2100 iterations & no; reference computed after the four predictions were hashed & the four runs, tests, comparison & (a), (b) passed \\
BGK, $\tau=1/\rho$, weight \eqref{eq:al_rule}, Mach 20--100 & yes: weight tested at Mach 20 against the reference; references examined before & the rule \eqref{eq:al_rule}, each run, tests, comparison & (a), (b) passed \\
BGK, $\tau=1/\rho$, bounded form, Mach 5--12 & no & the one change (form), tests, comparison & (a), (b) passed \\
Shakhov, Mach 15 & kept after the comparison & procedure, tests, comparison & (a) missed \\
Shakhov, Mach 20 (attempt C) & started after the comparison of attempt B & each attempt & (a), (b) passed \\
Shakhov, Mach 15 and 20, continuation from Mach 12 & references compared with earlier runs & settings, tests, comparison & (a), (b), (c) passed \\
Shakhov, Mach 30 to 50, both weights (limit of the method) & Mach 30, 40: examined before; Mach 50: prediction with the first weight hashed before the reference was read & settings, tests, comparison & (a), (b), (c) passed; far from the reference \\
Nitrogen, Mach 2, 5, 7, 8, 12 & yes, first DVM printed in training & nothing & (a) passed; (b) n.e. \\
Nitrogen, Mach 10 & no reference existed & solution frozen and hashed before the reference & (a) passed; (b) n.e. \\
Nitrogen, Mach 15 & reference peaks seen before the job ended & procedure and commands; prediction hashed & (a) passed; (b) n.e. \\
Nitrogen, Mach 5 ablation and runs from scratch & no & each run & see Table~\sref{tab:ablation} \\
Nitrogen, one network for Mach 2--8 & no & run, tests, freezing & (a), (b) passed \\
Boltzmann, Mach 2 (two runs) & no & each run & (a), (b), (c) passed \\
Boltzmann, Mach 3, first run & no & the run & (a), (b), (c) passed; failed line searches; continued by the second run \\
Boltzmann, Mach 3, second run & no & the one change & (c) missed; time limit \\
Boltzmann, Mach 5 & no; compared once with the DSMC after the three tests & the one change (operator) & (a), (b), (c) passed \\
Shakhov, two-scale nodes, Mach 8, 10 & Mach 8 DVM compared with earlier runs & each run & (a), (b), (c) passed \\
Boltzmann, Mach 8 & no; compared once with the DSMC after the three tests & the run, 800 iterations, tests & (a), (b), (c) passed \\
Boltzmann, Mach 10 & interim comparison at iteration 550 of 800; final state compared once & the run, 800 iterations, tests & (a), (b), (c) passed \\
\bottomrule
\end{tabularx}
\end{table}

The nitrogen method was developed at Mach 2, 5, 7, 8 and 12 while the distance from the first DVM solver was printed during training, and the three weights of \cref{sec:loss} were chosen in this way. The procedure was then tested at Mach 10 and 15, where nothing could be monitored (\cref{sec:n2_blind}); only the Mach 10 solution was frozen before any reference existed. The monatomic sequences were run afterwards, with the training procedure, the acceptance tests and the rule of comparison fixed before any training and the hashes of the code recorded; their training script contains no code that reads a reference, and no reference was present on the cluster. Each run was compared with its reference once, after its reference-free evaluation had been written, and each further attempt changed one thing, chosen from a diagnosis that used no reference and specified under a new name. The calculations with three velocity dimensions followed the same rule. There are two exceptions: the interim comparison of the Boltzmann stage at Mach 10 (\cref{sec:boltz_M8M10}), and the diagnosis of the passive distribution above Mach 20, which used the distribution of the DVM (\cref{sec:mono_m50}).

\section{Monatomic shocks}\label{sec:mono}

\subsection{Runs and procedure}\label{sec:mono_chain}

The monatomic shocks use the training procedure of the nitrogen shock (\cref{sec:n2}) without change, for the BGK ($\Pran=1$) and Shakhov ($\Pran=2/3$) models with $\tau=1/\rho$: the loss \eqref{eq:loss} with the three reduced distributions $g$, $k$, $l$ and five moments of the residual, a network of 7834 parameters, and the half-length $L=15M$. Mach 5 is trained from scratch, and Mach 7, 8, 10, 12, 15 and 20 follow by continuation (\cref{sec:optimiser}). The settings, tests (a) and (b) of \cref{sec:acceptance} and the rule of the comparison were fixed before any job was submitted (\cref{sec:information}; Supplementary Section~\sref{supp:mono_runs}), and each run was compared once with the DVM solution of the same model, on the window of \cref{sec:error}. The weights had been chosen for nitrogen with a reference available, and the Shakhov solution at Mach 20 of \cref{tab:mono_err} was obtained after earlier attempts had been compared (\cref{tab:information}). The Shakhov shocks at Mach 15 and 20 were later solved again from the accepted Mach 12 run with settings fixed in advance (\cref{sec:mono_shk_limit}), and the BGK continuation was extended to Mach 70 (\cref{sec:mono_m50}).

\begin{table}
\centering
\footnotesize
\setlength{\tabcolsep}{2.6pt}
\caption{Monatomic shocks, $\tau=1/\rho$: reference-free tests and relative $L^2$ difference \eqref{eq:error} between the network and the DVM solution of the same model, in per cent. $e_F$: largest flux error, test (a), limit $10^{-3}$; $r_{\mathrm{rel}}$: ratio of test (b), limit 10; Passed: the tests fixed in advance that the run passes (the Shakhov run at Mach 15 misses test (a) and is not accepted). The Shakhov entry at Mach 20 is attempt C of \cref{tab:mono_m20}. The fourth-order entries at Mach 12 and above are uncertain by up to a factor of about two: changes of the base that pass the tests move them by 5 to 80 per cent at Mach 20 (\cref{sec:mono_m20}). Density thickness, minimum heat flux and maximum stress of every run: Supplementary Table~\sref{tab:mono_free}.}
\label{tab:mono_err}
\begin{tabular}{@{}llrrrrrrrrrrl@{}}
\toprule
Model & $M$ & $10^4e_F$ & $r_{\mathrm{rel}}$ & $\rho$ & $u$ & $T$ & $q$ & $\sigma$ & $m_{xxx}$ & $R_{xx}$ & $\Delta$ & Passed \\
\midrule
BGK & 5 & 0.069 & 1.00 & 0.0026 & 0.0015 & 0.0019 & 0.012 & 0.016 & 0.047 & 0.052 & 0.066 & a, b \\
 & 7 & 0.13 & 1.00 & 0.0030 & 0.0027 & 0.0017 & 0.015 & 0.024 & 0.048 & 0.18 & 0.26 & a, b \\
 & 8 & 0.18 & 1.00 & 0.0022 & 0.0014 & 0.0021 & 0.017 & 0.0099 & 0.042 & 0.21 & 0.33 & a, b \\
 & 10 & 0.40 & 1.00 & 0.015 & 0.0047 & 0.0029 & 0.048 & 0.061 & 0.12 & 0.42 & 0.59 & a, b \\
 & 12 & 0.53 & 1.00 & 0.024 & 0.0067 & 0.0041 & 0.068 & 0.11 & 0.14 & 0.45 & 0.64 & a, b \\
 & 15 & 0.39 & 1.00 & 0.020 & 0.0095 & 0.0091 & 0.059 & 0.079 & 0.089 & 2.2 & 3.4 & a, b \\
 & 20 & 0.95 & 1.00 & 0.066 & 0.018 & 0.014 & 0.18 & 0.27 & 0.29 & 3.8 & 5.7 & a, b \\
\midrule
Shakhov & 5 & 0.52 & 1.00 & 0.0029 & 0.0016 & 0.0017 & 0.010 & 0.020 & 0.037 & 0.060 & 0.11 & a, b \\
 & 7 & 2.3 & 1.00 & 0.011 & 0.0083 & 0.0092 & 0.045 & 0.073 & 0.17 & 0.61 & 0.78 & a, b \\
 & 8 & 2.3 & 1.00 & 0.0091 & 0.0074 & 0.0084 & 0.034 & 0.067 & 0.17 & 0.55 & 0.77 & a, b \\
 & 10 & 3.3 & 1.00 & 0.012 & 0.010 & 0.012 & 0.049 & 0.090 & 0.16 & 1.6 & 2.2 & a, b \\
 & 12 & 8.6 & 1.01 & 0.016 & 0.014 & 0.020 & 0.086 & 0.084 & 0.17 & 3.5 & 5.0 & a, b \\
 & 15 & 19 & 1.04 & 0.049 & 0.058 & 0.11 & 0.41 & 0.24 & 0.86 & 4.7 & 6.7 & b \\
 & 20 & 0.82 & 0.94 & 0.021 & 0.018 & 0.038 & 0.13 & 0.11 & 0.24 & 2.3 & 3.5 & a, b \\
\bottomrule
\end{tabular}
\end{table}

\begin{figure}[tp]
\centering
\includegraphics[width=\textwidth]{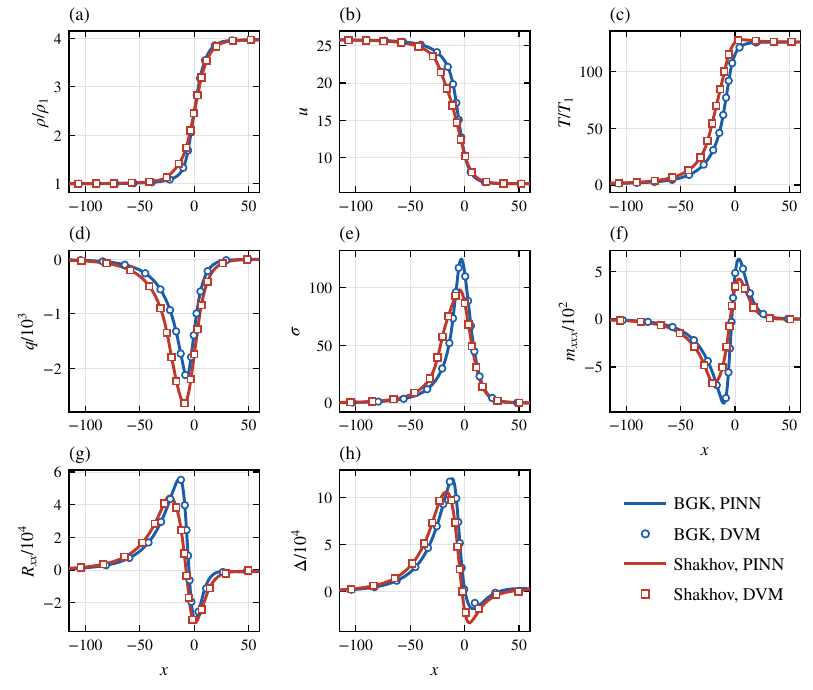}
\caption{Monatomic gas, Mach 20, BGK (blue) and Shakhov (red): (a) $\rho$, (b) $u$, (c) $T$, (d) $q$, (e) $\sigma$, (f) $m_{xxx}$, (g) $R_{xx}$, (h) $\Delta$. Lines: PINN; symbols: DVM of the same model.}
\label{fig:mono_profiles}
\end{figure}

\Cref{tab:mono_err} lists the tests and the differences from the DVM, \cref{fig:mono_profiles} shows the profiles at Mach 20, and Supplementary Fig.~\sref{fig:mono_errors} draws the differences against the Mach number. All BGK runs and the Shakhov runs to Mach 12 pass tests (a) and (b) and reproduce the fields up to third order within 0.3 per cent. The error moves to the fourth-order moments as the Mach number grows: for BGK $R_{xx}$ and $\Delta$ stay within 0.7 per cent up to Mach 12 and reach 3.8 and 5.7 per cent at Mach 20, and for Shakhov they reach 3.5 and 5.0 per cent at Mach 12. The Shakhov run at Mach 15 misses test (a) ($e_F=1.9\times10^{-3}$) and is not accepted; it is listed because it was kept after the comparison (\cref{sec:information}). Under a finer and wider velocity rule, test (c), no field changes by more than $2.6\times10^{-4}$ of its downstream scale. The maximum of the normal temperature $T_{xx}$, which the constancy of the mass and momentum fluxes fixes as a function of the density (Yen~\cite{Yen1966Overshoot}), is reproduced within $4\times10^{-5}$ of its value in every run; this overshoot therefore does not test the collision model (Supplementary Section~\sref{supp:mono_txx}).

The difference between the BGK and Shakhov shocks depends on the collision model, and the networks reproduce it. At Mach 20 the Shakhov shock is thicker ($\delta_\rho=25.2$ against 19.6), has a larger heat flux and a smaller stress, and its temperature overshoots the downstream value by 1.6 per cent (maximum 127.95 for the network and 127.96 for the DVM, $T_2=125.87$), whereas the BGK temperature of the DVM is monotone and that of the network exceeds $T_2$ by only $1.1\times10^{-4}$, far downstream. The thickness of the network is within 0.25 per cent of the DVM value in every run that passes the tests and 0.36 per cent low for the Shakhov run at Mach 15 (Supplementary Table~\sref{tab:mono_free}). Which model is closer to the Boltzmann equation is a different question; against DSMC of Maxwell molecules the Shakhov shock is the closer in thickness, but its overshoot is two to seven times too large (\cref{sec:validation}).

\subsection{The Shakhov shock at Mach 20}\label{sec:mono_m20}

\begin{table}
\centering
\footnotesize
\setlength{\tabcolsep}{3.5pt}
\caption{Runs at Mach 20 ($\tau=1/\rho$) that change the procedure, each fixed in advance, with their reference-free tests and their differences from the DVM in per cent. A: smaller step in Mach number; B: $\Delta v=0.25$; C: B continued by 2000 and by 4000 quasi-Newton iterations; D: B with a network of width 64; the last two runs of each model change the base \eqref{eq:mixture}. The reported results are marked $\ast$. Two runs whose base width followed the free path of each molecule (a collision-history base) collapsed ($e_F=0.20$ and 0.31, thickness 0.2) and are not listed; continuations of the reported BGK run are in Supplementary Table~\sref{tab:envelope_control}.}
\label{tab:mono_m20}
\begin{tabularx}{\textwidth}{@{}lLrrrrrrrrrr@{}}
\toprule
Model & Run & $10^4e_F$ & $r_{\mathrm{rel}}$ & $\rho$ & $u$ & $T$ & $q$ & $\sigma$ & $m_{xxx}$ & $R_{xx}$ & $\Delta$ \\
\midrule
Shakhov & Planned, $15\to20$ & 21 & 1.03 & 0.26 & 0.37 & 0.80 & 2.7 & 1.6 & 5.9 & 26 & 39 \\
 & A, $15\to17\to20$ & 21 & 1.03 & 0.20 & 0.21 & 0.46 & 1.6 & 0.94 & 3.5 & 17 & 24 \\
 & B & 2.6 & 0.96 & 0.24 & 0.33 & 0.71 & 2.5 & 1.4 & 5.4 & 31 & 46 \\
 & C, 2000 & 1.2 & 0.99 & 0.045 & 0.031 & 0.063 & 0.23 & 0.17 & 0.47 & 8.7 & 13 \\
 & C, 4000 $\ast$ & 0.82 & 0.94 & 0.021 & 0.018 & 0.038 & 0.13 & 0.11 & 0.24 & 2.3 & 3.5 \\
 & D & 2.7 & 0.95 & 0.21 & 0.33 & 0.71 & 2.4 & 1.4 & 5.3 & 20 & 29 \\
 & Third Maxwellian & 0.74 & 0.95 & 0.023 & 0.018 & 0.037 & 0.13 & 0.11 & 0.25 & 1.7 & 2.7 \\
 & Precursor base & 0.61 & 0.85 & 0.0072 & 0.011 & 0.022 & 0.078 & 0.050 & 0.17 & 1.2 & 1.7 \\
\midrule
BGK & Planned, $15\to20$ $\ast$ & 0.95 & 1.00 & 0.066 & 0.018 & 0.014 & 0.18 & 0.27 & 0.29 & 3.8 & 5.7 \\
 & Third Maxwellian & 1.1 & 1.00 & 0.071 & 0.020 & 0.018 & 0.19 & 0.29 & 0.34 & 3.6 & 5.4 \\
 & Precursor base & 0.66 & 1.00 & 0.043 & 0.012 & 0.012 & 0.11 & 0.19 & 0.29 & 6.9 & 10 \\
\bottomrule
\end{tabularx}
\end{table}

The Shakhov shock at Mach 20 was not solved with the settings fixed in advance, and the attempts of \cref{tab:mono_m20} show what the reference-free tests detect and what they miss; Supplementary Section~\sref{supp:mono_m20_attempts} gives the full account. The planned step and attempt A missed test (a), and a diagnosis without the reference located the flux error at the cold beam. Attempt B, with $\Delta v=0.25$, passes tests (a) and (b) and is still 2.5 per cent from the DVM in $q$ and 46 per cent in $\Delta$, which shows that the tests are not sufficient; its loss was still falling, and the residuals of the exact stress and heat-flux balances on its profiles were $1.4\times10^{-2}$ and $1.5\times10^{-2}$. Attempt C continued B for 4000 iterations and is the reported result (0.13 per cent in $q$ and $\sigma$, 2.3 and 3.5 per cent in $R_{xx}$ and $\Delta$, balance residuals $2.1\times10^{-3}$ and $1.4\times10^{-3}$), but not a blind one: it was started because the comparison had shown B to be wrong. Attempt D, B with a wider network, reproduced B in the fields up to third order (2.4 per cent in $q$).

Three changes of the base \eqref{eq:mixture} were also tried at Mach 20, and none was retained (Supplementary Section~\sref{supp:mono}). They show how weakly the fourth-order entries are determined: adding a third Maxwellian whose trained weight stayed negligible moved $R_{xx}$ and $\Delta$ by 25 per cent (Shakhov) and 5 per cent (BGK), and the precursor base, which also passes the tests, moved them by 50 per cent (Shakhov) and 80 per cent (BGK). Continuing the reported BGK run by 1300 iterations reduced the errors of $\rho$, $q$ and $\sigma$ three- to fivefold but left those of $R_{xx}$ and $\Delta$ unchanged (Supplementary Table~\sref{tab:envelope_control}), so their error is not simply unfinished optimisation; its cause, the weight of the passive distribution $l$, is identified in \cref{sec:mono_m50}. A second BGK continuation with another seed, compared once, gave 6.8 and 10 per cent in $R_{xx}$ and $\Delta$ at Mach 20 against 3.8 and 5.7 per cent, with all lower-order fields within a factor 1.5 of the first. Together with the constant-collision-time runs below, these results indicate that, with the weight $a_l=(T_1/T_2)^2$, the fourth-order entries at Mach 12 and above are uncertain by up to a factor of about two.

\subsection{Constant collision time}\label{sec:mono_tau1}

\begin{table}
\centering
\small
\setlength{\tabcolsep}{4.5pt}
\caption{BGK model with constant collision time $\tau=1$, the convention of the BGK calculations of \cite{Roohi2026KineticShockPINN}: relative $L^2$ difference from the DVM solution of the same model in per cent, and density thickness. All seven runs pass tests (a) and (b).}
\label{tab:mono_tau1}
\begin{tabular}{rrrrrrrrrrr}
\toprule
 & & & & & & & & & \multicolumn{2}{c}{$\delta_\rho$} \\
\cmidrule(lr){10-11}
$M$ & $\rho$ & $u$ & $T$ & $q$ & $\sigma$ & $m_{xxx}$ & $R_{xx}$ & $\Delta$ & PINN & DVM \\
\midrule
5 & 0.0056 & 0.0022 & 0.0039 & 0.049 & 0.032 & 0.12 & 0.28 & 0.69 & 11.206 & 11.198 \\
7 & 0.020 & 0.0068 & 0.0082 & 0.13 & 0.085 & 0.23 & 0.78 & 1.7 & 14.995 & 14.989 \\
8 & 0.023 & 0.0064 & 0.0080 & 0.13 & 0.10 & 0.14 & 0.97 & 2.0 & 16.934 & 16.947 \\
10 & 0.054 & 0.015 & 0.011 & 0.24 & 0.26 & 0.74 & 1.5 & 2.2 & 20.928 & 20.922 \\
12 & 0.078 & 0.020 & 0.013 & 0.27 & 0.48 & 1.1 & 3.1 & 5.1 & 24.940 & 24.939 \\
15 & 0.11 & 0.029 & 0.037 & 0.34 & 0.87 & 1.8 & 6.6 & 10 & 30.982 & 31.005 \\
20 & 0.14 & 0.035 & 0.049 & 0.64 & 0.75 & 1.8 & 18 & 29 & 41.076 & 41.166 \\
\bottomrule
\end{tabular}
\end{table}

The BGK shocks of the previous paper at Mach 7, 8 and 12 used a constant collision time, $\tau=1$, and needed a characteristic correction at Mach 8 and 12 \cite{Roohi2026KineticShockPINN}. A third continuation, from Mach 5 to 20 with $\tau=1$ as the only change (in the targets and in the factor $\chi$), gives the like-for-like comparison (\cref{tab:mono_tau1}). All seven runs pass tests (a) and (b), with $e_F$ between $10^{-5}$ and $6\times10^{-5}$ and $r_{\mathrm{rel}}=1.00$. Without any correction the network gives the heat flux and the stress within 0.13 per cent of the DVM at Mach 8 and, at Mach 12, within 0.27 and 0.48 per cent (\cref{tab:mono_tau1}), and the bulk fields within 0.14 per cent up to Mach 20; no comparison with the reference files of the previous paper was made. The fourth-order moments are worse than with $\tau=1/\rho$ at every Mach number, 5, 10 and 29 per cent in $\Delta$ at Mach 12, 15 and 20, although the tests are passed; no retry was made.

\subsection{BGK to Mach 100 and the weight of the passive distribution}\label{sec:mono_m50}

\begin{figure}[tp]
\centering
\includegraphics[width=\textwidth]{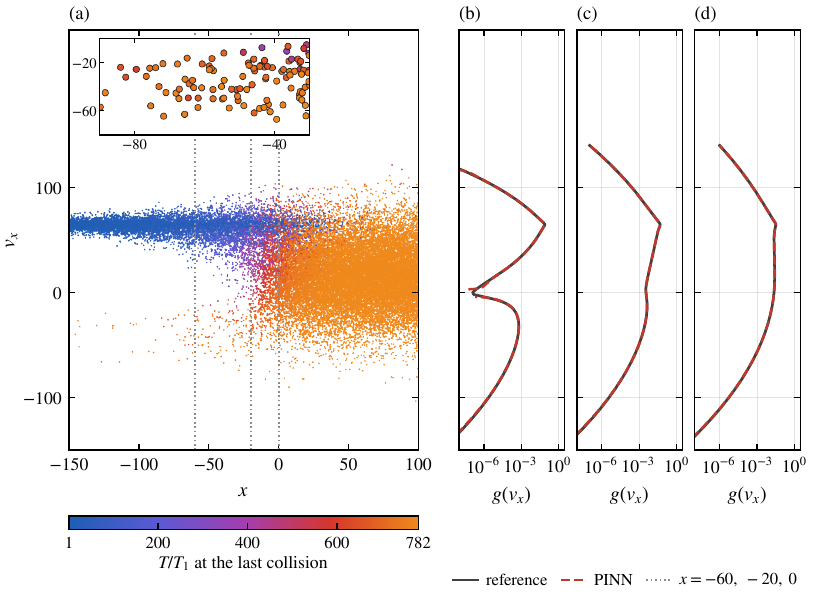}
\caption{BGK shock at Mach 50, $\tau=1/\rho$. (a) Molecules in the plane of position $x$ and velocity $v_x$: 30\,000 samples of the stationary distribution of the DVM fields, coloured by the temperature of the gas where each molecule last collided; the framed inset is a magnified view of the region $-90<x<-30$, $-80<v_x<0$ ahead of the shock, with the molecules that move upstream drawn larger; the scatter is rasterised at 400 dots per inch. (b), (c), (d) Reduced distribution $g(v_x)$ at $x=-60$, $-20$ and 0, the dotted lines of (a), with $v_x$ on the vertical axis as in (a). Solid: BGK distribution of the DVM fields; dashed: PINN (the blind prediction of \cref{tab:mono_m50}, $a_l=(T_1/T_2)^2$). The two curves separate visibly only in the minimum near $v_x=0$ of panel (b), where $g<10^{-6}$. Lengths in $\ell_0$ from the density midpoint, velocities in $\sqrt{RT_1}$. Supplementary Video 5 animates these data (\url{https://www.youtube.com/watch?v=_u03dhZOFhA}).}
\label{fig:molecules_M50}
\end{figure}

The BGK continuation was extended from Mach 20 to 30, 40 and 50 with four changes fixed in advance: the half-length $L=20M$, the scales of the inputs $\tanh(x/a)$ multiplied by $M/20$, the base width reset to $0.723M$ at the start of each step and then trained, and the parameters carried over from the previous step. Each step had 800 and 300 quasi-Newton iterations, and the three steps took 5.2 hours of one processor core. All three runs pass tests (a) and (b), and test (c) changes no field by more than $2.2\times10^{-4}$ of its downstream scale (\cref{tab:mono_m50}). At Mach 50, in a blind comparison, the bulk fields are within 0.1 per cent of the DVM and $q$, $\sigma$ and $m_{xxx}$ within 0.5 per cent, while $R_{xx}$ and $\Delta$ are 14 and 21 per cent off. \Cref{fig:molecules_M50} shows this shock through its molecules and compares the reduced distribution of the network with that of the DVM fields at three positions: over the velocities where the distribution exceeds $10^{-3}$ of its maximum the two differ by 0.7, 0.5 and 0.2 per cent (relative $L^2$) at $x=-60$, $-20$ and 0, and the share of molecules that move upstream is 2.12, 11.3 and 22.5 per cent in the network against 2.11, 11.3 and 22.5 per cent for the DVM fields. Supplementary Video 5 animates the same data (\url{https://www.youtube.com/watch?v=_u03dhZOFhA}).

\begin{table}
\centering
\footnotesize
\setlength{\tabcolsep}{3pt}
\caption{BGK model, $\tau=1/\rho$, continuation from Mach 20 to 50: reference-free tests and relative $L^2$ difference from the DVM solution of the same model, in per cent. All three runs pass tests (a) and (b). The Mach 50 prediction was saved and hashed before any Mach 50 reference file was read; the references at Mach 30 and 40 had been examined when the domain, inputs and base width were chosen.}
\label{tab:mono_m50}
\begin{tabular}{@{}rrrrrrrrrrr@{}}
\toprule
$M$ & $10^4e_F$ & $r_{\mathrm{rel}}$ & $\rho$ & $u$ & $T$ & $q$ & $\sigma$ & $m_{xxx}$ & $R_{xx}$ & $\Delta$ \\
\midrule
30 & 1.2 & 1.00 & 0.076 & 0.023 & 0.032 & 0.24 & 0.26 & 0.34 & 6.6 & 10 \\
40 & 1.6 & 1.00 & 0.093 & 0.030 & 0.044 & 0.30 & 0.32 & 0.41 & 8.9 & 13 \\
50 & 2.5 & 1.01 & 0.10 & 0.035 & 0.053 & 0.34 & 0.31 & 0.47 & 14 & 21 \\
\bottomrule
\end{tabular}
\end{table}

\paragraph{Cause}
$R_{xx}$ and $\Delta$ are the only fields to which the passive distribution $l$ contributes, and both are linear in $l$. At three positions in and behind the shock the two moments of the network were recomputed with the $l$ of the DVM in place of its own; this diagnosis uses the reference. At Mach 50, half a thickness behind the density midpoint, the error of $\Delta$ is $-1.079\times10^{6}$, of which $-1.080\times10^{6}$ comes from $l$, and the same holds at Mach 20 and at the other positions (Supplementary Section~\sref{supp:mono_al}). The loss hardly constrains $l$: with $a_l=(T_1/T_2)^2$ its part of $\mathcal{L}_{\mathrm{rel}}$ on the training points is 10 per cent of that of $g$ at Mach 20 and 1.7 per cent at Mach 50, a quantity that needs no reference. The weight had been made small because of the residual of $l$ far upstream (\cref{sec:loss}); behind the shock it is too small.

\paragraph{Test at Mach 20 and rule}
The weight was tested at Mach 20 by continuing the reported BGK run, each run being specified in advance and compared once. A tenfold $a_l$ changed nothing in 300 iterations. A hundredfold $a_l$ ($6.3\times10^{-3}$) with 1300 further iterations reduced the differences in $R_{xx}$ and $\Delta$ from 3.8 and 5.7 per cent to 0.65 and 1.1 per cent; the same 1300 iterations with the original weight left them at 3.8 and 5.8 per cent. The criterion fixed in advance (tests passed; $R_{xx}$ and $\Delta$ at most half of those of the reported run and of a second seed; no lower-order field more than 1.5 times worse) was met. The larger weight costs a little at lower order: against the run with the original weight and the same iterations, the differences in $T$ and $m_{xxx}$ are 1.3 and 1.9 times larger. For the higher Mach numbers the rule
\begin{equation}
  a_l=6.3\times10^{-3}\,\frac{T_2(20)}{T_2(M)},\qquad T_2(20)=125.87,
  \label{eq:al_rule}
\end{equation}
was fixed before the runs, because the ratio of the unweighted residual of $l$ to that of $g$ grows with $T_2$ (1600 at Mach 20 and 10\,800 at Mach 50, where $T_2$ is 126 and 782). It gives $a_l=2.8\times10^{-3}$, $1.6\times10^{-3}$, $1.0\times10^{-3}$, $5.2\times10^{-4}$ and $2.5\times10^{-4}$ at Mach 30, 40, 50, 70 and 100. The rule is thus chosen with a reference available at Mach 20, like the other weights of the loss (\cref{tab:information}).

\begin{table}
\centering
\footnotesize
\setlength{\tabcolsep}{2.6pt}
\caption{BGK model, $\tau=1/\rho$, Mach 20 to 100: reference-free tests, largest residual $b_q$ of the exact heat-flux balance on the profiles of the network (as a fraction of its largest term) and relative $L^2$ difference from the DVM solution of the same model, in per cent. First lines: the weight $a_l=(T_1/T_2)^2$ of \cref{tab:mono_m50}, continued to Mach 70 and 100. Other lines: the weight \eqref{eq:al_rule}; above Mach 20 these runs also have 2100 instead of 1100 quasi-Newton iterations per step. All runs pass tests (a) and (b). None of these comparisons is blind, except the Mach 100 run with $a_l=(T_1/T_2)^2$, whose prediction was hashed before its reference was read; the Mach 100 run with the weight \eqref{eq:al_rule} was compared after that reference had been opened.}
\label{tab:mono_al}
\begin{tabular}{@{}lrrrrrrrrrrrr@{}}
\toprule
$a_l$ & $M$ & $10^4e_F$ & $r_{\mathrm{rel}}$ & $10^3b_q$ & $\rho$ & $u$ & $T$ & $q$ & $\sigma$ & $m_{xxx}$ & $R_{xx}$ & $\Delta$ \\
\midrule
$(T_1/T_2)^2$ & 70 & 5.3 & 1.01 & 6.8 & 0.12 & 0.081 & 0.16 & 0.67 & 0.43 & 1.2 & 39 & 60 \\
 & 100 & 6.4 & 1.00 & 26 & 0.17 & 0.23 & 0.46 & 1.8 & 0.80 & 3.1 & 53 & 81 \\
\midrule
\eqref{eq:al_rule} & 20 & 0.45 & 1.00 & 0.48 & 0.014 & 0.0070 & 0.012 & 0.067 & 0.058 & 0.17 & 0.65 & 1.1 \\
 & 30 & 0.79 & 1.00 & 0.64 & 0.018 & 0.0082 & 0.013 & 0.081 & 0.057 & 0.13 & 0.76 & 1.2 \\
 & 40 & 1.9 & 1.00 & 0.94 & 0.024 & 0.011 & 0.017 & 0.11 & 0.071 & 0.15 & 0.96 & 1.6 \\
 & 50 & 2.3 & 1.01 & 1.0 & 0.019 & 0.0095 & 0.016 & 0.091 & 0.057 & 0.15 & 1.7 & 2.7 \\
 & 70 & 4.3 & 1.01 & 2.2 & 0.029 & 0.017 & 0.030 & 0.15 & 0.085 & 0.24 & 2.0 & 3.2 \\
 & 100 & 5.2 & 1.00 & 4.1 & 0.041 & 0.034 & 0.073 & 0.32 & 0.16 & 0.46 & 4.3 & 7.1 \\
\bottomrule
\end{tabular}
\end{table}

\begin{figure}[tp]
\centering
\includegraphics[width=\textwidth]{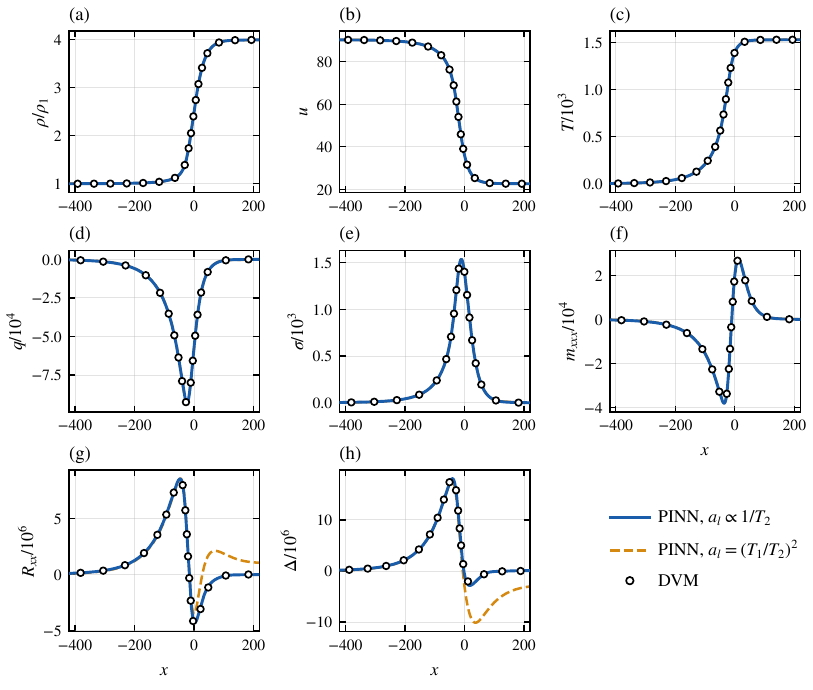}
\caption{BGK model, Mach 70: (a) $\rho$, (b) $u$, (c) $T$, (d) $q$, (e) $\sigma$, (f) $m_{xxx}$, (g) $R_{xx}$, (h) $\Delta$. Solid lines: PINN with the weight \eqref{eq:al_rule} of the passive distribution; dashed lines: PINN with $a_l=(T_1/T_2)^2$; symbols: DVM. The two networks differ visibly only in $R_{xx}$ and $\Delta$ behind the shock. The coordinate $x$ is in units of $\ell_0$.}
\label{fig:mono_M70}
\end{figure}

\begin{figure}[tp]
\centering
\includegraphics[width=\textwidth]{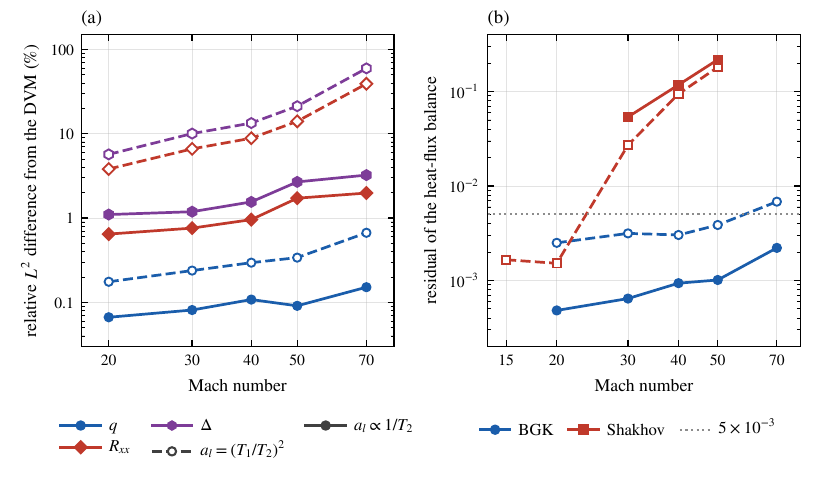}
\caption{Monatomic relaxation models above Mach 20, $\tau=1/\rho$. Open symbols and dashed lines: $a_l=(T_1/T_2)^2$; filled symbols and solid lines: the weight \eqref{eq:al_rule}, in both panels. (a) BGK: relative $L^2$ difference of $q$, $R_{xx}$ and $\Delta$ from the DVM. (b) Largest residual of the exact heat-flux balance evaluated on the profiles of the network, as a fraction of its largest term, for the BGK runs of (a) and the Shakhov runs of \cref{tab:mono_shk} (Mach 15 to 50); no reference is used. The dotted line is the level $5\times10^{-3}$ of \cref{sec:mono_shk_limit}.}
\label{fig:mono_highmach}
\end{figure}

\paragraph{Mach 30 to 70}
The continuation was repeated from the Mach 20 run with the larger weight, with the four changes of the first continuation and \eqref{eq:al_rule}; each step has 800 quasi-Newton iterations followed by 1300 with the weights of the final stage, 2100 in place of 1100, and the four steps took about 14 hours of one processor core, evaluation included. All runs pass tests (a) and (b) (\cref{tab:mono_al}), and test (c) changes no field by more than $3.1\times10^{-5}$ of its downstream scale. The differences in $R_{xx}$ and $\Delta$ fall by factors of 8 to 20: at Mach 70 they are 2.0 and 3.2 per cent instead of 39 and 60 (\cref{fig:mono_M70}), and at Mach 50 they are 1.7 and 2.7 per cent instead of 14 and 21 (\cref{tab:mono_al}, \cref{fig:mono_highmach}a). These runs differ from the first continuation in two things, the weight and the number of iterations of each step (2100 against 1100), and the two were separated at Mach 20, where the same additional iterations with the original weight left $R_{xx}$ and $\Delta$ unchanged, and at Mach 60 (below). The lower-order fields improve as well, by factors of 2.5 to 5, which the longer optimisation of each step may explain in part; at Mach 70 the bulk fields are within 0.03 per cent, $q$ and $\sigma$ within 0.15 per cent and $m_{xxx}$ within 0.24 per cent, and the density thickness is 67.86 against 67.93 for the DVM (\cref{fig:mono_M70}). The error that remains in the fourth-order moments grows slowly with the Mach number. These comparisons are not blind; the blind results of this paper for the BGK model are the Mach 50 prediction with the first weight and the four runs at Mach 60 below.

\paragraph{Mach 100}
The step from Mach 70 to 100 was made in both continuations with the settings of the steps below it ($L=2000$; in the second continuation $a_l=2.5\times10^{-4}$ by \eqref{eq:al_rule}). With the first weight the Mach 100 prediction was hashed before its reference was read; it passes tests (a), (b) and (c), but $R_{xx}$ and $\Delta$ are 53 and 81 per cent from the DVM and $q$ 1.8 per cent, and its heat-flux balance residual, $2.6\times10^{-2}$, is five times the level of \cref{sec:mono_shk_limit} (\cref{tab:mono_al}). With the weight \eqref{eq:al_rule} the step passes the three tests at the first attempt (test (c) $2.8\times10^{-6}$; 9.0 hours of one core, evaluation included), the bulk fields are within 0.073 per cent of the DVM, $q$, $\sigma$ and $m_{xxx}$ within 0.46 per cent, $R_{xx}$ and $\Delta$ within 4.3 and 7.1 per cent, the density thickness is 96.87 against 97.00 and the heat-flux balance residual is $4.1\times10^{-3}$. This comparison is not blind: it was made after the reference at Mach 100 had been opened for the run with the first weight, although nothing of the run was chosen after that.

\paragraph{Weight and iterations separated at Mach 60}
The runs above Mach 20 with the weight \eqref{eq:al_rule} changed two things at once. Both were varied at Mach 60, a Mach number without a reference at the time, in four runs fixed in advance, each started from the Mach 50 network of \cref{tab:mono_m50}: $a_l=(T_1/T_2)^2=7.9\times10^{-7}$ or $a_l=7.0\times10^{-4}$ by \eqref{eq:al_rule}, each with 800 and 300 iterations (1100, the first continuation) and with 1000 more (2100, the second). The four predictions were hashed on the cluster before the DVM solution at Mach 60 was computed, and each was compared once. All four pass tests (a), (b) and (c) (\cref{tab:mono_m60}). At equal iterations the weight reduces the differences in $R_{xx}$ and $\Delta$ by factors of 2.1 (1100 iterations) and 2.4 to 2.5 (2100); at equal weight the 1000 further iterations reduce them by 1.6 to 1.7 (first weight) and 2.0 (weight \eqref{eq:al_rule}). The two effects nearly multiply: together they reduce $R_{xx}$ and $\Delta$ 4.1 times, from 23 and 35 to 5.6 and 8.6 per cent, and about 60 per cent of this gain, on a logarithmic scale, comes from the weight and 40 per cent from the iterations. They act differently on the other fields: at equal iterations the weight alone makes $u$, $T$, $q$ and $m_{xxx}$ 1.4 to 2.0 times worse, as at Mach 20, whereas the further iterations improve every field at either weight. The combined gain at Mach 60 is smaller than along the second continuation (factors 7.8 at Mach 50 and 19 at Mach 70 in $\Delta$), which had carried $l$ with the larger weight from Mach 20 upward, while the four runs at Mach 60 start from a network whose $l$ was trained with the first weight; this explanation was not tested separately.

\begin{table}
\centering
\footnotesize
\setlength{\tabcolsep}{2.4pt}
\caption{BGK model, $\tau=1/\rho$, Mach 60, blind: four runs from the Mach 50 network of \cref{tab:mono_m50} with either weight of the passive distribution and either number of quasi-Newton iterations. Reference-free tests, largest residual $b_q$ of the exact heat-flux balance (as in \cref{tab:mono_al}), relative $L^2$ difference from the DVM solution in per cent, and density thickness $\delta_\rho$ (DVM: 58.241). All four runs pass tests (a), (b) and (c); test (c) changes no field by more than $7.8\times10^{-6}$ of its downstream scale. The predictions were hashed before the reference was computed.}
\label{tab:mono_m60}
\begin{tabular}{@{}lrrrrrrrrrrrrr@{}}
\toprule
$a_l$ & iterations & $10^4e_F$ & $r_{\mathrm{rel}}$ & $10^3b_q$ & $\rho$ & $u$ & $T$ & $q$ & $\sigma$ & $m_{xxx}$ & $R_{xx}$ & $\Delta$ & $\delta_\rho$ \\
\midrule
$(T_1/T_2)^2$ & 1100 & 4.0 & 1.01 & 4.3 & 0.11 & 0.047 & 0.078 & 0.41 & 0.35 & 0.63 & 23 & 35 & 58.116 \\
 & 2100 & 5.9 & 1.01 & 1.8 & 0.043 & 0.017 & 0.027 & 0.16 & 0.13 & 0.20 & 14 & 21 & 58.215 \\
\eqref{eq:al_rule} & 1100 & 4.0 & 1.01 & 7.2 & 0.089 & 0.077 & 0.15 & 0.64 & 0.32 & 1.1 & 11 & 17 & 57.938 \\
 & 2100 & 5.3 & 1.01 & 2.2 & 0.046 & 0.025 & 0.047 & 0.22 & 0.15 & 0.39 & 5.6 & 8.6 & 58.207 \\
\bottomrule
\end{tabular}
\end{table}

\begin{figure}[tp]
\centering
\includegraphics[width=\textwidth]{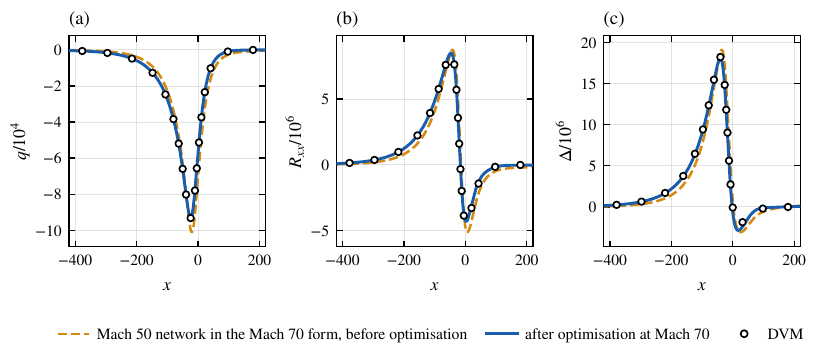}
\caption{BGK model, Mach 70: (a) $q$, (b) $R_{xx}$, (c) $\Delta$. Dashed: the Mach 50 network placed in the Mach 70 form, before any optimisation at Mach 70; solid: the network after the optimisation at Mach 70 (weight \eqref{eq:al_rule}); symbols: DVM. The coordinate $x$ is in units of $\ell_0$ from the density midpoint of each solution.}
\label{fig:transfer}
\end{figure}

\paragraph{What the optimisation at a new Mach number adds}
Strong shocks are nearly similar when scaled with the downstream state, so a network carried from one Mach number to the next might be accurate before any optimisation. This was checked for the step from Mach 50 to Mach 70. The Mach 50 network was placed in the Mach 70 form exactly as the continuation starts (new end states, half-length and input scales, base width reset, parameters unchanged) and evaluated without any iteration. Its differences from the DVM are 4.7, 2.9, 2.8, 15, 16, 20, 18 and 16 per cent in $\rho$, $u$, $T$, $q$, $\sigma$, $m_{xxx}$, $R_{xx}$ and $\Delta$, and its density thickness is 66.20 against 67.93; after the 2100 iterations of the step they are 0.029, 0.017, 0.030, 0.15, 0.085, 0.24, 2.0 and 3.2 per cent (\cref{fig:transfer}). The transferred network is therefore a starting point, not a solution: the accuracy of the reported fields is obtained by minimising the residual at the new Mach number.

\subsection{The Shakhov shock: Mach 20 with settings fixed in advance, and a limit above it}\label{sec:mono_shk_limit}

\begin{table}
\centering
\footnotesize
\setlength{\tabcolsep}{2.4pt}
\caption{Shakhov model, $\tau=1/\rho$, continuation from the accepted Mach 12 run with settings fixed in advance: reference-free tests, largest residual $b_q$ of the exact heat-flux balance (as in \cref{tab:mono_al}), relative $L^2$ difference from the DVM solution of the same model in per cent, and density thickness. All eight runs pass tests (a), (b) and (c). The runs at Mach 30 to 50 are not solutions: they are listed because they pass the tests and miss the reference. The Mach 50 prediction with the first weight was hashed before its reference was read.}
\label{tab:mono_shk}
\begin{tabular}{@{}lrrrrrrrrrrrrrr@{}}
\toprule
 & & & & & & & & & & & & & \multicolumn{2}{c}{$\delta_\rho$} \\
\cmidrule(l){14-15}
$a_l$ & $M$ & $10^4e_F$ & $r_{\mathrm{rel}}$ & $10^3b_q$ & $\rho$ & $u$ & $T$ & $q$ & $\sigma$ & $m_{xxx}$ & $R_{xx}$ & $\Delta$ & PINN & DVM \\
\midrule
$(T_1/T_2)^2$ & 15 & 0.40 & 0.99 & 1.7 & 0.039 & 0.012 & 0.014 & 0.079 & 0.20 & 0.34 & 2.5 & 3.5 & 19.06 & 19.05 \\
 & 20 & 0.96 & 1.02 & 1.5 & 0.027 & 0.015 & 0.029 & 0.11 & 0.15 & 0.29 & 3.8 & 5.4 & 25.21 & 25.21 \\
 & 30 & 2.8 & 1.03 & 27 & 0.26 & 0.53 & 1.2 & 3.9 & 1.9 & 8.6 & 25 & 36 & 37.04 & 37.60 \\
 & 40 & 5.5 & 1.00 & 94 & 0.79 & 1.8 & 3.7 & 12 & 7.0 & 25 & 38 & 45 & 46.99 & 50.04 \\
 & 50 & 7.0 & 1.00 & 181 & 1.2 & 2.5 & 5.0 & 16 & 9.8 & 33 & 42 & 53 & 56.01 & 62.49 \\
\midrule
\eqref{eq:al_rule} & 30 & 4.0 & 0.99 & 54 & 0.41 & 0.99 & 2.1 & 7.1 & 3.8 & 15 & 18 & 14 & 36.73 & 37.60 \\
 & 40 & 5.5 & 0.99 & 118 & 0.98 & 2.3 & 4.6 & 15 & 9.0 & 31 & 33 & 25 & 46.10 & 50.04 \\
 & 50 & 7.4 & 0.97 & 217 & 1.3 & 2.9 & 5.7 & 18 & 11 & 36 & 38 & 30 & 55.45 & 62.49 \\
\bottomrule
\end{tabular}
\end{table}

The accepted Shakhov runs of \cref{tab:mono_err} end at Mach 12, and the Mach 20 solution of \cref{sec:mono_m20} started from a run that misses test (a). The Shakhov continuation was therefore repeated from the accepted Mach 12 run with settings fixed before any job was submitted, taken from the BGK continuation above Mach 20 and from the diagnosis of \cref{sec:mono_m20}: $L=20M$, the scales of the inputs multiplied by $M/20$, the base width reset to $0.912M$ at each step, $\Delta v=0.25$ at the beam (110 to 149 nodes) and 3100 quasi-Newton iterations per step in four stages. The references at Mach 15 to 40 had been examined before, so these comparisons are not blind.

The steps to Mach 15 and to Mach 20 pass tests (a), (b) and (c) at the first attempt (\cref{tab:mono_shk}; 3.2 and 3.4 hours of one core). At Mach 15 the flux error is $4.0\times10^{-5}$, where the earlier run had $1.9\times10^{-3}$, and every field is closer to the DVM than in that run. At Mach 20 the bulk fields are within 0.03 per cent, $q$ and $\sigma$ within 0.15 per cent, $m_{xxx}$ within 0.3 per cent and $R_{xx}$ and $\Delta$ within 3.8 and 5.4 per cent, and the thickness is 25.213 against 25.207. Up to third order this run agrees with attempt C of \cref{tab:mono_m20} within a factor 1.4; in $R_{xx}$ and $\Delta$ it is about 1.6 times farther from the DVM, within the uncertainty of these fields stated above. The Shakhov shock is thus solved to Mach 20 with settings fixed in advance.

Above Mach 20 the procedure fails, and the acceptance tests do not show it. The steps to Mach 30, 40 and 50 pass the three tests, yet the heat flux is 3.9, 12 and 16 per cent from the DVM and the shock is 1.5, 6 and 10 per cent too thin (at Mach 50, 56.0 against 62.5). The Mach 50 step is a blind test: its prediction was saved and hashed before the reference was read. Repeating the three steps with the weight \eqref{eq:al_rule} and unchanged iterations did not help: the heat flux is 7.1, 15 and 18 per cent off and the thickness at Mach 50 is 55.5. The error of the Shakhov model above Mach 20 therefore does not lie in the passive distribution, and its cause was not found. The Shakhov shock above Mach 20 is not solved in this paper.

The residual of the exact heat-flux balance, evaluated on the profiles of the network as in \cref{sec:mono_m20}, separates these runs without a reference (\cref{fig:mono_highmach}b). It is $1.7\times10^{-3}$ and $1.5\times10^{-3}$ of the largest term of the balance at Mach 15 and 20 and between $5\times10^{-4}$ and $2.2\times10^{-3}$ for the BGK runs with the weight \eqref{eq:al_rule} to Mach 70 ($4.1\times10^{-3}$ at Mach 100), but 0.027, 0.094 and 0.18 for the Shakhov runs at Mach 30, 40 and 50, and 0.054, 0.12 and 0.22 with the weight \eqref{eq:al_rule}. A level of $5\times10^{-3}$ was stated after the Shakhov runs at Mach 30 and 40 with the first weight had been compared and before the runs with the weight \eqref{eq:al_rule} and the blind comparison at Mach 50 were made; all of these exceed it tenfold or more. The diagnostic is not a test of the fourth-order moments: the BGK run at Mach 50 with $a_l=(T_1/T_2)^2$ has a residual of $3.9\times10^{-3}$ and is 21 per cent off in $\Delta$. Nor is the level sharp: the BGK run at Mach 70 with that weight has $6.8\times10^{-3}$, above the level, with $q$ within 0.7 per cent of the DVM (and the Mach 60 run with the weight \eqref{eq:al_rule} and 1100 iterations $7.2\times10^{-3}$ with $q$ within 0.64 per cent), whereas the Shakhov runs at Mach 30 to 50 exceed the level by factors of 5 to 43. The residual is proposed as a fourth reference-free check, which indicated the wrong heat flux and thickness of these runs where tests (a) to (c) did not.

\subsection{Summary of the monatomic results}

With the loss of \cref{sec:loss}, no reference in training and $\tau=1/\rho$, the network alone solves the BGK shock to Mach 100 and the Shakhov shock to Mach 20, with the fields up to third order within 0.5 per cent of the DVM in the accepted runs of \cref{tab:mono_err,tab:mono_m50} and, to Mach 20, of \cref{tab:mono_shk} and, with the weight \eqref{eq:al_rule}, to Mach 100. The fourth-order moments $R_{xx}$ and $\Delta$ are the most demanding fields: with the weight $a_l=(T_1/T_2)^2$, runs that pass the tests can differ from the reference by as much as 81 per cent in $\Delta$ (\cref{tab:mono_m20,tab:mono_tau1,tab:mono_al}). For the BGK model the cause is the passive distribution $l$, whose residual carried the weight $a_l=(T_1/T_2)^2$; with the weight \eqref{eq:al_rule} and 2100 instead of 1100 iterations per step $R_{xx}$ and $\Delta$ are within 3.2 per cent up to Mach 70 and 7.1 per cent at Mach 100, and at Mach 60 about 60 per cent of the gain comes from the weight. For the Shakhov model the runs above Mach 20, to Mach 50, pass the tests and miss the reference, with either weight, and the heat-flux balance indicates this without a reference. The cost is that of one processor core: 8.6 hours for the BGK continuation from Mach 5 to 20, 5.2 hours from 20 to 50 with the first weight about 14 hours from 20 to 70 with the weight \eqref{eq:al_rule} and 9.0 hours for its step to Mach 100 (\cref{sec:disc_cost}).

\section{Nitrogen shocks to Mach 15}\label{sec:n2}

\subsection{Runs and references}\label{sec:n2_runs}

For the Rykov-type model of nitrogen \cite{Rykov1975Model,Wu2015Polyatomic} the loss is \eqref{eq:loss} with the four reduced distributions $g$, $k$, $h$, $l$ and seven moments of the residual (\cref{sec:models_diff}), minimised by the stages of \cref{sec:optimiser}. The reference at every Mach number is the fourth-order DVM solution of the same model with the same constants (\cref{sec:refs_shock}). The training procedure was developed on this problem: at Mach 2, 5, 7, 8 and 12 the distance from a DVM solution was printed during training, and the three weights of \cref{sec:loss} were chosen with the reference available (\cref{tab:information}). The results at these Mach numbers are therefore those of a procedure tuned on them; Mach 10 and 15, where nothing could be monitored, test it (\cref{sec:n2_blind}). Mach 2 was trained from scratch with a network of width 32 (3789 parameters), the default weights, a uniform velocity rule and two seeds. The stronger shocks use width 48 (7981 parameters) and the continuation $5\to7\to8\to10\to12\to15$, at 0.6 to 2.7 hours per step on one processor core (one thread of a laptop processor for most steps; Supplementary Section~\sref{app:settings_shock}). Only the steps to Mach 8, 12 and 15 ran in the final form of the procedure (\ref{app:settings}; Supplementary Section~\sref{supp:n2_history}).

\begin{table}
\centering
\footnotesize
\setlength{\tabcolsep}{2.4pt}
\caption{Nitrogen shocks: largest flux error $e_F$ (test (a), limit $10^{-3}$) and relative $L^2$ difference \eqref{eq:error} between the network and the DVM solution of the same model, in per cent; at Mach 2 the seed is given in brackets. Window $|x|<20$ (Mach 2), $33.3$ (Mach 5) and $40$ (Mach 7 to 15). Reference: the fourth-order DVM solution at every Mach number ($\Delta x=0.05$ at Mach 2, 5 and 7, $\Delta x=0.025$ at Mach 8 to 15). Passed: every run passes test (a); test (b) was not evaluated for any run of this table (\cref{sec:acceptance}). $\dagger$: prediction fixed before any DVM solution of this Mach number existed. $\ddagger$: procedure fixed in advance, prediction hashed before the comparison. The fourth-order entries at Mach 12 and 15 carry the uncertainty discussed in \cref{sec:mono_m20}; the scatter between seeds was measured at Mach 2 and 5 only. Peak values and thicknesses: Supplementary Table~\sref{tab:n2_free}.}
\label{tab:n2_err}
\begin{tabular}{@{}lrrrrrrrrrrrl@{}}
\toprule
$M$ & $10^4e_F$ & $\rho$ & $u$ & $T_{\mathrm{tr}}$ & $T_{\mathrm{rot}}$ & $q_{\mathrm{tr}}$ & $q_{\mathrm{rot}}$ & $\sigma$ & $m_{xxx}$ & $R_{xx}$ & $\Delta$ & Passed \\
\midrule
2 (101) & 0.13 & 0.0067 & 0.0048 & 0.012 & 0.0085 & 0.059 & 0.088 & 0.036 & 0.11 & 0.12 & 0.13 & a \\
2 (202) & 0.22 & 0.0071 & 0.0049 & 0.0057 & 0.0094 & 0.074 & 0.14 & 0.043 & 0.12 & 0.16 & 0.15 & a \\
5 & 0.72 & 0.0043 & 0.0068 & 0.015 & 0.025 & 0.054 & 0.054 & 0.060 & 0.087 & 0.10 & 0.14 & a \\
7 & 0.84 & 0.0066 & 0.015 & 0.028 & 0.033 & 0.10 & 0.13 & 0.15 & 0.35 & 0.29 & 0.36 & a \\
8 & 1.4 & 0.010 & 0.026 & 0.049 & 0.059 & 0.19 & 0.28 & 0.23 & 0.75 & 0.66 & 0.58 & a \\
10$^{\dagger}$ & 2.0 & 0.016 & 0.049 & 0.077 & 0.020 & 0.29 & 0.17 & 0.44 & 0.61 & 0.87 & 1.3 & a \\
12 & 4.8 & 0.011 & 0.036 & 0.052 & 0.012 & 0.18 & 0.11 & 0.31 & 0.37 & 0.93 & 1.4 & a \\
15$^{\ddagger}$ & 6.9 & 0.014 & 0.033 & 0.048 & 0.020 & 0.15 & 0.15 & 0.26 & 0.43 & 1.8 & 2.7 & a \\
\bottomrule
\end{tabular}
\end{table}

\subsection{Differences from the discrete-velocity solutions}\label{sec:n2_results}

\begin{figure}[tp]
\centering
\includegraphics[width=\textwidth]{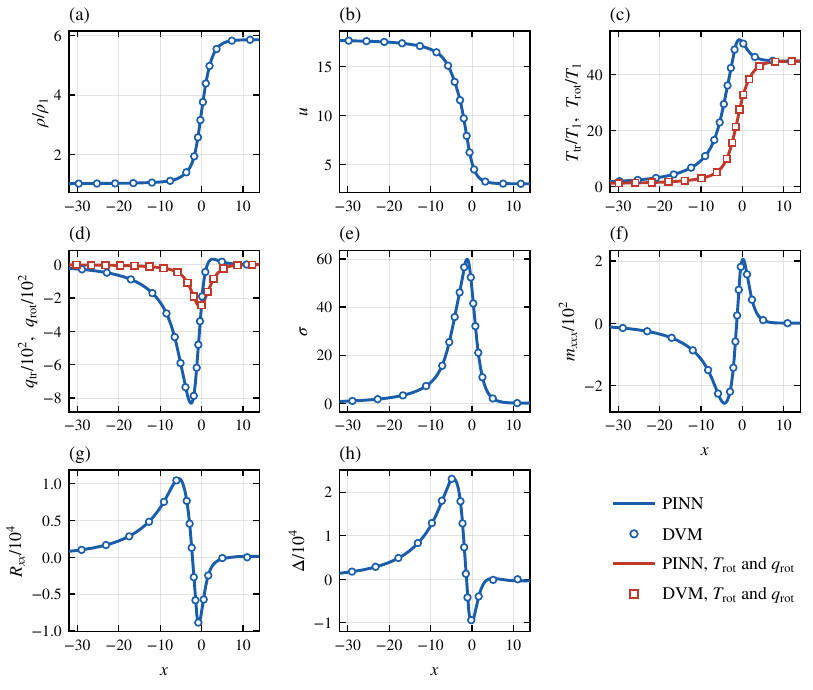}
\caption{Nitrogen (Rykov-type model), Mach 15: (a) $\rho$, (b) $u$, (c) translational (blue) and rotational (red) temperature, (d) translational and rotational heat flux, (e) $\sigma$, (f) $m_{xxx}$, (g) $R_{xx}$, (h) $\Delta$. Lines: PINN; symbols: independent DVM.}
\label{fig:n2_profiles}
\end{figure}

\Cref{tab:n2_err} and \cref{fig:n2_profiles} give the results, and Supplementary Fig.~\sref{fig:n2_errors} draws the differences against the Mach number. As for the monatomic gas, the difference grows with the order of the moment. From Mach 2 to 15 the density, the velocity and the two temperatures are within 0.08 per cent of the DVM, the two heat fluxes and the stress within 0.45 per cent and $m_{xxx}$ within 0.8 per cent; $R_{xx}$ and $\Delta$ are within 1.4 per cent up to Mach 12 and 1.8 and 2.7 per cent at Mach 15. The maximum of $T_{\mathrm{tr}}$ agrees within 0.02 per cent at every Mach number, and the density thickness of the network is 0.05 to 0.18 per cent below that of the DVM from Mach 2 to 12 and 0.03 per cent below at Mach 15 (Supplementary Table~\sref{tab:n2_free}). These are errors of the network: the error of the reference is at least 80 times smaller than any entry (\cref{sec:refs_shock}), while the networks have $e_F$ from $10^{-5}$ at Mach 2 to $7\times10^{-4}$ at Mach 15 and, up to Mach 12, errors of the balance of the rotational energy of 0.2 to 0.6 per cent of the largest $q_{\mathrm{rot}}$ (further remarks in Supplementary Section~\sref{supp:n2_history}). Every run passes test (a), and under a finer velocity rule the fields change by less than $3\times10^{-5}$ of their downstream scales up to Mach 12 and by $7\times10^{-5}$ at Mach 15; test (b) was not evaluated for these runs (\cref{sec:acceptance}).

For scale, the model differs more from a particle simulation of nitrogen than the network differs from the model: the density thickness of the DVM solution is 8.7 to 9.0 per cent below that of a DSMC calculation (variable hard spheres with Larsen--Borgnakke rotational exchange, Supplementary Section~\sref{supp:collision_supp}) at Mach 5, 8 and 12. Against measurements the model reproduces the density profiles of nitrogen shocks but, with its constant collision number, places the rotational rise of cold shocks about a third too close to the density (\cref{sec:validation}).

\subsection{Tests at Mach 10 and 15}\label{sec:n2_blind}

Two tests check whether the procedure, tuned at Mach 2 to 12, holds where nothing could be monitored. Only the first is blind, in the sense that no reference existed when the prediction was hashed; the protocols are given in full in Supplementary Section~\sref{supp:n2_blind}.

\emph{Mach 10.} The Mach 10 solution, produced as the step to Mach 12, was evaluated without a reference, hashed and given an expected accuracy, differences within the range of Mach 8 and 12, before any deterministic solution of the model at Mach 10 existed. A second DVM solver was then written from the model equations without access to the prediction (Supplementary Section~\sref{supp:n2_blind}), validated against the first at Mach 8 and 12, and run once with settings fixed beforehand. Against the fourth-order solutions now used as reference (\cref{tab:n2_err}), $T_{\mathrm{rot}}$, $q_{\mathrm{rot}}$, $m_{xxx}$, $R_{xx}$ and $\Delta$ fall inside the expected range, while $\rho$, $u$, $T_{\mathrm{tr}}$, $q_{\mathrm{tr}}$ and $\sigma$ lie 1.4 to 1.5 times above its upper end; the largest discrepancy of an extremum is 3.0 per cent, in the minimum of $\Delta$. The excess may come from the history of this run, which started from a Mach 8 solution that had stalled before the second weight was introduced; this was not tested. The expectation was therefore met for half of the fields, at an accuracy of the same order as at Mach 8 and 12; Mach 10 lies inside the monitored range, and the test does not show that the procedure could have been developed without the DVM.

\emph{Mach 15.} Mach 15 lies outside that range. Before the prediction was started and before any reference at Mach 15 existed, a protocol fixed the commands (one continuation step from Mach 12 with its settings: 800 and 300 iterations, $a_l=(T_1/T_2)^2=5.0\times10^{-4}$, $L=120$), the acceptance and the expectation: each field within about twice the largest difference at Mach 8, 10 and 12. The reference was finished, and its peaks seen, twenty minutes before the prediction job ended; this is therefore a prediction with a procedure fixed in advance and hashed before the comparison, not a blind test. All fields up to third order are at or below the upper ends of the expected ranges, and $R_{xx}$ and $\Delta$, 1.8 and 2.7 per cent, are about twice the Mach 12 values, at the edge of the expectation; the maximum of $T_{\mathrm{tr}}$ differs by 0.016 per cent, the thickness by 0.03 per cent and the minimum of $\Delta$ by 6.2 per cent, and $e_F=6.9\times10^{-4}$ is the largest of the continuation. The procedure thus extends one step beyond the Mach numbers on which it was tuned, with the fourth-order moments twice as far from the reference. Test (b) was not evaluated for any run of the nitrogen continuation. No nitrogen calculation above Mach 15 was made.

\subsection{Loss terms and weights}\label{sec:design_terms}

The experiments of this and the next two subsections were made for nitrogen between Mach 2 and 8. The differences quoted from the development, printed during training, are against the first DVM solver; those of the ablation are against the fourth-order solver.

\begin{figure}[tp]
\centering
\includegraphics[width=\textwidth]{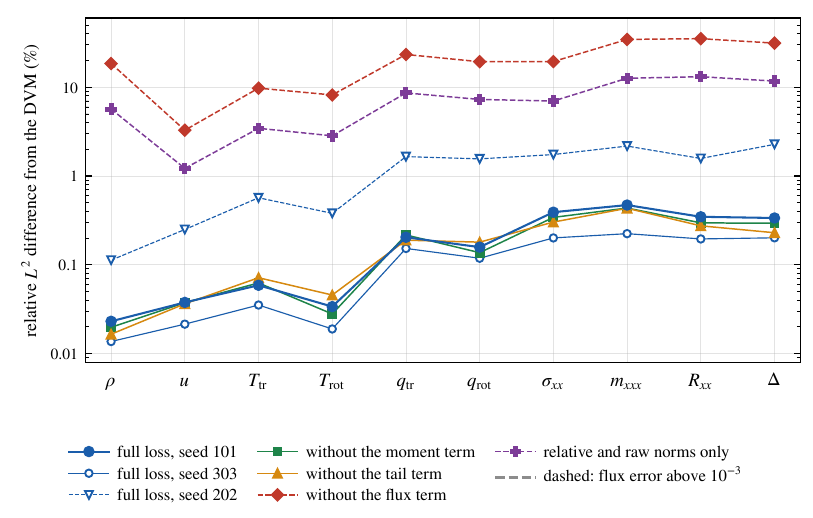}
\caption{Nitrogen, Mach 5, trained from scratch: difference from the fourth-order DVM, field by field, for the full loss (three seeds), with one term removed (flux, moment or tail), and with the flux, moment and tail terms all removed, so that the relative and raw norms and the pin remain. Dashed: runs whose largest flux error exceeds $10^{-3}$ and which therefore fail the reference-free flux test. The thin lines only join the symbols of one run. The numbers are in Supplementary Table~\sref{tab:ablation}.}
\label{fig:ablation}
\end{figure}

The ablation of \cref{fig:ablation} was specified before it was run: the Mach 5 shock trained from scratch with the default weights of \cref{sec:loss} (width 32), and seven runs that change one thing each, each compared once with the DVM (Supplementary Table~\sref{tab:ablation}). Two seeds of three with the full loss are within 0.014 to 0.47 per cent of the DVM over the ten fields; the third stalls with 90 per cent of its loss in the tail term, is 0.11 to 2.3 per cent off and is rejected by the flux test ($e_F=1.8\times10^{-3}$). Without the flux term the network converges to a different, thinner shock (thickness 2.37 instead of 3.33, peak of $T_{\mathrm{tr}}$ 7 per cent low), 3 to 35 per cent from the DVM; without the flux, moment and tail terms the differences are 1 to 13 per cent. Both failures show without a reference, as fluxes that vary by 16 and 5 per cent. Removing the moment term or the tail term alone changes nothing beyond the difference between two seeds. The ablation shows that the flux term is needed at Mach 5 with the default weights. It does not show that the moment and tail terms are needed, nor why removing the flux term alone does more harm than removing all three. It also exposes a weakness of test (b): the two good runs fail it by its letter ($r_{\mathrm{rel}}=265$ and 220), because one test point next to the inflow boundary holds 99.7 per cent of the term (without it, 0.78 and 0.74); the criterion was not changed.

\begin{figure}[tp]
\centering
\includegraphics[width=\textwidth]{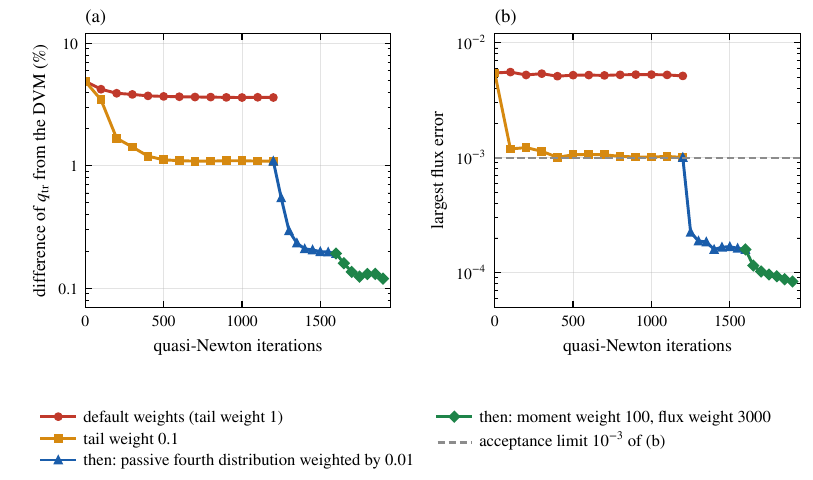}
\caption{Nitrogen, Mach 7, continuation from Mach 5: the quasi-Newton stage with the default weights and after three successive changes of the weights. (a) Difference of the translational heat flux from the first DVM solver (printed during training, never used by it). (b) Largest relative error of the fluxes of mass, momentum and energy; dashed line: the acceptance limit $10^{-3}$.}
\label{fig:weights}
\end{figure}

The evidence for the tail and moment terms concerns their weights and comes from Mach 7 (\cref{fig:weights}). Continued from Mach 5 with the default weights, the quasi-Newton stage stalls with $q_{\mathrm{tr}}$ 3.6 per cent from the DVM and $e_F=5.1\times10^{-3}$ after 1200 iterations, with the tail term dominated by the few hot molecules far upstream. Three changes, each found by decomposing the loss by position, velocity and distribution without a reference, remove the stall: $\lambda_T=0.1$ (1.1 per cent and $e_F=1.0\times10^{-3}$ after the same 1200 iterations); $a_l=(T_1/T_2)^2=0.01$, because $l$ then holds 82 and 83 per cent of $\mathcal{L}_{\mathrm{rel}}$ and $\mathcal{L}_{\mathrm{tail}}$, its residual far upstream being $(T_2/T_1)^2$ times more sensitive to the hot molecules than that of $g$ (0.19 per cent and $1.6\times10^{-4}$ after 400 further iterations); and a final stage with $\lambda_P=100$ and $\lambda_F=3000$ (0.12 per cent and $8.4\times10^{-5}$), which removes a one-signed error of the rotational-energy balance that accumulated along the shock (0.6 per cent of $q_{\mathrm{rot}}$ at Mach 8). The flux error falls in the same steps as the difference from the DVM, so that the progress is visible without the reference. The weights were nevertheless chosen while that difference was printed, which is why Mach 10 and 15 test the procedure; and since a run from scratch at Mach 7 with the default weights reaches 0.03 to 0.6 per cent (\cref{sec:design_points}), the stall is caused by this continuation, not by the Mach number.

\subsection{Collocation points in the shock layer}\label{sec:design_points}\label{sec:n2_scratch}

\begin{table}
\centering
\small
\caption{Nitrogen, training from scratch with the Mach 2 settings (width 32, default weights, uniform velocity rule), without and with the shock-layer set of collocation points: largest flux error, density thickness, and smallest and largest relative $L^2$ difference from the fourth-order DVM over the nine fields of \cref{tab:n2_err} other than $m_{xxx}$.}
\label{tab:n2_scratch}
\begin{tabular}{llrrrr}
\toprule
 & & & \multicolumn{2}{c}{$\delta_\rho$} & \\
\cmidrule(lr){4-5}
Collocation points & $M$ & $10^4e_F$ & PINN & DVM & Differences (per cent) \\
\midrule
Procedure of Mach 2 & 5 & 1.2 & 3.328 & 3.328 & 0.02 to 0.36 \\
 & 7 & 280 & 0.231 & 3.555 & 6 to 100 \\
 & 8 & 1800 & 0.066 & 3.702 & 9 to 100 \\
With the shock-layer set & 2 & 0.12 & 5.208 & 5.212 & 0.007 to 0.16 \\
 & 5 & 2.2 & 3.320 & 3.328 & 0.02 to 0.39 \\
 & 7 & 2.4 & 3.538 & 3.555 & 0.03 to 0.61 \\
 & 8 & 2500 & 0.159 & 3.702 & 7 to 100 \\
\bottomrule
\end{tabular}
\end{table}

The Mach 2 settings were repeated at Mach 5, 7 and 8 with only the Mach number changed, each run specified in advance and compared once (\cref{tab:n2_scratch}). Mach 5 succeeds. At Mach 7 the shock collapses into a density jump of thickness 0.23 (DVM: 3.56) between two fixed collocation points 0.31 apart (\cref{fig:points}b), 6 to 100 per cent from the DVM. The loss does not reveal the collapse, ending at $2.0\times10^{-6}$, below the $3.5\times10^{-5}$ of the correct solution described next; the tests do ($e_F=2.8\times10^{-2}$, and $\mathcal{L}_{\mathrm{rel}}$ at the test points 500 times its training value). Adding the shock-layer set of collocation points, which at this domain length adds weight and random points in $|x|<10$ but no fixed positions (Supplementary Section~\sref{supp:n2_points}), leaves Mach 2 and 5 unchanged and makes Mach 7 correct ($e_F=2.4\times10^{-4}$, $r_{\mathrm{rel}}=1.7$, thickness 3.538, differences 0.03 to 0.6 per cent). At Mach 8 the procedure from scratch collapses without and with the set (thickness 0.07 and 0.16 against 3.70, $e_F=0.18$ and 0.25), and Mach 8 is reached by continuation (\cref{sec:design_opt}).

Together with the ablation, in which one seed of three stalled at Mach 5, this shows the limits of training from scratch with the Mach 2 settings: it reaches Mach 7, it is not robust to the seed, and at Mach 5 and 7 it is two to six times further from the DVM than the continuation of \cref{tab:n2_err} in most fields. The results above Mach 7 rest on continuation and on the three weights. The reference-free tests detected every collapse recorded in this work; they are necessary and not sufficient (\cref{sec:acceptance}).

\subsection{The optimiser}\label{sec:design_opt}

\begin{figure}[tp]
\centering
\includegraphics[width=0.98\textwidth]{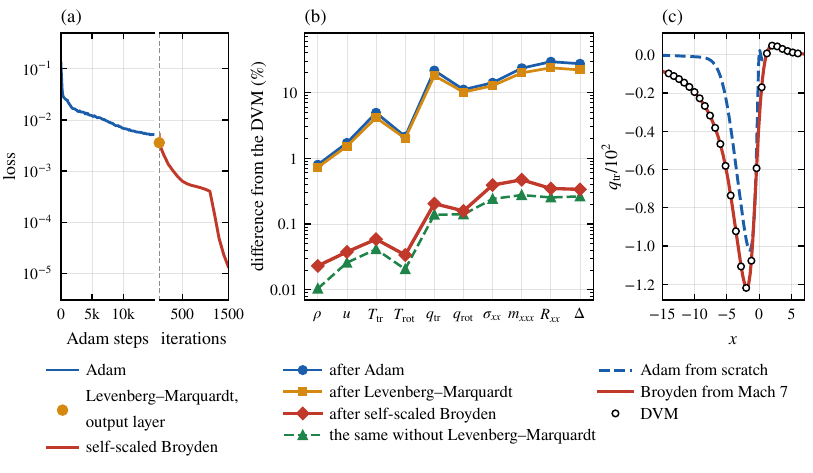}
\caption{The optimiser sequence. (a) Nitrogen, Mach 5, from scratch: loss against Adam steps (left part, steps) and against the iterations of the self-scaled Broyden method (right part, iterations; the two parts have different scales); the single marker shows the loss after the Levenberg--Marquardt solve of the output layer. (b) The same run: difference from the fourth-order DVM after each stage, and for a run without the Levenberg--Marquardt stage; the thin lines only join the symbols of one state. (c) Nitrogen, Mach 8: translational heat flux after Adam from scratch and after 600 quasi-Newton iterations started from the Mach 7 solution; symbols: DVM.}
\label{fig:optimiser}
\end{figure}

\Cref{fig:optimiser}(a,b) follows the baseline of the ablation through its three stages. Adam brings the loss to $5.2\times10^{-3}$ in 15\,000 steps (40 minutes on one core), leaving $q_{\mathrm{tr}}$ 22 per cent and the fourth-order moments 27 to 29 per cent from the DVM ($e_F=1.1\times10^{-2}$). The Levenberg--Marquardt solve of the output layer lowers the loss to $3.6\times10^{-3}$ in 4 minutes without improving the fields (18 per cent in $q_{\mathrm{tr}}$). The self-scaled Broyden method then lowers the loss 260-fold, to $1.4\times10^{-5}$, in 1500 iterations (42 minutes) and brings every field within 0.5 per cent; without the Levenberg--Marquardt stage the run ends at $5.9\times10^{-6}$ and 0.010 to 0.28 per cent, within the difference between two seeds. The accuracy therefore comes from the quasi-Newton stage, although that method was not compared with L-BFGS in this code and part of its gain may come from the fixed collocation points and the frozen envelope (two other alternatives: Supplementary Section~\sref{supp:n2}).

For strong shocks the starting point decides (\cref{fig:optimiser}c). At Mach 8, Adam from scratch leaves $\rho$ 2.8 per cent and $q_{\mathrm{tr}}$ 42 per cent from the DVM after 25\,000 steps ($e_F=3.7\times10^{-2}$), and the three-stage procedure from scratch collapses; the quasi-Newton method alone, started from the Mach 7 solution, reaches 7.1 per cent in $q_{\mathrm{tr}}$ after 50 iterations and 0.44 per cent after 600. The transfer works because the inputs \eqref{eq:inputs} and the base \eqref{eq:mixture} follow the Rankine--Hugoniot states (the Mach 5 parameters evaluated at Mach 7 give $\rho$ within 0.7 per cent and the heat flux within 11 per cent before any training), and Adam does not exploit it: from the Mach 8 solution, 12\,000 Adam steps at Mach 12 stayed at 24 per cent in $q_{\mathrm{tr}}$, where 800 quasi-Newton iterations from the Mach 10 solution gave 0.24 per cent. From Mach 7 or 8 upward every solution of \cref{sec:mono,sec:n2} is obtained by continuation (two paths to one Mach number: Supplementary Section~\sref{supp:n2}).

\subsection{Temperature rise far upstream}\label{sec:n2_weak}

The loss measures residuals relative to the local population and moments of the residual in downstream units, and it does not see the heating of the cold gas far ahead of the shock once the rise is below about $10^{-2}\,T_2$. For the Mach 8 solution the rise $T_{\mathrm{tr}}-T_1$ agrees with the DVM within 3 per cent down to 30 lengths ahead of the shock and is 18 per cent low at 40, and the rise of the rotational temperature is 42 per cent low at 30 lengths ahead, where it is $0.018\,T_1$ (further values in Supplementary Section~\sref{supp:n2_weak}); at 50 lengths ahead the network's $T_{\mathrm{rot}}-T_1$ even has the wrong sign. Neither profiles normalised with the downstream state nor the entries of \cref{tab:n2_err} show this error, and an application that needs the far precursor in relative terms is not served by the present loss. The model itself is least physical there: $6\,\ell_0$ ahead of the density midpoint its translational temperature has risen 1.6 to 3.5 times more than in DSMC and molecular dynamics at Mach 7 and 10 with $T_1=300$\,K, a property of its velocity-independent relaxation time \cite{Wu2015Polyatomic} (Supplementary Section~\sref{supp:val_n2}).

\subsection{One network for a range of Mach numbers}\label{sec:n2_param}

\begin{table}
\centering
\footnotesize
\setlength{\tabcolsep}{4.5pt}
\caption{One network with the Mach number as an input, trained on the equation alone for nitrogen from Mach 2 to 8 with the interval $5.5\le M\le6.5$ excluded from every stage: relative $L^2$ difference from the fourth-order DVM solution in per cent. Tests (a) and (b) are passed at 49 Mach numbers.}
\label{tab:n2_param}
\begin{tabular}{lrrrrrrrrrr}
\toprule
$M$ & $\rho$ & $u$ & $T_{\mathrm{tr}}$ & $T_{\mathrm{rot}}$ & $q_{\mathrm{tr}}$ & $q_{\mathrm{rot}}$ & $\sigma$ & $m_{xxx}$ & $R_{xx}$ & $\Delta$ \\
\midrule
2 & 0.13 & 0.085 & 0.15 & 0.22 & 0.65 & 1.5 & 0.56 & 3.5 & 2.5 & 5.2 \\
3 & 0.023 & 0.021 & 0.067 & 0.091 & 0.29 & 0.61 & 0.21 & 0.51 & 0.51 & 0.51 \\
4 & 0.029 & 0.038 & 0.057 & 0.042 & 0.25 & 0.24 & 0.32 & 0.67 & 0.63 & 0.63 \\
5 & 0.026 & 0.040 & 0.058 & 0.034 & 0.21 & 0.28 & 0.37 & 0.60 & 0.73 & 0.63 \\
6 (held out) & 0.030 & 0.046 & 0.066 & 0.032 & 0.22 & 0.28 & 0.43 & 0.57 & 0.69 & 0.73 \\
7 & 0.031 & 0.050 & 0.086 & 0.037 & 0.35 & 0.28 & 0.51 & 0.64 & 0.72 & 0.96 \\
8 & 0.032 & 0.068 & 0.13 & 0.041 & 0.62 & 0.30 & 0.64 & 1.0 & 1.0 & 1.8 \\
\bottomrule
\end{tabular}
\end{table}

All results so far use one network per Mach number. One run, specified in advance, tested whether a single network can represent the range from Mach 2 to 8: the Mach number enters as a seventh input, every constant of a single-Mach run follows it, the network has 8031 parameters, and the loss is the mean of \eqref{eq:loss} over Mach numbers (Supplementary Section~\sref{supp:n2_param}). The interval $5.5\le M\le6.5$ was excluded from every stage. Tests (a) and (b) are passed at 49 Mach numbers, including that interval ($e_F\le2.0\times10^{-4}$, $r_{\mathrm{rel}}\le1.09$). At the held-out Mach 6 the bulk fields are within 0.07 per cent of the DVM, the heat fluxes and the stress within 0.43 per cent and the higher moments within 0.73 per cent, and no entry exceeds the larger of its two neighbours at Mach 5 and 7 (\cref{tab:n2_param}): no loss of accuracy inside the excluded interval is detected. Covering the range costs accuracy: the differences are 0.7 to 7 times those of single-Mach networks of the same size at Mach 5, 7 and 8. Mach 2 is the weak end, with differences 10 to 40 times those of the single-Mach network, up to 3.5 per cent in $m_{xxx}$ and 5.2 per cent in $\Delta$, which the two tests ($e_F=1.7\times10^{-4}$ and $r_{\mathrm{rel}}=0.90$ at Mach 2) did not flag; the cause was not investigated. This is one run with one seed, and only Mach 6 tests the excluded interval.

\section{The Boltzmann equation for hard spheres}\label{sec:boltz}

The relaxation models need a few moments of the distribution; the collision integral \eqref{eq:boltzmann} needs the distribution itself in three velocity dimensions. The form of $f$, the six terms of the loss and the optimiser are those of \cref{sec:method} with the changes of \cref{sec:models_diff}; \ref{app:collision} outlines how the collision integral is evaluated, and Supplementary Section~\sref{supp:collision_supp} gives the evaluation, its tests and the Monte Carlo references. The loss contains the Boltzmann equation, the analytic end states and conservation. Every Boltzmann solution below was obtained by minimising it with the quasi-Newton method, started from a network solution of the Shakhov model in the same three velocity variables; no reference was present on the cluster, and each run was compared once, after its acceptance tests; the one exception, an interim comparison at Mach 10, is stated in \cref{sec:boltz_M8M10}. Solutions are reported at Mach 2, 3, 5, 8 and 10.

\subsection{Evaluation of the collision integral}\label{sec:boltz_residual}

\begin{figure}[tp]
\centering
\includegraphics[width=\textwidth]{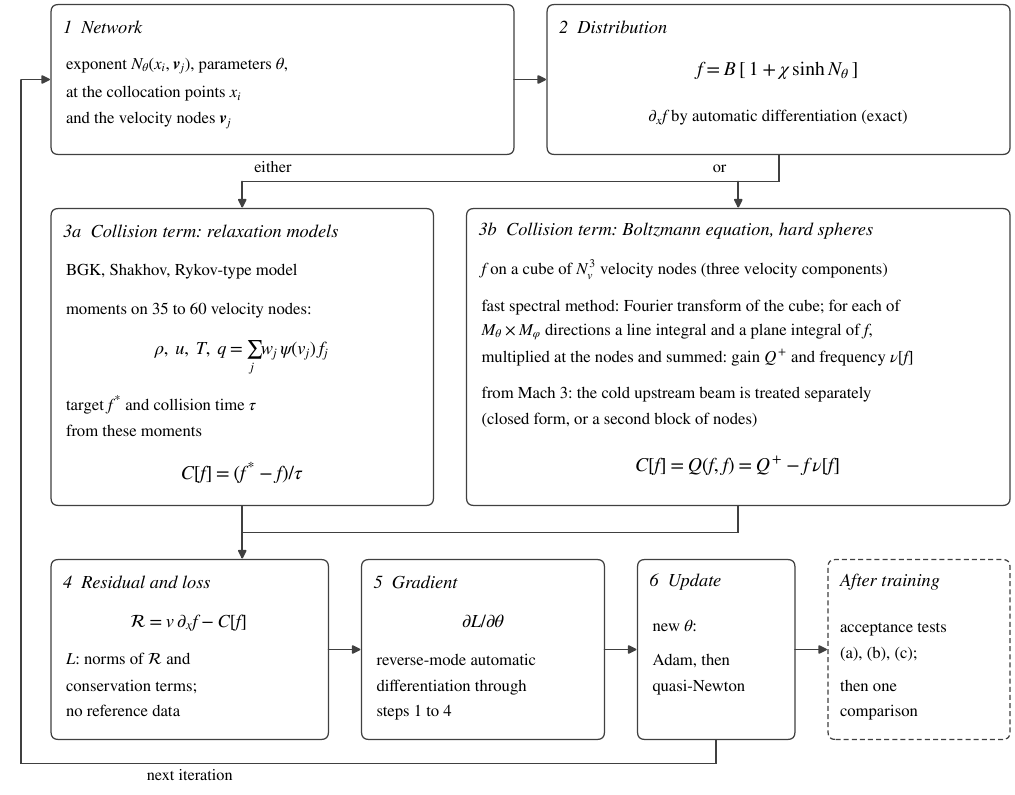}
\caption{One optimisation step. Steps 1 and 2 and steps 4 to 6 are the same for all models; the collision term $C[f]$ is evaluated either from a few moments of $f$ (3a, relaxation models) or by the fast spectral method on a cube of velocity nodes (3b, Boltzmann equation for hard spheres). The gradient of step 5 passes through all earlier steps, the Fourier transforms of step 3b included. The acceptance tests and the comparison with a reference (dashed frame) follow the last iteration and are not part of the loss.}
\label{fig:algorithm}
\end{figure}

\Cref{fig:algorithm} shows one optimisation step for both kinds of collision term. In every case there is one unknown, the network: the fast spectral method is not a separate solver of the Boltzmann equation, but the rule by which the collision integral of the distribution given by the network is evaluated inside the residual. In each step the network gives the exponent $\Nth$ at the collocation points and at the velocity nodes (step 1), and the distribution \eqref{eq:form} follows with its exact derivative in $x$ (step 2). For the relaxation models the collision term needs only a few moments of $f$ on a one-dimensional set of velocity nodes (step 3a). For the Boltzmann equation the values of $f$ on a cube of velocity nodes are passed to the fast spectral method, which returns $Q(f,f)$ at the same nodes (step 3b). The residual and the loss are then formed in the same way (step 4), the gradient of the loss with respect to the parameters is obtained by reverse-mode automatic differentiation through all steps (step 5), and the optimiser updates the parameters (step 6).

The residual $\Res=v\,\partial_xf-Q(f,f)$ is collocated on a cube of $N_v^3$ velocity nodes, and $Q$ is evaluated by the fast spectral method \cite{MouhotPareschi2006Fast}: for each of $M_\theta\times M_\varphi$ directions, a line integral and a plane integral of $f$ are obtained by two inverse Fourier transforms of the cube and multiplied at the nodes. The network has 7637 parameters, the quasi-Newton stages use 160 fixed collocation points (Supplementary Section~\sref{supp:boltz_details}), and the three tests of \cref{sec:acceptance} were fixed in advance for every run. Five discretisations of the collision term were tried. The uniform operator was used at Mach 2 and failed at Mach 5 (\cref{sec:boltz_uniform}); the operator with the cold beam split off and its collisions evaluated in closed form was used at Mach 3 and 5; the operator on two-scale nodes passed its operator tests from Mach 8 to 20, was verified in training at Mach 2 and was used at Mach 8 and 10 (\cref{sec:boltz_M8M10}); and two further operators failed their operator tests and were never used for a result (Supplementary Section~\sref{supp:collision_supp}).

\subsection{Results at Mach 2, 3 and 5}\label{sec:boltz_results}

The Mach 2 solution was reached in three stages on $48^3$ nodes with $L=30$: the Shakhov model with $\tau=1/\rho$ from scratch, the Shakhov model with the hard-sphere viscosity, and the Boltzmann equation with $8\times16$ directions (1500 iterations). The two Shakhov stages are within 0.001 to 0.03 and 0.002 to 0.07 per cent of the DVM solutions of their models. The Mach 3 solution started from a Shakhov solution obtained on $64^3$ nodes by continuation from Mach 2, and used the split operator on $48^3$ nodes of step 1.0 with $12\times24$ directions and $L=40$. Its first run ended at iteration 377 of 1500 on failed line searches, passed the three tests and was compared as specified; the cause, a gradient that omitted the dependence of the loss on the beam polynomial, was found without a reference (Supplementary Section~\sref{supp:collision_split}). A second run with this one change, started from the final model of the first, was stopped by its time limit at iteration 1408. The Mach 5 solution started from the Shakhov solution with $\tau=1/\rho$ at Mach 5 on $64^3$ nodes (half-width 37.6, step 1.18) and used the split operator on the same cube with $16\times32$ directions and $L=50$. The settings were fixed before the job was submitted; the one change from the failed run of \cref{sec:boltz_uniform} is the collision operator. The run completed its 1500 quasi-Newton iterations in 29.7 hours of one A100 card, in ten allocations between which the cluster preempted the job (Supplementary Section~\sref{supp:boltz_M5}). One evaluation of the loss with its gradient took 21 to 25\,s (32\,s on one card), against 5.0\,s at Mach 3, roughly in proportion to the product of the numbers of nodes and of directions, 2.4 and 1.8 times those of Mach 3.

\begin{figure}[tp]
\centering
\includegraphics[width=\textwidth]{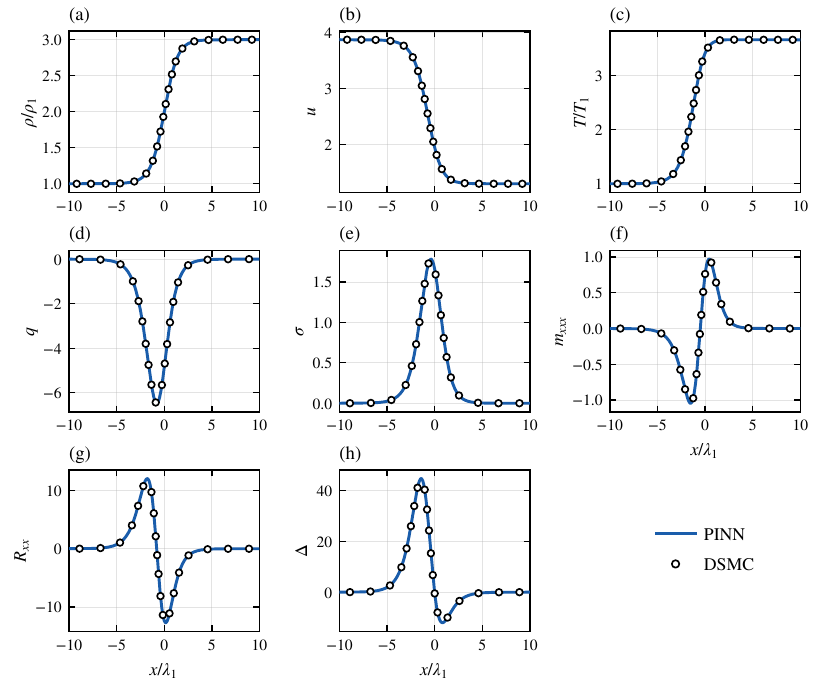}
\caption{Boltzmann equation for hard spheres, Mach 3, second run of \cref{tab:boltz_results} (not accepted: misses test (c)): (a) $\rho$, (b) $u$, (c) $T$, (d) $q$, (e) $\sigma$, (f) $m_{xxx}$, (g) $R_{xx}$, (h) $\Delta$. Line: PINN; symbols: DSMC. The coordinate $x$ is in upstream mean free paths.}
\label{fig:boltz}
\end{figure}

\begin{figure}[tp]
\centering
\includegraphics[width=\textwidth]{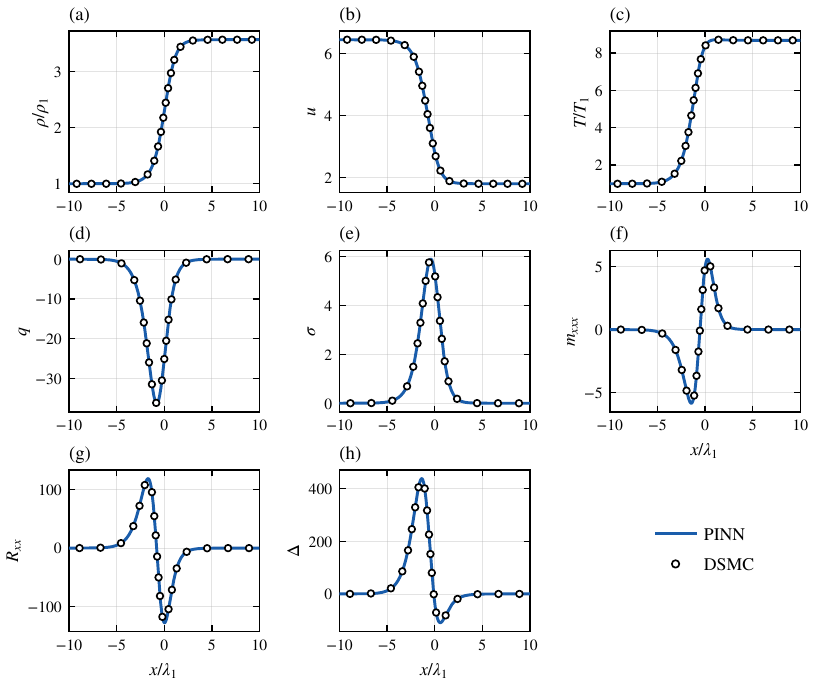}
\caption{Boltzmann equation for hard spheres, Mach 5, split operator with $16\times32$ directions on $64^3$ velocity nodes (\cref{tab:boltz_results}): (a) $\rho$, (b) $u$, (c) $T$, (d) $q$, (e) $\sigma$, (f) $m_{xxx}$, (g) $R_{xx}$, (h) $\Delta$. Line: PINN; symbols: DSMC. The coordinate $x$ is in upstream mean free paths.}
\label{fig:boltz_M5}
\end{figure}

\begin{table}
\centering
\small
\setlength{\tabcolsep}{3.2pt}
\caption{Hard spheres: flux error $e_F$ of test (a), status under the three tests fixed in advance, and relative $L^2$ difference \eqref{eq:error} between the network and the DSMC, in per cent. The lines marked s.e. give the root-mean-square (r.m.s.) difference from the DSMC in units of its standard error (s.e.), which was evaluated at Mach 2, 5, 8 and 10. Window: $|x|\le20$, 40, 18, 17 and $18\,\lambda_1$ at Mach 2, 3, 5, 8 and 10; at Mach 3 the end values of the DSMC are used outside its profile, $-28\le x\le23$. The last two lines give, for scale, DVM solutions of the Shakhov and BGK models with the hard-sphere viscosity against the same DSMC.}
\label{tab:boltz_results}
\begin{tabularx}{\textwidth}{@{}Lcccccccccc@{}}
\toprule
Solution & $10^5e_F$ & Tests & $\rho$ & $u$ & $T$ & $q$ & $\sigma$ & $m_{xxx}$ & $R_{xx}$ & $\Delta$\\
\midrule
\multicolumn{11}{@{}l}{Mach 2}\\
Uniform cube, $48^3$, $8\times16$ & 1.4 & passed & 0.014 & 0.012 & 0.015 & 0.13 & 0.11 & 0.55 & 0.27 & 0.47\\
\quad the same, s.e. & & & 1.3 & 1.3 & 1.3 & 1.7 & 1.1 & 1.1 & 1.8 & 6.8\\
Two-scale nodes, $64^3+18^3$ & 0.68 & passed & 0.014 & 0.013 & 0.015 & 0.14 & 0.11 & 0.55 & 0.26 & 0.31\\
\quad the same, s.e. & & & 1.4 & 1.4 & 1.3 & 1.6 & 1.1 & 1.1 & 1.5 & 1.6\\
Two-scale against uniform & & & 0.003 & 0.003 & 0.003 & 0.05 & 0.03 & 0.08 & 0.09 & 0.38\\
\midrule
\multicolumn{11}{@{}l}{Mach 3, split operator, $48^3$, $12\times24$}\\
First run (377 iterations) & 14 & passed & 0.039 & 0.031 & 0.041 & 0.36 & 0.37 & 0.93 & 0.80 & 0.76\\
Second run (1408 iterations) & 0.81 & (c) missed & 0.006 & 0.003 & 0.005 & 0.07 & 0.07 & 0.26 & 0.13 & 0.18\\
\midrule
\multicolumn{11}{@{}l}{Mach 5, split operator, $64^3$, $16\times32$}\\
1500 iterations & 8.4 & passed & 0.0087 & 0.017 & 0.035 & 0.20 & 0.12 & 0.26 & 0.31 & 0.39\\
\quad the same, s.e. & & & 4.4 & 17 & 38 & 31 & 18 & 14 & 26 & 30\\
\midrule
\multicolumn{11}{@{}l}{Mach 8, two-scale nodes, $80^3+26^3$, $8\times16$ (\cref{sec:boltz_M8M10})}\\
800 iterations & 4.3 & passed & 0.0093 & 0.0077 & 0.013 & 0.054 & 0.051 & 0.19 & 0.14 & 0.10\\
\quad the same, s.e. & & & 3.3 & 8.8 & 22 & 6.5 & 3.9 & 5.0 & 19 & 14\\
\midrule
\multicolumn{11}{@{}l}{Mach 10, two-scale nodes, $80^3+26^3$, $8\times16$ (\cref{sec:boltz_M8M10}; comparison not blind)}\\
800 iterations & 8.8 & passed & 0.011 & 0.015 & 0.024 & 0.15 & 0.14 & 0.34 & 0.30 & 0.26\\
\quad the same, s.e. & & & 4.9 & 24 & 89 & 44 & 16 & 15 & 72 & 200\\
\midrule
\multicolumn{11}{@{}l}{Mach 2, relaxation models against the DSMC}\\
Shakhov & & & 0.75 & 1.1 & 1.6 & 10 & 8.8 & 61 & 54 & 20\\
BGK & & & 2.2 & 2.5 & 4.2 & 27 & 14 & 89 & 62 & 24\\
\bottomrule
\end{tabularx}
\end{table}

\Cref{tab:boltz_results} gives the differences from the DSMC in the measure \eqref{eq:error}, and \cref{fig:boltz,fig:boltz_M5} show the eight fields at Mach 3 and 5. At Mach 2 the comparison is limited by the scatter of the reference. The root-mean-square (r.m.s.) difference is 1.1 to 1.8 standard errors of the DSMC in seven fields for the uniform cube, and 1.1 to 1.6 in all eight for the two-scale nodes; in $\rho$ to $R_{xx}$ the two networks differ from each other three to seven times less than either differs from the DSMC, and the 0.5 per cent in $m_{xxx}$ is the relative size of the DSMC scatter in a small field. In $\Delta$ the uniform cube carries an error of its own, 0.4 per cent (6.8 standard errors), which the two-scale nodes remove. The density thickness is $3.922\,\lambda_1$ against 3.902 for the DSMC (3.912 with the fit of Supplementary Section~\sref{supp:val_hs}), so the network is 0.3 to 0.5 per cent thicker; the thickness of the DSMC itself depends on the fit by up to 0.4 per cent. Against the tables of Ohwada's deterministic solution \cite{Ohwada1993Shock} the network differs by $5.6\times10^{-4}$, $3.4\times10^{-4}$ and $7.2\times10^{-4}$ of the jumps in $\rho$, $u$ and $T$ (r.m.s.), as the DSMC does (\cref{sec:validation}).

At Mach 3 the second run is within 0.003 to 0.006 per cent of the DSMC in $\rho$, $u$ and $T$, 0.07 per cent in $q$ and $\sigma$ and 0.13 to 0.26 per cent in the higher moments (thickness 2.808 against $2.806\,\lambda_1$); how much of this is statistical was not assessed. Against the table of Dodulad and Tcheremissine \cite{DoduladTcheremissine2013Shock}, at its 37 points, the network differs by 0.0011, 0.0020 and 0.0046 (r.m.s.) in density, temperature and longitudinal temperature, and the DSMC by 0.0012, 0.0021 and 0.0046, so that the three decimals of the table limit this comparison; against Ohwada's tables \cite{Ohwada1993Shock} the differences are $2.2\times10^{-4}$, $1.6\times10^{-4}$ and $3.0\times10^{-4}$ of the jumps in $\rho$, $u$ and $T$, again those of the DSMC. The distribution function was checked with the first run, which is 3.6 to 9.0 times farther from the DSMC in the moments than the reported run, so this check is qualitative. The first run reproduces Ohwada's $f$ at three positions and four radial speeds to 0.006 of the upstream maximum, the reading error of his figure; the reported network gives 0.009 in the same comparison, the difference lying at the beam, where the network is evaluated between its velocity nodes of step 1.0, and its marginal agrees with the DSMC to $2\times10^{-4}$ on average. The marginal of the first run $\int f\,\dd v_y\dd v_z$ agrees with the DSMC at 29 positions to $3\times10^{-4}$ of its maximum ($2\times10^{-4}$ for the Mach 2 network) (\cref{sec:validation}; Supplementary Section~\sref{supp:val_hs}). By the rule of \cref{sec:acceptance} the second run is not accepted: it ended on its time limit, and while it passes tests (a) and (b) ($e_F=8.1\times10^{-6}$, $r_{\mathrm{rel}}=1.01$) it misses test (c) in $\Delta$, which changes on the finer cube by $1.12\times10^{-3}$ of its downstream scale against the limit $10^{-3}$ (other fields at most $8.1\times10^{-4}$). The first run passed all three tests (test (c) with $5.7\times10^{-5}$) but ended on failed line searches with a gradient that was not consistent with its objective, which is why it was continued rather than reported; it is 3.6 to 9.0 times farther from the DSMC. The moments of the second run are thus at the limit of what a cube of step 1.0 integrates.

At Mach 5 the run passes the three tests: $e_F=8.4\times10^{-5}$, the kinetic term at the test points is 1.01 times its training value, and on the finer cube the fields change by at most $9.3\times10^{-4}$ of their downstream scales, in $\Delta$. The largest flux error lies inside the shock, and five units from the inflow boundary $T-1=2.1\times10^{-4}$, so that the domain leaves no boundary layer like that of the Shakhov model at Mach 10, where this excess was 0.70 (\cref{sec:boltz_twoscale}). The differences from the DSMC are 0.009 to 0.035 per cent in $\rho$, $u$ and $T$, 0.12 and 0.20 per cent in $\sigma$ and $q$, and 0.26 to 0.39 per cent in the third- and fourth-order moments. The density thickness is $2.387\,\lambda_1$ against 2.389 for the DSMC ($-0.068$ per cent); both solutions place the maximum of the temperature, 0.40 per cent above $T_2$, $1.2\,\lambda_1$ behind the density midpoint and agree on it to five digits, and the extrema of $q$, $\sigma$, $R_{xx}$ and $\Delta$ agree within 0.02 to 0.12 per cent.

These differences exceed the statistical scatter of the DSMC (4 to 38 of its standard errors in r.m.s., the standard errors being about $10^{-4}$ of the scales of the fields). They are systematic and, in every field but the density, largest 3 to 6 mean free paths ahead of the density midpoint, where they reach $9\times10^{-4}$ of $T_2$ in the temperature and $5\times10^{-3}$ of $\rho_2T_2^2$ in $\Delta$ (Supplementary Fig.~\sref{fig:boltz_M5_diff}). They are 1.3 to 8 times the change of the network on the finer cube, and 0.8 to 2.3 times the change of the DSMC when its cell and time step are halved (\cref{sec:refs_dsmc}), that is, of the size of the discretisation error of the reference. The DSMC domain begins closer to the shock, at $-30\,\lambda_1$, than that of the network ($L=50\,\lambda_1$), and a domain that is short upstream is known to distort the precursor (\cref{sec:boltz_twoscale}); a second DSMC calculation with the inflow boundary at $-60\,\lambda_1$ therefore tested this. It differs from the first by 0.5 to 1.7 joint standard errors, and the difference between the network and the DSMC in the precursor is the same against both ($3.4\times10^{-4}$ of $T_2$ in the temperature, r.m.s.), so that the position of the inflow boundary of the DSMC is not its cause; together with the far-upstream excess of temperature of the networks (\cref{sec:boltz_M8M10}) this attributes the precursor difference to the network. Far ahead of the shock, where the DSMC is at the upstream state within its standard error, the network also shows a weak heating, $0.7\times10^{-3}$ to $1.4\times10^{-3}$ of $T_1$ between 28 and 12 mean free paths ahead; this is below $2\times10^{-4}$ of $T_2$, which the loss does not resolve (\cref{sec:n2_weak}).

The Boltzmann equation is therefore solved here at Mach 2, to within the scatter of the DSMC, and at Mach 5, to within 0.4 per cent (relative $L^2$) in each of the eight fields, with differences of the size of the discretisation error of the reference; at Mach 3 the best run is within 0.3 per cent of the DSMC but misses one test fixed in advance by 12 per cent and ended on its time limit. For comparison, the last two lines of \cref{tab:boltz_results} show how far the relaxation models are from the hard-sphere solution: with the same viscosity, the Shakhov and BGK models differ from the hard-sphere result at Mach 2 by 10 and 27 per cent in the heat flux and by 20 to 89 per cent in the third- and fourth-order moments, two orders of magnitude more than the network differs from the DSMC.

\subsection{Why a uniform cube fails from Mach 5}\label{sec:boltz_uniform}

The first Mach 5 calculation, on a uniform cube of $64^3$ nodes (step 1.18) with $8\times16$ directions, did not converge: after 1500 iterations the loss had stalled at $1.8\times10^{-3}$ with $e_F=4.9\times10^{-2}$, and the run was not compared with any reference. Two causes were found from the operator alone. On that cube the cold upstream Maxwellian is not an equilibrium of the discrete operator: its collision term is $1.7\times10^{-3}$ of its loss term, against $2.8\times10^{-6}$ on the cube of Mach 2, so that the loss has a floor set by the velocity grid and the network distorts the solution to lower it. And $8\times16$ directions do not integrate the collisions between the beam and the hot gas: on a cube fine enough for the beam their error is $9\times10^{-6}$ at Mach 2 and $2.6\times10^{-3}$ at Mach 5. The split operator removes the first cause, since the narrow Maxwellian of the beam never passes through a transform on the coarse cube, and $16\times32$ directions remove the second; with both, the Mach 5 shock is solved on the same cube (\cref{sec:boltz_results}). On a uniform cube the two requirements would lead to $104^3$ nodes and $16\times32$ directions at Mach 5 and, with the analytic beam, to an estimated 5.6 minutes per quasi-Newton iteration at Mach 8. The two-scale nodes remove the growth of both numbers with the Mach number (\cref{fig:grids}). Deterministic solvers met the same requirements: with a uniform velocity step of $1.5\sqrt{RT_1}$ at Mach 8, the nodes of the discrete-velocity solution of Malkov et al.\ \cite{Malkov2015Deterministic} miss the maximum of the beam by 11 per cent, and the cost of the solver of Dodulad and Tcheremissine \cite{DoduladTcheremissine2013Shock} grows like $M^2$, a cluster being needed above Mach 10.

\subsection{Two-scale nodes and a preliminary Shakhov stage}\label{sec:boltz_twoscale}

Before the Boltzmann operator is used on the two-scale nodes at a Mach number, the Shakhov model with $\tau=1/\rho$ is solved on the same nodes, by continuation in the Mach number with 1500 quasi-Newton iterations per step, and compared once with the DVM solution of the same model (length unit $\ell_0$). \Cref{tab:boltz_twoscale} records the three sequences of runs that were made.

\begin{table}
\centering
\small
\setlength{\tabcolsep}{2.8pt}
\caption{Shakhov model on the two-scale velocity nodes: tests of \cref{sec:acceptance} ($e_F$: flux error, limit $10^{-3}$; $r_{\mathrm{rel}}$: ratio of the losses, limit 10; $c$: largest change of a field on the finer velocity grid, limit $10^{-3}$), the excess temperature $\vartheta_5=T-1$ five units from the inflow boundary, and relative $L^2$ difference \eqref{eq:error} from the DVM in per cent. Sequence T: $64^3+18^3$ nodes, $L=40$, 160 collocation points. Sequence F: $80^3+26^3$ nodes, $L=40$, 160 collocation points. Sequence G: the nodes of F, $L=15M$, 320 collocation points. A dash marks a run that was not compared. Accepted: T at Mach 3 and 5, F at Mach 8, G at Mach 8 and 10.}
\label{tab:boltz_twoscale}
\footnotesize
\begin{tabular}{@{}lccccccccccccc@{}}
\toprule
Sequence & $M$ & $10^5e_F$ & $r_{\mathrm{rel}}$ & $10^5c$ & $\vartheta_5$ & $\rho$ & $u$ & $T$ & $q$ & $\sigma$ & $m_{xxx}$ & $R_{xx}$ & $\Delta$\\
\midrule
T & 3 & 1.1 & 0.97 & 0.98 & & 0.003 & 0.001 & 0.002 & 0.015 & 0.021 & 0.029 & 0.021 & 0.021\\
T & 5 & 5.3 & 1.03 & 9.5 & 0.010 & 0.004 & 0.011 & 0.020 & 0.040 & 0.026 & 0.059 & 0.071 & 0.064\\
T & 8 & 60 & 1.06 & 300 & 0.21 & 0.027 & 0.075 & 0.087 & 0.18 & 0.20 & 1.1 & 2.1 & 3.5\\
F & 8 & 62 & 0.92 & 9.6 & 0.18 & 0.031 & 0.090 & 0.14 & 0.24 & 0.18 & 0.30 & 0.42 & 0.96\\
F & 10 & 161 & 1.19 & 51 & 0.70 & -- & -- & -- & -- & -- & -- & -- & --\\
G & 8 & 1.9 & 1.00 & 15 & 0.00006 & 0.002 & 0.005 & 0.007 & 0.023 & 0.034 & 0.050 & 0.030 & 0.033\\
G & 10 & 4.4 & 1.00 & 39 & 0.00033 & 0.004 & 0.009 & 0.008 & 0.031 & 0.067 & 0.078 & 0.041 & 0.092\\
\bottomrule
\end{tabular}
\end{table}

Sequence T was accepted to Mach 5. At Mach 8 it missed test (c) in $\Delta$ and, as specified, the next Mach number was not accepted on its basis; its run at Mach 10 reached its time limit with $e_F=1.2\times10^{-3}$. Sequence F took the finer grid of that test as its training grid. At Mach 8 it met the three tests, with 0.96 per cent in $\Delta$ instead of 3.5; at Mach 10 it missed test (a), with a flux error that was no longer falling, and was not compared. The cause was found in the stored profiles, without a reference. The largest flux errors lay in a layer next to the inflow boundary ($1.61\times10^{-3}$ at $x=-39.8$), not in the shock, and five units from that boundary the temperature still exceeded the upstream value by 0.70 at Mach 10 and 0.18 at Mach 8 (Supplementary Section~\sref{supp:boltz_twoscale}). The form \eqref{eq:form} imposes the upstream equilibrium at $x=-L$, and $L=40$ cuts off the hot molecules that run upstream; the cut appears as a boundary layer that the kinetic equation does not allow. The velocity grid was not the cause, since test (c) was passed. The remedy is the rule of the reduced calculations, $L=15M$, with 320 collocation points.

\begin{figure}[tp]
\centering
\includegraphics[width=\textwidth]{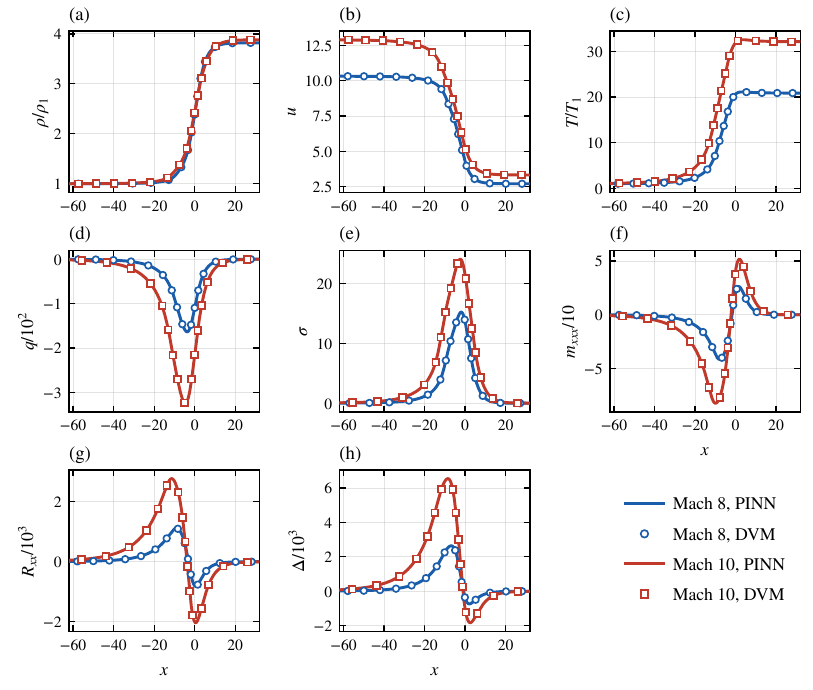}
\caption{Shakhov model solved by the PINN on the two-scale three-dimensional velocity nodes (sequence G of \cref{tab:boltz_twoscale}), at Mach 8 (blue) and Mach 10 (red); panels as in \cref{fig:mono_profiles}. Lines: PINN; symbols: one-dimensional Shakhov DVM.}
\label{fig:twoscale}
\end{figure}

With this one change sequence G was accepted at Mach 8 and at Mach 10 (\cref{fig:twoscale}): the largest flux errors lie inside the shock, and the differences from the DVM are 0.002 to 0.05 per cent at Mach 8 and 0.004 to 0.09 per cent at Mach 10. At Mach 8 they are 5 to 30 times smaller than in sequence F, whose largest error was therefore the short domain. With twice the parameters (15\,177) and about half a million velocity nodes per collocation point in place of fifty, these solutions are closer to the DVM in $\Delta$ than the reduced calculations of \cref{sec:mono} (0.03 and 0.09 per cent against 0.77 and 2.2 per cent for the reduced Shakhov runs at Mach 8 and 10, on the same window $|x|\le40$), with 1500 to 2000 quasi-Newton iterations per step instead of 800.

\subsection{Results at Mach 8 and 10}\label{sec:boltz_M8M10}

The Boltzmann stages at Mach 8 and 10 start from the Shakhov solutions of sequence G, mapped to mean free paths with the function of $x$ kept ($L=95.5$ and $119.4\,\lambda_1$), and use the same nodes ($80^3+26^3$), $8\times16$ directions, 320 collocation points and the network of 15\,177 parameters. The loss contains the Boltzmann equation, the analytic end states and conservation, and it is minimised by 800 quasi-Newton iterations; this number was fixed in advance from the measured cost, about 100 seconds per iteration (21.9 hours of one A100 card at Mach 8). The settings and the three tests were fixed before each job was submitted, and the Mach 10 job could start only after the Mach 8 stage had passed its tests. Both stages pass: at Mach 8 $e_F=4.3\times10^{-5}$, the kinetic term at the test points is 1.00 times its training value and the fields change by at most $4.5\times10^{-6}$ of their downstream scales on the finer nodes; at Mach 10 the three numbers are $8.8\times10^{-5}$, 1.01 and $1.3\times10^{-5}$.

\begin{figure}[tp]
\centering
\includegraphics[width=\textwidth]{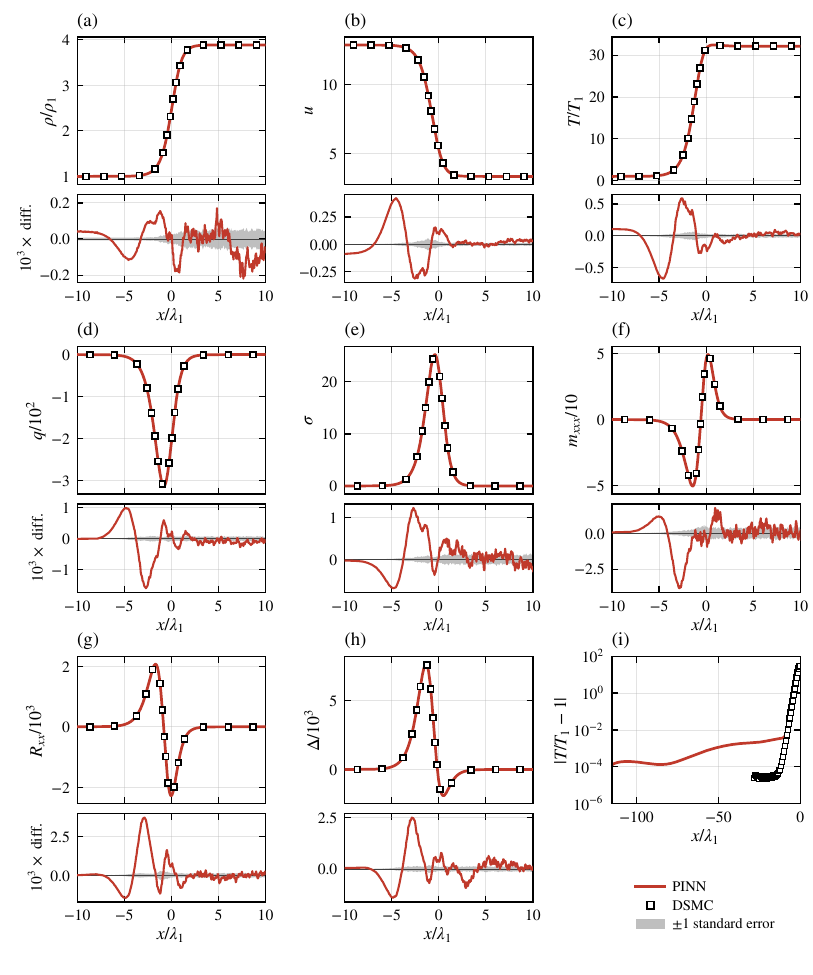}
\caption{Boltzmann equation for hard spheres at Mach 10 (comparison not blind): panels (a) to (h) and the difference panels as in \cref{fig:boltz_M8}; the DSMC has halved cell and time step. (i) Far-upstream temperature, $|T/T_1-1|$, on a logarithmic axis: the network keeps an excess of $2\times10^{-3}$ to $3.5\times10^{-3}$ between 28 and 12 mean free paths ahead of the density midpoint, where the DSMC is within $10^{-4}$ of $T_1$; the profiles (a) to (h) cannot show this.}
\label{fig:boltz_M10}
\end{figure}

At Mach 8 the network was compared once with the DSMC (\cref{tab:boltz_results}; \cref{fig:boltz_M8} in the introduction shows the profiles and, under each, the difference from the DSMC with its standard error, and \cref{fig:boltz_M10} the same at Mach 10). On $|x|\le10\,\lambda_1$ the pointwise difference at Mach 8 is at most $0.6\times10^{-3}$ of the scale of the field in $\rho$, $u$, $T$, $q$ and $\sigma$ and at most $2.0\times10^{-3}$ in the third- and fourth-order moments; at Mach 10 the two bounds are $1.6\times10^{-3}$ and $3.9\times10^{-3}$. The standard error of the DSMC is $0.6\times10^{-3}$ or less, and in most fields the largest difference lies 1 to 5 mean free paths ahead of the density midpoint. At Mach 8 the relative $L^2$ differences are 0.008 to 0.013 per cent in $\rho$, $u$ and $T$, 0.05 per cent in $q$ and $\sigma$ and 0.10 to 0.19 per cent in the third- and fourth-order moments; the density thickness is $2.2652\,\lambda_1$ against 2.2650, and the extrema of $q$, $\sigma$, $T$ and $\Delta$ agree within 0.02 per cent. Against the table of Dodulad and Tcheremissine \cite{DoduladTcheremissine2013Shock} the network differs by $2.8\times10^{-4}$, $7.5\times10^{-4}$ and $1.4\times10^{-3}$ of the jumps (r.m.s. over its 37 points) in density, temperature and longitudinal temperature, and the DSMC by $2.8\times10^{-4}$, $8.5\times10^{-4}$ and $1.5\times10^{-3}$, while network and DSMC differ from each other about six times less.

At Mach 10 the reference is a DSMC calculation with the cell and the time step halved (24 ensembles). This comparison is not blind. Before the run ended, a copy of its state at iteration 550 was compared with the DSMC, at the author's request, while the job continued unchanged; the differences were then 0.014 to 0.38 per cent. The final state, after the 800 iterations fixed in advance, was compared once: 0.011 to 0.024 per cent in $\rho$, $u$ and $T$, 0.15 and 0.14 per cent in $q$ and $\sigma$, and 0.26 to 0.34 per cent in the third- and fourth-order moments; the thickness is $2.2384\,\lambda_1$ against 2.2363 ($+0.09$ per cent), and the maximum of the temperature, 0.97 per cent above $T_2$, agrees to five digits. Nothing in the training depended on the interim look, but the result cannot count as a prediction.

In units of the standard error of the DSMC the differences are resolved at both Mach numbers (3 to 22 standard errors at Mach 8, 5 to 200 at Mach 10; \cref{tab:boltz_results}), and, as at Mach 5, they are systematic and largest in the precursor. They go together with a slow excess of temperature far upstream that the networks keep from Mach 5 on and the DSMC does not have. Between 28 and 18 mean free paths ahead of the density midpoint $T-T_1$ is $0.7\times10^{-3}$ to $1.1\times10^{-3}$ of $T_1$ at Mach 5, $0.7\times10^{-3}$ to $0.8\times10^{-3}$ at Mach 8 and $2.1\times10^{-3}$ to $2.8\times10^{-3}$ at Mach 10, where the DSMC is within $4\times10^{-5}$ of $T_1$; the networks at Mach 2 and 3 have no such tail. Relative to the temperature jump this is $1.4\times10^{-4}$ or less, below what the three tests resolve (limits of $10^{-3}$ of the downstream scales). A likely origin, not demonstrated, is the Shakhov solution from which each stage starts, whose precursor is much longer because its viscosity is proportional to $T$; the loss weights its remnant lightly once the local distribution is the cold beam. The accuracies stated above therefore refer to the shock; far upstream the temperature of these networks is accurate to about $3.5\times10^{-3}\,T_1$.

For the Mach 10 shock F.~G. Tcheremissine provided a solution by his conservative projection method (private communication \cite{TcheremissinePrivate2026}). His data are not reproduced here; only differences are quoted. The reduced density and temperature of the network differ from his by 0.49 and 0.68 per cent of the jumps (r.m.s.), the heat flux by 0.51 per cent of his peak value, and the stress, formed from his density and longitudinal temperature, by 1.3 per cent of its peak; for the DSMC the four numbers are 0.49, 0.68, 0.49 and 1.3. The distribution $f(v_x,v_r\approx0)$ at four positions, from the inflow boundary to $1.8\,\lambda_1$ behind the density midpoint, differs from his by 0.6 to 0.9 per cent of its peak (r.m.s.) for the network and by 0.6 to 1.1 per cent for the DSMC. Network and DSMC thus differ from the third solution by the same amounts and agree with each other more closely than either agrees with it, at Mach 8 and at Mach 10.

\subsection{Distribution functions}\label{sec:boltz_f}

\begin{figure}[tp]
\centering
\includegraphics[width=\textwidth]{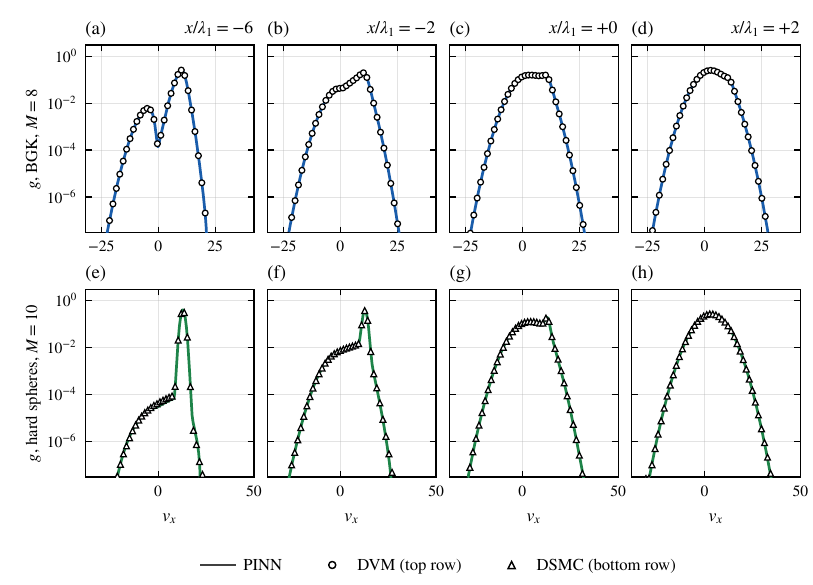}
\caption{Reduced distribution $g(v_x)=\int f\,\dd v_y\dd v_z$ of the networks (lines) and of the references (symbols) at four positions from the density midpoint, stated above the columns. (a--d) BGK, Mach 8, DVM; (e--h) hard spheres, Mach 10, DSMC. For the BGK model $x$ is converted with $\ell_0=0.795882\,\lambda_1$ (\cref{sec:units}); for hard spheres the network is averaged over the velocity bins of the DSMC. The Shakhov model and hard spheres at Mach 8 are in Supplementary Fig.~\sref{fig:S_f_all}.}
\label{fig:f_validation}
\end{figure}

The moments compared so far are integrals of the distribution; \cref{fig:f_validation} and Supplementary Fig.~\sref{fig:S_f_all} compare the distribution itself. The networks were evaluated, after their acceptance and their comparison in the moments, at four positions through the shock where reference distributions are stored: the DVM distribution for the BGK and Shakhov models at Mach 8, and the histogram of $v_x$ of the DSMC for hard spheres at Mach 8 and 10. The r.m.s. difference over the velocities, in units of the largest value of the reference at the position, is $0.5\times10^{-4}$ to $6\times10^{-4}$ for the BGK model (largest pointwise difference $2.1\times10^{-3}$), $0.9\times10^{-4}$ to $5\times10^{-4}$ for the Shakhov model ($2.0\times10^{-3}$), $0.7\times10^{-4}$ to $1.6\times10^{-4}$ for hard spheres at Mach 8 ($8\times10^{-4}$) and $0.8\times10^{-4}$ to $1.8\times10^{-4}$ at Mach 10 ($1.3\times10^{-3}$). At Mach 8 the agreement extends over the six decades drawn, including the molecules that move upstream ahead of the shock: six mean free paths ahead, their share is 3.24 per cent in the BGK network against 3.23 in the DVM, 7.14 against 7.12 per cent for the Shakhov model, and $2.66\times10^{-4}$ against $2.69\times10^{-4}$ for hard spheres at Mach 8. At Mach 10 the network gives $2.25\times10^{-4}$ against $2.60\times10^{-4}$ for the DSMC, a difference of 13 per cent in a population of $2\times10^{-4}$ of the density that is resolved statistically (19 standard errors r.m.s. at this position, 1 to 4 at the others). The Mach 5 network, which uses the split operator, is less accurate in this population: six mean free paths ahead it has 0.022 per cent of the molecules moving upstream against 0.031 per cent in the DSMC, and eight mean free paths ahead about a third of the DSMC value. The sparse upstream-moving molecules of the hard-sphere gas are thus reproduced at Mach 8, within 13 per cent at Mach 10 and only in order of magnitude at Mach 5.

\begin{figure}[tp]
\centering
\includegraphics[width=\textwidth]{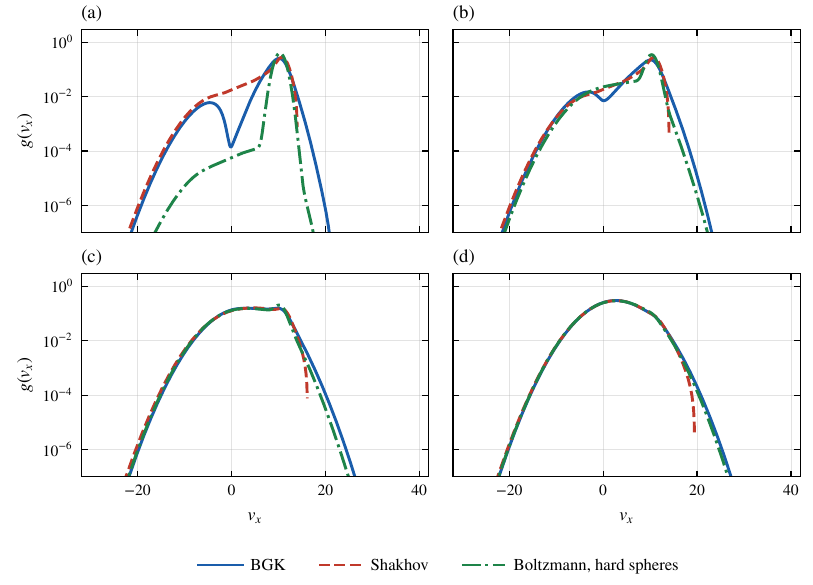}
\caption{Reduced distribution $g(v_x)$ of the three monatomic models at Mach 8, from the networks: BGK and Shakhov with $\tau=1/\rho$ (viscosity proportional to $T$) and the Boltzmann equation for hard spheres (viscosity proportional to $T^{1/2}$), for the same upstream viscosity. (a) Six upstream mean free paths ahead of the density midpoint; (b), (c), (d) where the normalised density $(\rho-\rho_1)/(\rho_2-\rho_1)$ of each solution is 0.1, 0.5 and 0.9. Negative values of the Shakhov distribution are not drawn.}
\label{fig:f_models}
\end{figure}

\begin{figure}[tp]
\centering
\includegraphics[width=\textwidth]{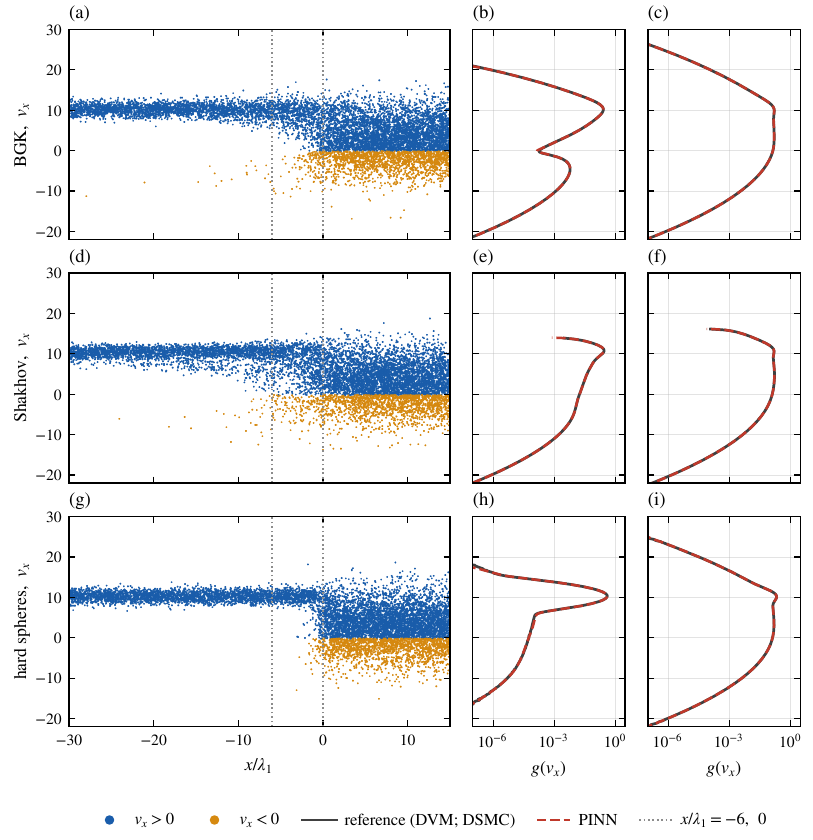}
\caption{The three monatomic models at Mach 8 for the same upstream viscosity: BGK (a--c), Shakhov (d--f), Boltzmann equation for hard spheres (g--i). Left: molecules in the plane of position and velocity $v_x$, 9000 samples of the stored distribution of each model in the window drawn (DVM for BGK and Shakhov, the network for hard spheres; for the Shakhov model samples of its positive part), blue for $v_x>0$ and orange for $v_x<0$; the scatter is rasterised at 400 dots per inch. Middle and right: reduced distribution $g(v_x)$ at $x/\lambda_1=-6$ and 0, the dotted lines of the left panels, with $v_x$ on the vertical axis. Solid: reference (DVM; DSMC for hard spheres); dashed: PINN. Six mean free paths ahead of the density midpoint 3.2, 7.1 and 0.027 per cent of the molecules move upstream in the three models. Negative values of the Shakhov distribution are not drawn on the logarithmic axis. Velocities in $\sqrt{RT_1}$. Supplementary Video 3 animates these data (\url{https://www.youtube.com/watch?v=pMRke3ZgKbI}); Supplementary Video 2 shows the same at Mach 5 (\url{https://www.youtube.com/watch?v=Zow8LDTRQ5M}).}
\label{fig:molecules_models}
\end{figure}

\Cref{fig:f_models} places the three models side by side at Mach 8; the viscosity laws differ, so the comparison is not like for like, and a study of the models is outside the scope of this paper. Where the normalised density is the same (panels b to d) the three distributions nearly coincide in the hot part, and the share of molecules that move upstream is the same within one percentage point (7 to 8, 21 to 22 and 26.5 per cent at the three positions). They differ in the cold beam, which hard spheres keep sharper at the density midpoint, and in the fast forward tail, where the Shakhov distribution turns negative. At a fixed distance ahead of the shock (panel a) the models differ by orders of magnitude, 3.2 and 7.1 per cent of the molecules moving upstream for BGK and Shakhov against 0.027 per cent for hard spheres, because the shocks of the relaxation models with $\tau=1/\rho$ are 2.9 and 3.7 times thicker in upstream mean free paths. \Cref{fig:molecules_models} shows the same three solutions through their molecules, with the reference and the network distributions at two positions; the corresponding figure at Mach 5 is Supplementary Fig.~\sref{fig:S_molecules_M5}. Supplementary Videos 2 and 3 animate these data (Supplementary Section~\sref{supp:video}).

\subsection{Higher Mach numbers}\label{sec:boltz_progress}

Boltzmann stages above Mach 10 are not results of this paper. On the two-scale nodes the hot cube has to grow with the Mach number ($144^3$ nodes at Mach 15), and the cost of one stage grows with it (\cref{sec:disc_cost}); the Boltzmann equation above Mach 10 is therefore open here.

\section{Comparison with published solutions and measurements}\label{sec:validation}

The solutions of Sections~\ref{sec:mono} to \ref{sec:boltz} are now compared with independent published solutions and, for nitrogen, with experiments and molecular simulations (Supplementary Section~\sref{supp:validation}). Network and reference agree within 0.4 per cent in every quantity used here, so larger differences from the literature come from the model equation or from the published result; the far precursor, where the networks are less accurate (\cref{sec:n2_weak}), is not used.

\begin{figure}[tp]
\centering
\includegraphics[width=0.8\textwidth]{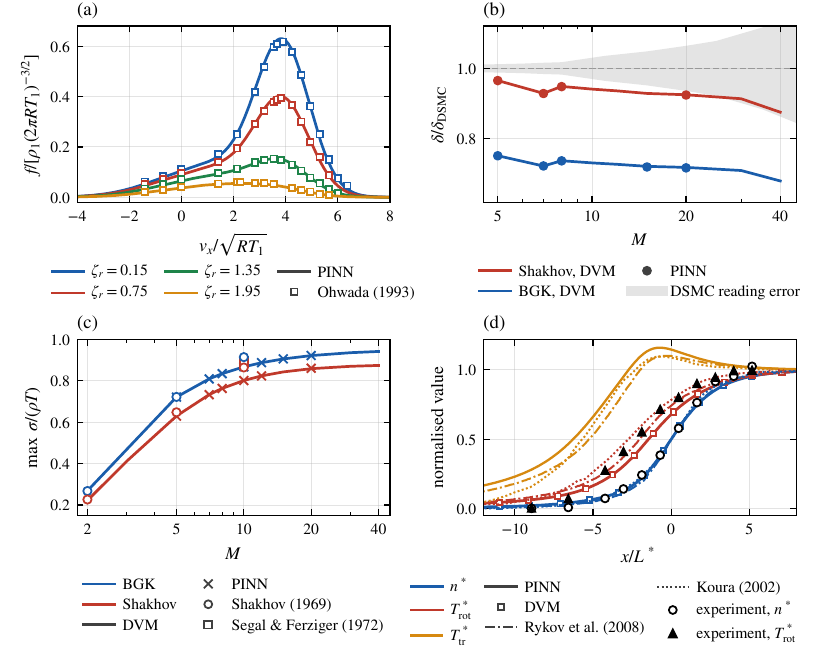}
\caption{(a) Hard spheres, Mach 3: $f$ at $x=-0.51\,\lambda_1$ for four radial speeds $\zeta_r$ (unit $\sqrt{2RT_1}$); squares: Ohwada \cite{Ohwada1993Shock}, Fig.~9. (b) $\tau=1/\rho$: thickness relative to the Maxwell-molecule DSMC of \cite{LumpkinChapman1992Burnett}, interpolated in $\ln M$; band: its reading error. (c) Maximum of $\sigma/(\rho T)$ against \cite{Shakhov1969Shock} (circles) and \cite{SegalFerziger1972Shock} (squares; Supplementary Section~\sref{supp:val_classical}). (d) Nitrogen at the Mach 7.0 conditions of \cite{RobbenTalbot1966Nitrogen} (symbols); dash-dot: R-model with $Z(T)$ \cite{RykovTitarevShakhov2008Diatomic}; dotted: trajectory DSMC \cite{Koura2002Rotational}. The Shakhov network at Mach 15, which misses test (a), is not drawn.}
\label{fig:validation}
\end{figure}

\paragraph{Hard spheres}
Against the tables of Ohwada's deterministic solution \cite{Ohwada1993Shock} at Mach 2 and 3 the networks differ as our DSMC does, by the rounding of the tables (\cref{sec:boltz_results}). The first Mach 3 run (\cref{sec:boltz_results}) reproduces Ohwada's $f$ to 0.006 of the upstream maximum, the reading error of his Fig.~9 (\cref{fig:validation}a).

\paragraph{Relaxation models}
Shakhov's solutions for $\mu\propto T$ \cite{Shakhov1969Shock} agree with ours within 2.9 per cent at Mach 2 and 5, but at Mach 10 his maxima of $\sigma/(\rho T)$ are 5.4 (BGK) and 7.8 per cent (Shakhov) higher (\cref{fig:validation}c); network and DVM agree within 0.04 per cent there, and the cause of the difference is not established. With $\tau=1/\rho$ the viscosity is that of Maxwell molecules, whose stress and heat flux the Shakhov model relaxes at the Boltzmann rates, but $R_{xx}$ and $\Delta$ at $p/\mu$ instead of $\tfrac76\,p/\mu$ and $\tfrac23\,p/\mu$ \cite{Struchtrup2005}. From Mach 5 to 20 the Shakhov shocks are 3.5 to 7.5 per cent thinner than the DSMC of Lumpkin and Chapman \cite{LumpkinChapman1992Burnett} (\cref{fig:validation}b), about the spread between DSMC studies (9.8 to $11.2\,\ell_0$ at Mach 8 \cite{Bird1970Strong,FisckoChapman1989Hypersonic,LumpkinChapman1992Burnett}), and their temperature overshoot, 1.4 to 1.7 per cent of $T_2-T_1$ above Mach 10, is two to seven times the digitised DSMC value of $0.5\pm0.3$ per cent at Mach 11 \cite{LumpkinChapman1992Burnett}. The BGK shocks are 25 to 28 per cent thinner, without overshoot.

\paragraph{Nitrogen}
In the length unit of each experiment (Supplementary Section~\sref{supp:val_n2}), and after a shift of 0.13 to $0.17\,L^*$ fitted to each profile, the density is reproduced: the r.m.s. difference in $n^*=(\rho-\rho_1)/(\rho_2-\rho_1)$ is 0.016 from the Mach 10 shock-tube profile of Alsmeyer \cite{Alsmeyer1976Density} and 0.03 to 0.04 from the free-jet shocks of Robben and Talbot \cite{RobbenTalbot1966Nitrogen} ($T_1\approx28$ and 9\,K), as for trajectory DSMC \cite{Koura2002Rotational} and the R-model with a temperature-dependent collision number \cite{RykovTitarevShakhov2008Diatomic}. The rotational rise is not: the rotational temperature leads the density by 1.44 and $1.6\,L^*$ at Mach 7 and 12 to 15, $L^*=\mu(T^*)/(\rho_1u_1)$ (\cref{fig:validation}d), against 2.31 and $2.51\,L^*$ from the measured populations at Mach 7.0 and 12.9, 2.4 to $2.9\,L^*$ in trajectory DSMC and 2.0 and $2.45\,L^*$ for the R-model. Since the measured lead depends on how it is extracted (1.9 to $2.7\,L^*$ at Mach 7), the model's lead is about 24 to 47 per cent short. This is attributed to the constant $Z$, a rotational collision number of 6.7 (\cref{sec:equations}), which molecular dynamics gives at 1000 to 2000\,K but not at the 100\,K or less where these shocks begin to relax (3.2 at 100\,K \cite{ValentiniZhangSchwartzentruber2012Rotational}); at $T_1=300$\,K the same model with $Z=2.226$ \cite{Wu2015Polyatomic} is closer to DSMC and molecular dynamics. Far upstream the model also heats the gas too early (\cref{sec:n2_weak}).

The networks thus reproduce independent solutions of their equations, and the differences from experiment are due to the models; runs of the nitrogen model with a temperature-dependent $Z$, which would test the attribution, were not made. Measured distribution functions \cite{HoltzMuntz1983Argon,PhamVanDiep1989Hypersonic} show the features of ours, but a quantitative test needs the experimental Mach numbers and gases (Supplementary Section~\sref{supp:val_vdf}).

\section{Discussion}\label{sec:discussion}

\subsection{Cost}\label{sec:disc_cost}

The cost is best compared by the work per collocation point, which does not depend on the hardware. A relaxation model needs about fifty velocity nodes and a target built from three or four moments. The Boltzmann residual on $48^3$ nodes (Mach 2 and 3) or $64^3$ nodes (Mach 5) needs two to five thousand times as many nodes and, for each, a sum over hundreds of directions, obtained from $2M_\theta M_\varphi+2$ transforms of the cube per collocation point: 258 at Mach 2, 578 at Mach 3 and 1026 at Mach 5; the two-scale nodes of Mach 8 and 10 have half a million nodes and 258 transforms. This is why the Boltzmann results stop at Mach 10 on a graphics card, while the relaxation models reach Mach 15 to 70 on one processor core. The wall-clock times of \cref{tab:disc_cost} are indicative only, because the hardware differs between lines (times per iteration: Supplementary Section~\sref{supp:disc_times}). The dense quasi-Newton method stores a square matrix of the size of the parameter vector (0.49~GB for the 7834 parameters of the monatomic shock network), one reason for keeping the networks small.

\begin{table}
\centering
\footnotesize
\caption{Recorded wall-clock times, indicative only. ``One core'' is one processor core of a cluster, ``laptop'' one thread of a laptop on which other jobs ran at the same time, and A100 one NVIDIA A100 graphics card. For the Boltzmann stage at Mach 5, which the cluster preempted nine times, it is the sum of its ten allocations (Supplementary Table~\sref{tab:boltz_M5_alloc}). The last three lines are the reference solvers; DVM times at Mach 20 and 50 and the DSMC time at Mach 5 were not recorded. Only the reported runs are listed: failed and development attempts (for example the Shakhov attempts at Mach 20 of \cref{tab:mono_m20} and the search for the nitrogen weights) and the Shakhov stage that precedes each Boltzmann stage are not included.}
\label{tab:disc_cost}
\begin{tabularx}{\textwidth}{@{}Lrll@{}}
\toprule
Calculation & Parameters & Hardware & Time\\
\midrule
BGK shock, Mach 5 from scratch & 7834 & one core & 2.1 h\\
BGK shock, each further Mach number to 20 & 7834 & one core & 0.9 to 1.3 h\\
BGK shock, Mach 5 to 20 in all & 7834 & one core & 8.6 h\\
BGK shock, Mach 20 to 50 in all & 7834 & one core & 5.2 h\\
BGK shock, Mach 20 to 70 with the weight \eqref{eq:al_rule}, evaluation included & 7834 & one core & about 14 h\\
BGK shock, Mach 70 to 100 with the weight \eqref{eq:al_rule}, evaluation included & 7834 & one core & 9.0 h\\
Shakhov shock, Mach 12 to 15; 15 to 20 (\cref{sec:mono_shk_limit}) & 7834 & one core & 3.2 h; 3.4 h\\
Shakhov shock, continuation from Mach 5 to 15 & 7834 & one core & 7.1 h\\
Shakhov shock, Mach 20, 800 + 4000 iterations, without the final stage & 7834 & one core & 4.6 h\\
Nitrogen shock, Mach 5 with all its stages & 7981 & laptop & 2.7 h\\
Nitrogen shock, Mach 8; Mach 12 & 7981 & laptop & 0.7 h; 0.9 h\\
Nitrogen shock, Mach 15 & 7981 & one core & 1 h\\
Boltzmann stage, Mach 2 & 7637 & A100 & 1.9 h\\
Boltzmann stages, Mach 3 & 7637 & A100 & 2.4 h and 6.9 h\\
Boltzmann stage, Mach 5 & 7637 & A100 & 29.7 h\\
Shakhov on two-scale nodes, Mach 8; Mach 10 & 15\,177 & A100 & 3.9 h; 2.9 h\\
Boltzmann stage, Mach 8 (800 iterations) & 15\,177 & A100 & 21.9 h\\
Boltzmann stage, Mach 10 (800 iterations, resumed after interruptions) & 15\,177 & A100 & not recorded\\
\midrule
DVM, Shakhov shock, Mach 3 & & one core & under 1 min per run\\
DVM, BGK shock, Mach 70, four grids to $\Delta x=0.025$ & & one core & 2.5 h\\
DSMC, hard spheres, Mach 3 & & 48 members & 47 core-hours\\
\bottomrule
\end{tabularx}
\end{table}

The networks cost more than the reference solvers: a DVM solution of a shock takes minutes, and hours at Mach 70, where the networks take hours. Speed is not the purpose of the study. It shows that a network trained on the equation alone, without data, reaches the accuracies of \cref{tab:mono_err,tab:mono_al,tab:boltz_results} for strong shocks, which elements it needs, and how the range of validity of each solution is found without a reference. What the differentiable, mesh-free representation offers is discussed in \cref{sec:disc_provides}; one use is shown here, a network with the Mach number as an input (\cref{sec:n2_param}).

\subsection{What the network solution provides}\label{sec:disc_provides}

A discrete-velocity solution also contains the distribution, as a table on a grid of positions and velocities, and a Monte Carlo simulation contains it as samples. What is specific to the network is the form in which the solution is held, and it has four consequences that the present results support.

The solution is a closed-form function of position and velocity, \eqref{eq:form}, defined by about eight thousand numbers for the reduced equations and 15\,177 for the Boltzmann equation (files of 67 and 129 kilobytes). It can be evaluated at any point and differentiated exactly, so every term of the kinetic equation, and hence its residual, can be computed at points that were not used to obtain it. Test (b) of \cref{sec:acceptance} does this at 1523 test points, where the residual norm is between 0.94 and 1.03 times its value at the training points in the runs of \cref{tab:mono_err,tab:mono_m50,tab:mono_al,tab:mono_shk} that pass the tests, and with the parameters, which are to be deposited with the data of this paper, it can be repeated at points of the reader's choice. A table can be checked in this way only through an interpolation; the reference of the Mach 70 shock, for comparison, is a table of 134\,001 positions and 2663 velocities for each of three distributions, about $10^9$ numbers.

Moments of any order can be formed on any velocity rule. Test (c) uses this: on a finer and wider rule the fields up to fourth order change by at most $3.1\times10^{-4}$ of their scales in the runs of the same tables and by at most $1.3\times10^{-5}$ for the Boltzmann stages at Mach 8 and 10.

The function has no statistical noise, which matters for the tails and the high moments of the Boltzmann equation. The DSMC references use 24 to 48 ensembles, with standard errors of order $10^{-4}$ of the scales of the fields at Mach 5 (\cref{sec:refs_dsmc}), and they give the distribution as a histogram; the network gives it smoothly over six decades (\cref{fig:f_validation}). The absence of noise is not the absence of error: the errors of the network are systematic, they carry no error bar, and two of them, the fourth-order moments of the relaxation models at high Mach numbers and the far-upstream temperature of the Boltzmann networks, were found only by comparison with a reference.

Finally, the form states the solution relative to a physical model: the factor $1+\chi\sinh\Nth$ is the departure of the distribution from a mixture of the two end-state Maxwellians, the ansatz of Mott-Smith \cite{MottSmith1951Shock}, at every position and velocity (\cref{fig:factormap}).

These properties come at the accuracy and cost stated in the tables. The differences of the network from the discrete-velocity solutions are $10^{-5}$ to a few per cent, where the monatomic solutions are converged to $4\times10^{-8}$ or better, and for hard spheres at Mach 5 its differences from the DSMC are one to two times the change of the DSMC when its cell and time step are halved; part of them, the precursor difference and the far-upstream excess of temperature, belongs to the network. The cost is higher: the step from Mach 50 to Mach 70 of the BGK continuation took 4.5 hours of one core after all the lower Mach numbers, against 2.5 hours for the discrete-velocity solution of the Mach 70 shock on its four grids, and the Boltzmann stage at Mach 8 took 21.9 hours of a graphics card after its Shakhov stage (\cref{tab:disc_cost}).

\subsection{Failure modes}\label{sec:disc_failures}

\begin{table}
\centering
\scriptsize
\setlength{\tabcolsep}{3pt}
\caption{Failure modes met in this work: symptom and case, the indicator that showed it without a reference, cause and remedy, and the section where it is described. The remedies were chosen with the reference available where \cref{tab:information} says so. A fuller account is in Supplementary Section~\sref{supp:failures}.}
\label{tab:failures}
\begin{tabularx}{\textwidth}{@{}>{\raggedright\arraybackslash}p{0.27\textwidth}>{\raggedright\arraybackslash}p{0.30\textwidth}L>{\raggedright\arraybackslash}p{0.06\textwidth}@{}}
\toprule
Symptom (case) & Indicator without a reference & Cause; remedy & Section \\
\midrule
Shock collapsed between two collocation points (nitrogen, Mach 7, 8, from scratch) & $e_F=0.028$ to 0.25; $\mathcal{L}_{\mathrm{rel}}$ at test points 500 times its training value & layer under-sampled; shock-layer points, continuation & \ref{sec:design_points} \\
Boundary layer at the inflow (Shakhov, two-scale nodes, Mach 10) & largest flux error at the inflow; $T-1=0.70$ five units from it & domain cuts off upstream-running molecules; $L=15M$ & \ref{sec:boltz_twoscale} \\
Adam far from the solution (nitrogen, Mach 8, 12) & $e_F=3.7\times10^{-2}$ & Adam for strong shocks; quasi-Newton continuation & \ref{sec:design_opt} \\
Failed line searches (Boltzmann, Mach 3, first run) & directional derivative five times smaller than predicted & quantity recomputed without its derivative; differentiate it & \ref{sec:boltz_results} \\
Stall of the tail term (nitrogen, Mach 5, 7) & tail term 90 per cent of the loss, held far upstream & few hot molecules dominate; $\lambda_T=0.1$ & \ref{sec:design_terms} \\
Passive distribution holding the loss (nitrogen, Mach 7) & $l$ holds 82 per cent of $\mathcal{L}_{\mathrm{rel}}$ & residual of $l$ over-sensitive; $a_l=(T_1/T_2)^2$ & \ref{sec:design_terms} \\
One-signed rotational-energy error (nitrogen, Mach 8) & moment $P_7$ of the residual & moments weakly weighted; final stage & \ref{sec:design_terms} \\
Flux error in the precursor (Shakhov, Mach 20) & $e_F=2.1\times10^{-3}$, 67 lengths ahead & quadrature at the beam; $\Delta v=0.25$ & \ref{sec:mono_m20} \\
Unfinished optimisation (Shakhov, Mach 20, B; passes the tests) & loss falling; stress and heat-flux balances off by $1.5\times10^{-2}$ & too few iterations; 4000 more & \ref{sec:mono_m20} \\
Loss floor of the operator (Boltzmann, Mach 5, uniform cube) & loss stalled, $e_F=4.9\times10^{-2}$; $Q(\M_1)\ne0$ & cube coarse for the beam, few directions; split beam, two-scale nodes & \ref{sec:boltz_uniform} \\
Fourth-order moments wrong behind the shock (BGK, Mach 50, 70; passes the tests) & $l$ holds 1.7 per cent of the part of $g$ in $\mathcal{L}_{\mathrm{rel}}$ at Mach 50 & weight of $l$ too small; \eqref{eq:al_rule} & \ref{sec:mono_m50} \\
Heat flux and thickness wrong (Shakhov, Mach 30 to 50; passes the tests) & heat-flux balance off by 0.03 to 0.22 & not found; none & \ref{sec:mono_shk_limit} \\
\bottomrule
\end{tabularx}
\end{table}

\Cref{tab:failures} groups the failures by cause. The collocation points, the domain and the velocity nodes have to follow the physical scales of the solution, and the residual at test points is the check. A quasi-Newton method needs the exact gradient of a fixed function. The weights of the loss decide where a stalled optimiser spends its effort; each stall was located by decomposing the loss by region, distribution and moment, without a reference, although the remedy was chosen with the reference available. And the velocity quadrature or the discrete collision operator has to resolve the cold beam. Every failure has an indicator that needs no reference; for the unfinished Shakhov run at Mach 20, for the fourth-order moments of the BGK model and for the Shakhov runs above Mach 20 it was read only after the comparison had shown the run to be wrong (\cref{sec:mono_m20,sec:mono_m50,sec:mono_shk_limit}), and for the last of these the cause is not known. Several remedies (the shock-layer collocation points, the length of the domain, the treatment of the beam) carry over to other kinetic solvers that minimise a residual.

\subsection{Limits and open questions}\label{sec:disc_limits}

The acceptance tests are necessary and not sufficient. One Shakhov run at Mach 20 passed them and was 46 per cent from the DVM in $\Delta$; the exact stress and heat-flux balances of the model, evaluated on its profiles, had residuals above $10^{-2}$, which suggests them as a further test. The Shakhov runs at Mach 30 to 50 are a second case: they pass the three tests with either weight of the passive distribution, are 4 to 18 per cent from the DVM in the heat flux (16 per cent in the blind test), and have heat-flux balance residuals of 0.03 to 0.22, against $4.1\times10^{-3}$ or less in the second BGK continuation, with the weight \eqref{eq:al_rule}, to Mach 100 and $1.7\times10^{-3}$ or less in the Shakhov runs at Mach 15 and 20; the Shakhov model above Mach 20 is a limit of the present method whose cause was not found (\cref{sec:mono_shk_limit}). Much of the method was developed with a reference available (\cref{tab:information}), so the results at Mach 5 to 8 for nitrogen are not predictions. The independent checks are the nitrogen tests at Mach 10 and 15 (\cref{sec:n2_blind}) and the blind BGK predictions at Mach 50, 60 and, with the first weight, 100 (\cref{sec:mono_m50}); the BGK results with the weight \eqref{eq:al_rule} to Mach 100, including the accurate Mach 100 solution, and the Boltzmann result at Mach 10 are not blind.

The fourth-order moments are the least certain fields. Most results are single runs; changes of the base that pass the tests moved $R_{xx}$ and $\Delta$ by 5 to 80 per cent at Mach 20 (\cref{sec:mono_m20}), with a constant collision time the same procedure passes its tests and is 18 and 29 per cent off in $R_{xx}$ and $\Delta$ at Mach 20 (\cref{sec:mono_tau1}), and at Mach 50, 70 and 100 they are 14 and 21, 39 and 60, and 53 and 81 per cent off with the weight $a_l=(T_1/T_2)^2$ (\cref{sec:mono_m50}). With that weight the fourth-order entries at Mach 12 and above are uncertain by up to a factor of about two; a second seed at Mach 20 differed by a factor 1.8. For the BGK model the cause is the small weight of the passive distribution: with the weight \eqref{eq:al_rule} and 2100 instead of 1100 iterations per step $R_{xx}$ and $\Delta$ are within 3.2 per cent up to Mach 70 and 7.1 per cent at Mach 100, in comparisons that are not blind and with one seed; the two changes were separated at Mach 20 and, in a blind test with one seed, at Mach 60 (\cref{sec:mono_m50}). The rule was not applied to the Shakhov runs at Mach 20 and below or to nitrogen, whose fourth-order entries keep the earlier weight. Negative values of $f$ were measured for the BGK network at Mach 20 (0.89 per cent of the nodes, \cref{sec:design_form}) and, after the fact, for the Boltzmann networks up to Mach 8, in which a negative $f$ enters $Q(f,f)$: the most negative value is $7\times10^{-8}$ of the largest $f$ at the same position and the negative part at most $3\times10^{-6}$ of the local density (Supplementary Section~\sref{supp:boltz_M8M10}). For the Mach 10 network, evaluated on a grid in $(v_x,v_r)$, the two numbers are $9.5\times10^{-8}$ and $3.0\times10^{-6}$. The weak rise of the temperature far upstream is not resolved in relative terms (\cref{sec:n2_weak}). The models have limits of their own, which the networks inherit: with a constant collision number the nitrogen model places the rotational rise of cold shocks about a third too close to the density, and the Shakhov model overpredicts the temperature overshoot of Maxwell molecules two- to sevenfold (\cref{sec:validation}).

For the Boltzmann equation the results end at Mach 10 (\cref{sec:boltz_M8M10}). The networks from Mach 5 on keep a far-upstream excess of temperature of $10^{-3}$ to $3.5\times10^{-3}\,T_1$ that the DSMC does not have; six mean free paths ahead of the density midpoint their share of molecules moving upstream is 1 per cent (Mach 8), 13 per cent (Mach 10) and more than a quarter (Mach 5) below that of the DSMC (\cref{sec:boltz_f}); and the remaining points on the discrete operator are listed in Supplementary Section~\sref{supp:disc_limits}. A positive form $\Bmix\exp(\chi\Nth)$, tested for the BGK model at Mach 20 with a term that divides the residual by $|f|$, passed the tests and has no negative values, but with the same number of iterations on the equation, after a fit of its initial factor, it was less accurate than the signed form (4.3 and 6.6 per cent in $R_{xx}$ and $\Delta$ against 3.8 and 5.8, and 0.14 against 0.090 per cent in $m_{xxx}$), so the signed form is kept. Finally, the shocks are one-dimensional in space, and the form rests on analytic knowledge of where the molecules come from, the two end states. Wall-bounded flows, in which the molecules come from walls seen along straight rays, are outside this paper and will be reported separately. A flow without such structure, an unsteady flow, or a geometry in which the wall of origin is not explicit is outside what has been shown.

\section{Conclusions}\label{sec:conclusions}

A neural network trained by minimising a loss that contains only the kinetic equation, its analytic boundary data and conservation statements derived from it can solve strong normal shock waves, and its solutions remain accurate beyond the bulk fields $\rho$, $u$ and $T$. The bulk fields alone do not characterise such a solution: at Mach 20 they are within 0.1 per cent of the discrete-velocity solutions while the fourth-order moments differ by several per cent. The third- and fourth-order moments therefore belong in the assessment of any kinetic solver of this kind.

The form of the distribution should carry what is known analytically: the end states and the exact inflow. The correction left to the network is here signed and unbounded; a bounded correction, tested with the same procedure to Mach 12, was also accepted but gave less accurate fourth-order moments, so that this choice is supported but not shown to be necessary. The loss has to reach the sparse hot molecules that carry the higher moments, and in the one ablation made the constancy of the fluxes was the term whose removal destroyed the solution. The collocation points have to resolve the shock layer; otherwise the shock collapses between two points, which the reference-free tests reveal. In these calculations the accuracy was reached in the quasi-Newton stage and by continuation in the Mach number, not by Adam.

With these elements the network alone, on a single processor core and with fewer than eight thousand parameters, solved the BGK shock to Mach 100. At Mach 50 a prediction saved before the reference was read gave the bulk fields within 0.1 per cent and the fourth-order moments within about 20 per cent; this error was traced to the weight of the passive reduced distribution in the loss, and with a weight inversely proportional to the downstream temperature, tested at Mach 20, and 2100 instead of 1100 iterations per step the fourth-order moments are within 3.2 per cent up to Mach 70 and 7.1 per cent at Mach 100, in comparisons that are not blind; a blind test at Mach 60 attributes about 60 per cent of the gain to the weight and 40 per cent to the longer optimisation. It solved the Shakhov shock to Mach 20 with settings fixed in advance. Above Mach 20 the Shakhov solutions leave their range of validity although the three acceptance tests are passed (heat flux 4 to 18 per cent from the reference at Mach 30 to 50, 16 per cent in the blind test at Mach 50); the residual of the exact heat-flux balance, which needs no reference, marks this boundary: it is twelve times or more its largest value in the accurate runs. The nitrogen model was solved to Mach 15; at Mach 10, in a test fixed in advance, half of the fields fell outside the error range expected from Mach 8 and 12. For the Boltzmann equation of hard spheres the solutions at Mach 2, 5, 8 and 10 pass all tests, and at Mach 5, 8 and 10 each of the eight reported fields, which include two fourth-order moments, is within 0.4 per cent of the DSMC in the relative $L^2$ norm (0.19 per cent at Mach 8, compared once; the Mach 10 comparison was not blind). Stronger shocks are limited by the cost of the collision integral in three velocity dimensions.

These solvers take hours where a discrete-velocity code takes minutes to hours. Their interest lies in a differentiable, mesh-free representation of the distribution: a closed-form function whose residual can be evaluated at any point and whose moments can be formed on any velocity rule (\cref{sec:disc_provides}), and of which this paper gives one further use, a network with the Mach number as an input, at a cost in accuracy. Several weights of the loss were chosen with a reference available (\cref{tab:information}). The Shakhov model above Mach 20, the fourth-order moments of the Shakhov and nitrogen models, the far-upstream temperature rise, the Boltzmann equation at high Mach numbers and multidimensional flows remain open (\cref{sec:discussion}).

\section*{Declaration of competing interest}
The author declares no competing financial interests or personal relationships that could have appeared to influence the work reported in this paper.

\section*{Data availability}
The data that support the findings of this study are available from the author upon reasonable request. The trained networks, the reference solutions and the scripts that produce the figures will be made public in a GitHub repository after acceptance of the article.

\section*{Acknowledgements}
The computations were performed on the Unity Research Computing Platform; the author thanks the Unity team for the computing resources. The author thanks the Department of Mechanical and Industrial Engineering of the University of Massachusetts Amherst for its financial support.

\section*{Declaration of generative AI and AI-assisted technologies in the manuscript preparation process}
During the preparation of this work the author used Claude (Anthropic) to assist in developing and checking the computer codes and in preparing figures and text. After using this tool, the author reviewed and edited the content as needed and takes full responsibility for the content of the published article.

\section*{Supplementary material}
Supplementary material associated with this article contains five videos and further figures and tables of the shock calculations, an account of the failures, the settings and times of every run, the discrete-velocity reference solvers, the evaluation of the hard-sphere collision integral with the Monte Carlo references, and the data, units and tables of the comparison with published solutions and measurements. Supplementary Videos 1, 3 and 5 animate the data of \cref{fig:molecules}, \cref{fig:molecules_models} and \cref{fig:molecules_M50}, Video 2 those of Supplementary Fig.~\sref{fig:S_molecules_M5}, and Video 4 shows the nitrogen shock at Mach 5; they are described in Supplementary Section~\sref{supp:video}; public copies are at \url{https://www.youtube.com/watch?v=9hSQn6-cPD8} (Video 1, BGK shock at Mach 20, slow-motion version), \url{https://www.youtube.com/watch?v=Zow8LDTRQ5M} (Video 2, three collision models at Mach 5), \url{https://www.youtube.com/watch?v=pMRke3ZgKbI} (Video 3, three collision models at Mach 8), \url{https://www.youtube.com/watch?v=VJcGbaIG-d0} (Video 4, nitrogen at Mach 5) and \url{https://www.youtube.com/watch?v=_u03dhZOFhA} (Video 5, BGK shock at Mach 50, discrete-velocity solution and network together).

\appendix
\section{Reduced equations and moments}\label{app:reduced}

This appendix gives the targets of the reduced equations \eqref{eq:reduced_kinetic} and the moment formulas as they are implemented in the network codes and in the reference solvers. Throughout, $c=v-u$ and
\begin{equation}
  M_1(\Theta)=\frac{\rho}{\sqrt{2\pi\Theta}}\exp\Bigl(-\frac{c^2}{2\Theta}\Bigr)
  \label{eq:m1}
\end{equation}
is the Maxwellian of density $\rho$, velocity $u$ and temperature $\Theta$ integrated over the transverse plane. The moments of the transverse Maxwellian used below are $\langle r^2\rangle=2\Theta$, $\langle r^4\rangle=8\Theta^2$ and $\langle r^6\rangle=48\Theta^3$.

\paragraph{BGK and Shakhov models}
With $a=(1-\Pran)\,q\,c/(5\rho T^3)$, integration of \eqref{eq:shakhov_target} gives the targets of $g$, $k$ and $l$,
\begin{equation}
\begin{gathered}
  g^*=M_1(T)\bigl[1+a\,(c^2-3T)\bigr],\qquad
  k^*=2T\,M_1(T)\bigl[1+a\,(c^2-T)\bigr],\\
  l^*=8T^2M_1(T)\bigl[1+a\,(c^2+T)\bigr],
\end{gathered}
\label{eq:targets_mono}
\end{equation}
and $\tau=T^{\omega-1}/\rho$ with $\omega=1$. $\Pran=1$ gives the BGK model.

\paragraph{Nitrogen model}
With $M_t=M_1(T_{\mathrm{tr}})$, $M_r=M_1(T)$, $T=(3T_{\mathrm{tr}}+2T_{\mathrm{rot}})/5$, $a_t=q_{\mathrm{tr}}\,c/(15\rho T_{\mathrm{tr}}^3)$ and $a_r=\omega_0\,q_{\mathrm{tr}}\,c/(15\rho T^3)$, integration of \eqref{eq:rykov_targets} gives
\begin{equation}
\begin{aligned}
  g_t&=M_t\bigl[1+a_t(c^2-3T_{\mathrm{tr}})\bigr], & g_r&=M_r\bigl[1+a_r(c^2-3T)\bigr],\\
  k_t&=2T_{\mathrm{tr}}M_t\bigl[1+a_t(c^2-T_{\mathrm{tr}})\bigr], & k_r&=2T\,M_r\bigl[1+a_r(c^2-T)\bigr],\\
  l_t&=8T_{\mathrm{tr}}^2M_t\bigl[1+a_t(c^2+T_{\mathrm{tr}})\bigr], & l_r&=8T^2M_r\bigl[1+a_r(c^2+T)\bigr],\\
  h_t&=T_{\mathrm{rot}}\,g_t+(1-\delta_{\mathrm{R}})\frac{q_{\mathrm{rot}}\,c}{\rho T_{\mathrm{tr}}}M_t, &
  h_r&=T\,g_r+\omega_1(1-\delta_{\mathrm{R}})\frac{q_{\mathrm{rot}}\,c}{\rho T}M_r ,
\end{aligned}
\label{eq:targets_n2}
\end{equation}
and the target of each reduced distribution is
\begin{equation}
  \varphi^*=\varphi_t+\frac{\varphi_r-\varphi_t}{Z},\qquad \varphi\in\{g,k,h,l\},\qquad \tau=\frac{T_{\mathrm{tr}}^{\omega-1}}{\rho}.
  \label{eq:targets_n2_sum}
\end{equation}

\paragraph{End states and envelope}
The equilibrium distributions of a state $(\rho,u,T)$, which enter the base \eqref{eq:mixture} for the two Rankine--Hugoniot states, are $(g,k,h,l)=M_1(T)\,(1,\allowbreak\ 2T,\allowbreak\ T,\allowbreak\ 8T^2)$. The local part of the envelope \eqref{eq:envelope} is $M^{\mathrm{loc}}=M_1(T_{\mathrm{tr}})\,(1,\allowbreak\ 2T_{\mathrm{tr}},\allowbreak\ T_{\mathrm{rot}},\allowbreak\ 8T_{\mathrm{tr}}^2)$ with the current $\rho$, $u$, $T_{\mathrm{tr}}$ and $T_{\mathrm{rot}}$ of the network. For the monatomic gas $h$ is absent and $T_{\mathrm{tr}}=T$.

\paragraph{Moments}
All integrals over $v$ are evaluated with the velocity rule of \cref{sec:points}. With $e=c^2g+k$,
\begin{equation}
\begin{gathered}
  \rho=\int g\,\dd v,\qquad \rho u=\int v\,g\,\dd v,\qquad 3\rho T_{\mathrm{tr}}=\int e\,\dd v,\qquad \rho T_{\mathrm{rot}}=\int h\,\dd v,\\
  q_{\mathrm{tr}}=\frac12\int c\,e\,\dd v,\qquad q_{\mathrm{rot}}=\int c\,h\,\dd v,\qquad
  \sigma=\int\Bigl(c^2g-\frac{e}{3}\Bigr)\dd v,
\end{gathered}
\label{eq:reduced_moments}
\end{equation}
and the third- and fourth-order moments \eqref{eq:higher_moments} are
\begin{equation}
\begin{gathered}
  m_{xxx}=\int c^3g\,\dd v-\frac65\,q_{\mathrm{tr}},\qquad
  \Delta=\int\bigl(c^4g+2c^2k+l\bigr)\,\dd v-15\rho T_{\mathrm{tr}}^2,\\
  R_{xx}=\int\Bigl(\frac23c^4g+\frac13c^2k-\frac13l\Bigr)\dd v-7T_{\mathrm{tr}}\sigma .
\end{gathered}
\label{eq:reduced_higher}
\end{equation}
The fluxes of mass, momentum and energy in \eqref{eq:lflux} and their upstream values are
\begin{equation}
\begin{gathered}
  J_1=\int v\,g\,\dd v,\qquad J_2=\int v^2g\,\dd v,\qquad J_3=\int v\Bigl[\frac12\bigl(v^2g+k\bigr)+h\Bigr]\dd v,\\
  \bigl(J^{\mathrm{RH}}_1,J^{\mathrm{RH}}_2,J^{\mathrm{RH}}_3\bigr)=\Bigl(u_1,\ u_1^2+1,\ u_1\Bigl[\frac{u_1^2}{2}+\frac{\gamma}{\gamma-1}\Bigr]\Bigr),
\end{gathered}
\label{eq:fluxes}
\end{equation}
with $\gamma/(\gamma-1)=5/2$ for the monatomic gas and $7/2$ for nitrogen.

\paragraph{Moments of the residual}
Let $\Res_g$, $\Res_k$, $\Res_h$ be the residuals \eqref{eq:residual} of $g$, $k$, $h$ and $e_{\Res}=c^2\Res_g+\Res_k$. The quantities $P_k$ of \eqref{eq:lmom} are
\begin{equation}
\begin{gathered}
  P_1=\frac{\int\Res_g\,\dd v}{\rho_2T_2^{1/2}},\qquad
  P_2=\frac{\int v\,\Res_g\,\dd v}{\rho_2T_2^{1/2}u_1},\qquad
  P_3=\frac{\int\bigl[\frac12(v^2\Res_g+\Res_k)+\Res_h\bigr]\dd v}{\rho_2T_2^{1/2}u_1^2},\\
  P_4=\frac{\int\bigl(c^2\Res_g-e_{\Res}/3\bigr)\,\dd v}{\rho_2T_2^{3/2}},\qquad
  P_5=\frac{\int c\,e_{\Res}\,\dd v}{2\rho_2T_2^{2}},\\
  P_6=\frac{\int c\,\Res_h\,\dd v}{\rho_2T_2^{2}},\qquad
  P_7=\frac{\int\Res_h\,\dd v}{\rho_2T_2^{3/2}} .
\end{gathered}
\label{eq:projections}
\end{equation}
$P_1$, $P_2$ and $P_3$ are the balances of mass, momentum and energy, that is the derivatives of the three fluxes; $P_4$ and $P_5$ are the balances of the stress and of the translational heat flux, and $P_6$ and $P_7$, present only for nitrogen, those of the rotational heat flux and of the rotational energy, $\dd(\rho uT_{\mathrm{rot}}+q_{\mathrm{rot}})/\dd x=\rho\,(T-T_{\mathrm{rot}})/(Z\tau)$. For the monatomic gas $\Res_h=0$ and $k=1,\dots,5$. The peculiar velocity $c$ in \eqref{eq:projections} uses the current velocity of the network, and the gradient of the loss is taken through it. The checks of this implementation are listed in Supplementary Section~\sref{supp:reduced_checks}.

\section{Implementation and settings}\label{app:settings}

All network codes use double precision (64-bit floating point) on the processor and on the graphics card and are written in Python with PyTorch. The calculations on the cluster used Python 3.12.11, PyTorch 2.10.0 built for CUDA 12.8, NumPy 2.5.2 and SciPy 1.18.1, and the stages on a graphics card ran on NVIDIA A100 cards; the figures were drawn with Matplotlib 3.11.2. The update of the self-scaled Broyden method with its safeguards and line search, the Adam and Levenberg--Marquardt stages and the loss on three-dimensional velocity nodes are written out in Supplementary Sections~\sref{app:optimisers} to \sref{supp:threed}. The settings of every run are given in Supplementary Sections~\sref{app:settings_shock} and \sref{app:settings_boltz}, and the settings and wall-clock times of every stage in Supplementary Tables~\sref{tab:set_mono} to \sref{tab:B_boltz}.

\section{Units and collision frequency of the hard-sphere calculations}\label{app:collision}

\label{app:collision_units}The upstream density, the upstream temperature and the gas constant are one, velocities are measured in $\sqrt{RT_1}$ and lengths in the upstream mean free path $\lambda_1=1/(\sqrt2\,\pi d^2n_1)$, with $d$ the molecular diameter and $n_1$ the upstream number density. The steady equation in one space dimension is $v\,\partial f/\partial x=Q(f,f)$ with the collision integral \eqref{eq:boltzmann}, $\bv=(v,v_y,v_z)$. Scattering of hard spheres is isotropic with the total cross-section $\pi d^2$, and the factor $1/\sqrt2$ of \eqref{eq:boltzmann} is $\pi d^2n_1\lambda_1$. The integral is the difference of a gain and a loss, $Q=Q^+-f\,\nu[f]$, with the collision frequency
\begin{equation}
  \nu[f](\bv)=\frac{1}{\sqrt2}\int_{\Rset^3}|\bv-\bm{w}|\,f(\bm{w})\,\dd\bm{w}.
  \label{eq:app_nu}
\end{equation}
With the hard-sphere viscosity of \cref{sec:units} the length unit of the relaxation models is $\ell_0=0.795882\,\lambda_1$. A solution of a relaxation model with $\tau=1/\rho$ that starts a Boltzmann stage is mapped to the hard-sphere units by this factor: the position seen by the network and the width $w$ of the base \eqref{eq:mixture} are rescaled, so that the stage starts from the same profile expressed in mean free paths.

The integral is written in the form of Carleman and evaluated on a periodic cube of $N_v^3$ nodes by the fast spectral method \cite{MouhotPareschi2006Fast}: for each of $M_\theta\times M_\varphi$ directions on a half sphere, a line integral and a plane integral of $f$ are obtained by inverse Fourier transforms of the cube and multiplied at the nodes. No conservation correction is applied; conservation is imposed by the loss. The gradient with respect to the parameters is obtained by reverse-mode automatic differentiation through all steps. At Mach 3 and 5 the cold beam is split off as the upstream Maxwellian times a polynomial taken from the network, and its collisions are evaluated in closed form or on a fine auxiliary cube; on the two-scale nodes the gain of the collisions between the beam and the hot gas is written in Radon form. The DSMC references use the domains $-30\le x\le30$, $-30\le x\le25$ and $-30\le x\le20$ at Mach 2, 3 and 5, and $-30\le x\le20$ at Mach 8 and 10. The discrete operator, its tests and the DSMC references are described in Supplementary Section~\sref{supp:collision_supp}.

\bibliographystyle{elsarticle-num}
\bibliography{references}

\end{document}